\documentclass[trackchanges,twocolumn]{aastex7}

\usepackage{amsmath}

\graphicspath{{./}{figures/}}

\begin{document}

\title{MIAO-ALMA: Shocks and Protostellar Outflows in 70~$\mu$m-dark clumps with $L/M$ $<$ 1~$L_{\odot}$/$M_{\odot}$}

\author[0000-0001-5461-1905]{Shuting~Lin}
\affiliation{Department of Astronomy, Xiamen University, Zengcuo'an West Road, Xiamen, 361005}
\email{linst@stu.xmu.edu.cn}

\author[0000-0002-4707-8409]{Siyi~Feng}
\affiliation{Department of Astronomy, Xiamen University, Zengcuo'an West Road, Xiamen, 361005}
\email[show]{syfeng@xmu.edu.cn}

\author[0000-0001-6106-1171]{Junzhi~Wang}
\affiliation{School of Physical Science and Technology, Guangxi University, Nanning 530004}
\email{junzhiwang@gxu.edu.cn}

\author[0000-0003-1275-5251]{Shanghuo Li}
\affiliation{School of Astronomy and Space Science, Nanjing University, Nanjing 210093}
\affiliation{Key Laboratory of Modern Astronomy and Astrophysics (Nanjing University), Ministry of Education, Nanjing 210093}
\email{shanghuo.li@gmail.com}

\author[0000-0001-9798-9852]{Dan~Miao}
\affiliation{Department of Astronomy, Xiamen University, Zengcuo'an West Road, Xiamen, 361005}
\email{miaodan@stu.xmu.edu.cn}

\author[0000-0002-7299-2876]{Zhi-Yu~Zhang}
\affiliation{School of Astronomy and Space Science, Nanjing University, Nanjing 210093}
\affiliation{Key Laboratory of Modern Astronomy and Astrophysics (Nanjing University), Ministry of Education, Nanjing 210093}
\email{zzhang@nju.edu.cn}

\author[0000-0003-4603-7119]{Sheng-Yuan~Liu}
\affiliation{Institute of Astronomy and Astrophysics, Academia Sinica, 11F of Astronomy-Mathematics Building, AS/NTU No. 1, Section 4, Roosevelt Road}
\email{syliu@asiaa.sinica.edu.tw}

\author[0000-0002-3297-4497]{Nami Sakai}
\affiliation{RIKEN Cluster for Pioneering Research, 2-1 Hirosawa, Wako-shi, Saitama 351-0198, Japan}
\email{nami.sakai@riken.jp}

\author[0000-0001-5950-1932]{Fengwei Xu}
\affiliation{Max Planck Institute for Astronomy, K\"onigstuhl 17, 69117 Heidelberg, Germany}
\email[]{fengwei@mpia.de, fengweilookuper@gmail.com}

\author[0000-0003-2300-2626]{Hauyu~Baobab~Liu}
\affiliation{Department of Physics, National Sun Yat-Sen University, No. 70, Lien-Hai Road, Kaohsiung City 80424}
\affiliation{Center of Astronomy and Gravitation, National Taiwan Normal University, Taipei 116}
\email{hyliu.nsysu@mail.nsysu.edu.tw}

\author[0000-0002-1700-090X]{Henrik Beuther}
\affiliation{Max Planck Institute for Astronomy, K\"onigstuhl 17, 69117 Heidelberg, Germany}
\email{beuther@mpia.de}

\author[0000-0003-3010-7661]{Di Li}
\affiliation{New Cornerstone Science Laboratory, Department of Astronomy, Tsinghua University, Beijing, China}
\email{dili@tsinghua.edu.cn}

\author[0000-0003-2384-6589]{Qizhou Zhang}
\affiliation{Center for Astrophysics $\vert$ Harvard \& Smithsonian, 60 Garden Street, Cambridge, MA, 02138, USA}
\email{qzhang@cfa.harvard.edu}

\author[0000-0002-5286-2564]{Tie Liu}
\affiliation{Shanghai Astronomical Observatory, Chinese Academy of Sciences, 80 Nandan Road, Shanghai 200030, Peoples Republic of China}
\email{liutie@shao.ac.cn}

\author[0000-0002-7125-7685]{Patricio~Sanhueza}
\affiliation{Department of Astronomy, School of Science, The University of Tokyo, 7-3-1 Hongo, Bunkyo, Tokyo 113-0033, Japan}
\email{patosanhueza@gmail.com}
           
\author[0000-0002-9148-1625]{Olli Sipil\"a}
\affiliation{Max-Planck-Institut f\"ur Extraterrestrische Physik, Giessenbachstrasse 1, 85748 Garching, Germany}
\email{osipila@mpe.mpg.de}

\author[0000-0002-8149-8546]{Ken'ichi Tatematsu}
\affiliation{Nobeyama Radio Observatory, National Astronomical Observatory of Japan, National Institutes of Natural Sciences, 462-2 Nobeyama, Minamimaki, Minamisaku, Nagano 384-1305, Japan}
\affiliation{Astronomical Science Program, Graduate Institute for Advanced Studies, SOKENDAI, 2-21-1 Osawa, Mitaka, Tokyo 181-8588, Japan}
\email{kenichi.tatematsu@nifty.com}

\author[0000-0002-3972-1978]{Jaime E. Pineda}
\email{jpineda@mpe.mpg.de}
\affiliation{Max Planck Institute for Extraterrestrial Physics, Gie{\ss}enbachstra{\ss}se 1, D-85748 Garching bei M{\"u}nchen, Germany}

\author[0000-0003-2619-9305]{Xing Lu}
\email{xinglu@shao.ac.cn}
\affiliation{Shanghai Astronomical Observatory, Chinese Academy of Sciences, 80 Nandan Road, Shanghai, 200030}
\affiliation{State Key Laboratory of Radio Astronomy and Technology, A20 Datun Road, Chaoyang District, Beijing, 100101}

\linenumbers

\begin{abstract}
To investigate the initial conditions of high-mass star-forming regions, we use SiO~(2–1) emission to trace early shock-related kinematics toward sixteen 70~$\mu$m-dark and massive clumps with luminosity-to-mass ratios ($L/M$) $< 1\,L_{\odot}/M_{\odot}$, as part of the Multiwavelength Line-Imaging Survey of the 70 $\mu$m-dark and bright clouds (MIAO) project. Using ALMA observations at a spatial resolution of $\sim$0.06 pc and a velocity resolution of 0.21 km s$^{-1}$, we identify a total of thirty-seven outflows with a variety of morphologies. 
Outflow parameters were derived by integrating the HCO$^+$~(1–0) line wings, excluding the quiescent dense core component traced by H$^{13}$CO$^+$~(1–0).
We find that outflow masses and velocities show moderate positive correlations with the masses of their driving cores.
Owing to the high sensitivity of our observations, which yield longer projected outflow lengths compared to previous studies, the derived outflow dynamical ages span $\sim10^{3}$–$10^{5}$~yr.
We detect six narrow-linewidth (0.6–1.4 km s$^{-1}$) and three broad ($>$ 2 km s$^{-1}$) SiO~(2-1) features not associated with outflows driven by clearly identified protostars. Lacking coincident 3 mm dust continuum cores, their origins may be young  outflows from undetected low-mass protostars, dissipating shocks, cloud–cloud collisions, or projection effects when the outflows lie close to the plane of the sky.
The detection of these shocks and outflows in such extremely young environments demonstrates that protostellar activity has already begun.

\end{abstract}




\section{Introduction}
\label{sec1:intro}
Interstellar shocks on intra-Galactic (sub-parsec) scales can arise from various processes, including protostellar outflows and jets \citep[e.g.,][]{Schilke1997A&A...321..293S,Lee2007ApJ...670.1188L,Lefloch2012ApJ...757L..25L,Frank2014prpl.conf..451F,Bally2016ARA&A..54..491B}, stellar/disk winds \citep[e.g.,][]{Snell1980ApJ...239L..17S,Shu1994ApJ...429..781S,
Hirota2017NatAs...1E.146H}, cloud–cloud collisions (CCC) \citep{Fukui2014ApJ...780...36F,Wu2017ApJ...835..137W,Cosentino2020MNRAS.499.1666C}, or turbulent dissipation \citep[e.g.,][]{Godard2009A&A...495..847G,Lesaffre2013A&A...550A.106L}. 
During shock processes, the gas and dust temperatures and densities can increase by several orders of magnitude, while the post-shock gas cools down to temperatures of several tens to a few hundred Kelvin, which are higher than those of the quiescent interstellar medium
\citep[e.g.,][]{Draine1983ApJ...264..485D,Gusdorf2008A&A...482..809G,Flower2010MNRAS.406.1745F,Lefloch2012ApJ...757L..25L}.
Because silicon is largely locked in the interiors of dust grains, SiO is efficiently produced and released into the gas phase through the energetic processes described above \citep[e.g.,][]{Martin-Pintado1992A&A...254..315M,Mikami1992ApJ...392L..87M,Caselli1997A&A...322..296C,Schilke1997A&A...321..293S}. Consequently, SiO is widely used as a tracer of shocks, and its line profiles provide valuable diagnostics of such energetic processes.

Protostellar outflows and jets are key mechanisms for angular momentum transport during the star formation process \citep{Shu1987ARA&A..25...23S,Frank2014prpl.conf..451F}, and they also serve as evolutionary diagnostics for identifying regions in which protostars have already formed
\citep[e.g.,][]{Beuther2002A&A...383..892B,Zhang2005ApJ...625..864Z, Arce2007prpl.conf..245A,
Qiu2008ApJ...685.1005Q,Feng2016ApJ...828..100F,
Li2019ApJ...878...29L,Towner2024ApJ...960...48T,Xu2024ApJS..270....9X,Beuther2025ARA&A..63....1B}. 
Understanding the initial stages of star formation, particularly in high-mass star-forming regions, remains challenging due to their short evolutionary timescales, large distances, high extinction, and clustered mode of star formation, which together hinder detailed observational diagnostics \citep{Zinnecker2007ARA&A..45..481Z,Motte2018ARA&A..56...41M,Beuther2025ARA&A..63....1B}.

Previous outflow detections have mainly been reported in relatively evolved star-forming regions, where the host clumps are already luminous ($\gtrsim$ $10^3\,L_\odot$) and the associated molecular outflows are strong and extended \citep[e.g.,][]{Zhang2005ApJ...625..864Z,Arce2007prpl.conf..245A,Maud2015MNRAS.453..645M}.
In such regions, outflows commonly exhibit well-defined morphologies and kinematic signatures, including bipolar \citep[e.g.,][]{Snell1980ApJ...239L..17S, Wang2011ApJ...735...64W, PPVI2014prpl.conf.....B, Kong2019ApJ...874..104K}, monopolar \citep[e.g.,][]{FernandezLopez2013ApJ...778...72F, Jhan2022ApJ...931L...5J}, and multipolar \citep[e.g.,][]{Makin2018ApJS..234....8M} structures, as well as explosive outflows \citep[e.g.,][]{Zapata2011ApJ...726L..12Z, Zapata2019MNRAS.486L..15Z,Zapata2020ApJ...902L..47Z,Fernandez2021ApJ...913...29F, Issac2025AJ....169..324I}, some of which are accompanied by ``Hubble-like" structure, indicating a linear increase of velocity with distance from the driving source \citep[e.g.,][]{Arce2001ApJ...551L.171A,Cheng2019ApJ...877..112C,Nony2020A&A...636A..38N,Li2020ApJ...903..119L,
Morii2021ApJ...923..147M,Tafoya2021ApJ...913..131T}.
However, it remains unclear whether such morphological and kinematic properties differ at the earliest stages of massive star formation.
Although an increasing number of extremely young outflows have been reported in dense and early-stage environments \citep[e.g.,][]{Pillai2019A&A...622A..54P, Li2019ApJ...886..130L, Li2020ApJ...903..119L, Busch2020A&A...633A.126B, Liu2021ApJ...921...96L,Tafoya2021ApJ...913..131T, Towner2024ApJ...960...48T,Izumi2024ApJ...963..163I}, outflows at the earliest evolutionary stages remain poorly characterized.

In 2016, \citet[][]{Feng2016ApJ...828..100F} detected a bipolar outflow with a dynamical timescale of $\sim$10$^4$ years in a canonical ‘starless core candidate’ towards the 70 $\mu$m-dark clump G28.34+0.06, with luminosity-to-mass ratios ($L/M$) $<$ 1 $L_{\odot}/M_{\odot}$ and high deuteration \citep{Feng2019ApJ...883..202F}, marking it as a region pushing the limits of young high-mass star formation, with known signatures of star formation at the earliest evolutionary stage.

\setcounter{footnote}{0} 

To test whether such extremely young outflows are unique to G28.34+0.06 or are a common feature of low-$L/M$ environments,
we carried out the Multiwavelength Line-Imaging Survey of the 70 $\mu$m-dArk and bright clOuds (MIAO) project\footnote{In the MIAO project, ``dark'' clumps are defined as sources with high dust extinction at wavelengths shorter than 70~$\mu$m, a bolometric luminosity-to-mass ratio of $L/M < 1$ $L_{\odot}/M_{\odot}$, and bright 870~$\mu$m emission. In contrast, ``bright'' clumps exhibit strong emission at both 70~$\mu$m and 870~$\mu$m.} \citep{Feng2020ApJ...901..145F}, using the IRAM-30m, VLA, and ALMA. The MIAO-ALMA survey targets a sample of sixteen 70 $\mu$m-dark clumps at 3 mm with kinematic distances of 2.4–4.7 kpc (see Table \ref{tab:clumps}), where the field is centered on the coldest ($T_{\rm dust} < 15$ K) and densest ($n_{\rm H_2} > 10^{5}$ cm$^{-3}$) regions of each clump. IRAM-30m observations (project code: 115-17) reveal parsec-scale SiO~(2–1) emission associated with these clumps.
Our 3~mm observations target the SiO~(2–1) transition, which has an upper-level energy of $E_u$/k$_{\rm B}$ = 6.3~K and can effectively trace shock events not only associated with high-temperature outflows. This project is complementary to the ASHES survey \citep{Sanhueza2019ApJ...886..102S,Morii2023ApJ...950..148M,Morii2024ApJ...966..171M}, which uses the higher-excitation SiO~(5–4) transition to trace outflows in 39 massive (200 - 5000$\,M_{\odot}$) and cold (15 - 21~K) clumps.        

In the MIAO-ALMA sample, G34.78$-$0.57 is located in the interaction zone between a supernova remnant and an H\textsc{ii} region. Owing to its unique environment, it is presented separately in Xie et al. (submitted) and is therefore not included in this work.
In this paper, we focus on shock-related kinematics traced by SiO~(2–1) and HCO$^+$~(1-0) emission at a linear resolution of 0.03 $\text{-}$ 0.06~pc from the MIAO-ALMA project. 
The observations and data reduction procedures are described in Section \ref{sec:obser}. The observational results are presented in Section \ref{sec:result}. The properties of the outflows, as well as the possible origins of the narrow-linewidth SiO~(2-1) and broad-linewidth Gaussian SiO~(2-1) components, are discussed in Section \ref{sec:discuss}. A summary of our main conclusions is given in Section \ref{sec:summary}.

\section{Observation AND DATA REDUCTION}
\label{sec:obser}
This work uses ALMA Band 3 data from 2019 to 2025, including Cycle 6 (Project ID: 2018.1.00101.S; PI: S. Feng) with the 12m array, 7m array, and total power (TP) antennas, and Cycle 11 (Project ID: 2024.1.01696.S; PI: S. Lin) with 7m array and TP. To ensure consistent sensitivity, Cycle 11 7m array and TP data were added due to the lower sensitivity of Cycle 6 TP data compared to the 12m Array.
Four spectral windows (SPWs) were configured to observe SiO~(2–1) at  86.847 GHz, H$^{13}$CO$^+$~(1–0) at 86.749 GHz, HCN~(1–0) at 88.632 GHz, and HCO$^+$~(1–0) at 89.189 GHz. These SPWs were configured with a channel width of 61.035~kHz, corresponding to a spectral resolution of 0.21~km~s$^{-1}$ at 87.5~GHz. Two broad spectral windows were configured at central frequencies of 86.2~GHz and 87.5~GHz, each with a bandwidth of 937.5~MHz and spectral resolutions of 244.144~kHz and 488.281~kHz, respectively.
The details of the observations and the continuum images from line-free channels will be presented in a forthcoming paper.

Data calibration was performed using the Common Astronomy Software Application (CASA, \citealt{CASA2022PASP..134k4501C}) software package versions 5.6.1 and 6.6.1. 
No self-calibration was applied.
We used MIRIAD\footnote{\href{https://www.astro.umd.edu/~teuben/miriad/}{https://www.astro.umd.edu/$\sim$teuben/miriad}} to combine the data from the 12~m array, the 7m, and the TP in the \textit{uv} domain. The TP data were first converted into pseudo-visibilities using the \texttt{uvmodel} task, and subsequently combined with the interferometric data prior to imaging.  Imaging was then performed using the \texttt{clean} task in MIRIAD, with no mask applied and a cleaning threshold of 1$\sigma$.
All images were produced using Briggs's robust weighting of 0.5, for a balance between angular resolution and point-source sensitivity. The resulting images have a pixel size of  0.4$^{\prime\prime}$.
In this work, we focus on the SiO~(2-1), HCO$^+$~(1-0), and H$^{13}$CO$^+$~(1-0) lines. On average, the synthesized beam is $\sim$ \(3\farcs8 \times 2\farcs8\), and the rms noise level is about 4 mJy~beam$^{-1}$ (corresponding to $\sim$ 64~mK) per channel of 0.21~km~s$^{-1}$. The list of targeted sources with the observation beam and rms are summarized in Table \ref{tab:clumps}.

\begin{table*}[hbt!]
    \centering
    \scriptsize
    \caption{Clump Properties and Observational Parameters.}
    \label{tab:clumps}
    \setlength{\tabcolsep}{0.3pt}
    \begin{tabular}{cccccccccccc}
    \hline
    \hline
    Source & $M_{\rm clump}$\tablenotemark{\tiny  \textcolor{blue}{a}} & $L_{\rm bol}$\tablenotemark{\tiny  \textcolor{blue}{a}} & $d_{\rm kin}$\tablenotemark{\tiny  \textcolor{blue}{b}} & $V_{\rm sys}$ & $\chi_{\rm HCO^+}$\tablenotemark{\tiny  \textcolor{blue}{g}} & $\sigma_{\rm SiO}$\tablenotemark{\tiny  \textcolor{blue}{h}} & $\theta_{\rm SiO}$ (P.A.) &  $\sigma_{\rm HCO^+}$\tablenotemark{\tiny  \textcolor{blue}{h}} & $\theta_{\rm HCO^+}$ (P.A.) &  $\sigma_{\rm H^{13}CO^+}$\tablenotemark{\tiny  \textcolor{blue}{h}} & $\theta_{\rm H^{13}CO^+}$ (P.A.) \\
    & ($M_{\odot}$) & ($L_{\odot}$) &  (kpc) & (km s$^{-1}$) & (10$^{-10}$) & (mJy beam$^{-1}$) & ($\prime \prime$ $\times$ $\prime \prime$, $^{\circ}$) & (mJy beam$^{-1}$) & ($\prime \prime$ $\times$ $\prime \prime$, $^{\circ}$) & (mJy beam$^{-1}$) & ($\prime \prime$ $\times$ $\prime \prime$, $^{\circ}$)\\
    \hline
  G11.10-0.11  & 398 & 108.4 & 2.85 & 29.1\tablenotemark{\tiny  \textcolor{blue}{c}} & 1.1 & 4.2 & 3.6$^{\prime \prime}$ $\times$2.7$^{\prime \prime}$ (-79.8$^{\circ}$) & 4.2& 4.1$^{\prime \prime}$ $\times$3.0$^{\prime \prime}$ (-79.1$^{\circ}$) & 4.2 & 4.1$^{\prime \prime}$ $\times$3.1$^{\prime \prime}$ (-79.6$^{\circ}$) \\
  G11.38+0.81 & 224.7 & 49.7 & 2.78 & 27.0\tablenotemark{\tiny  \textcolor{blue}{d}} & 1.2 & 4.2 & 3.6$^{\prime \prime}$ $\times$2.7$^{\prime \prime}$ (-78.7$^{\circ}$) & 4.4& 4.2$^{\prime \prime}$ $\times$3.0$^{\prime \prime}$ (-80.1$^{\circ}$) & 4.8 & 3.8$^{\prime \prime}$ $\times$2.8$^{\prime \prime}$ (-80.8$^{\circ}$)\\
  G12.95-0.25 & 627.3 &	98.2 & 2.92 & 34.5\tablenotemark{\tiny  \textcolor{blue}{e}} & 1.1 & 4.1 & 3.6$^{\prime \prime}$ $\times$2.7$^{\prime \prime}$ (-78.7$^{\circ}$) & 4.3 & 4.1$^{\prime \prime}$ $\times$3.1$^{\prime \prime}$ (-78.0$^{\circ}$) & 4.5 & 4.9$^{\prime \prime}$ $\times$3.7$^{\prime \prime}$ (-76.8$^{\circ}$)\\
  G12.97-0.24 & 477.2 & 104.8 & 2.92 & 35.3\tablenotemark{\tiny  \textcolor{blue}{f}} & 1.1 & 4.0 & 3.6$^{\prime \prime}$ $\times$2.7$^{\prime \prime}$ (-78.4$^{\circ}$) & 4.4 & 3.8$^{\prime \prime}$ $\times$2.8$^{\prime \prime}$ (-78.5$^{\circ}$) & 3.9 & 4.6$^{\prime \prime}$ $\times$3.4$^{\prime \prime}$ (-76.8$^{\circ}$)\\
  G14.18-0.23 & 466 & 230.3 & 3.86 & 39.9\tablenotemark{\tiny  \textcolor{blue}{e}} & 1.1 & 4.1 & 3.5$^{\prime \prime}$ $\times$2.7$^{\prime \prime}$ (-78.5$^{\circ}$) & 4.4 & 3.9$^{\prime \prime}$ $\times$2.9$^{\prime \prime}$ (-78.4$^{\circ}$) & 3.9 & 4.9$^{\prime \prime}$ $\times$3.7$^{\prime \prime}$ (-76.3$^{\circ}$) \\
  G14.69-0.22 & 450.6 & 109.1 & 3.87 & 37.5\tablenotemark{\tiny  \textcolor{blue}{c}} & 1.1 & 3.9 & 3.5$^{\prime \prime}$ $\times$2.7$^{\prime \prime}$ (-78.2$^{\circ}$) & 4.4 & 3.8$^{\prime \prime}$ $\times$2.8$^{\prime \prime}$ (-78.8$^{\circ}$) & 3.5 & 5.3$^{\prime \prime}$ $\times$4.2$^{\prime \prime}$ (-75.0$^{\circ}$)\\
  G14.23-0.18 & 286.4 & 139.2 & 3.86 & 37.6\tablenotemark{\tiny  \textcolor{blue}{c}} & 1.1 & 4.1 & 3.6$^{\prime \prime}$ $\times$2.7$^{\prime \prime}$ (-78.1$^{\circ}$) & 4.3 & 4.3$^{\prime \prime}$ $\times$3.2$^{\prime \prime}$ (-77.2$^{\circ}$) & 4.3 & 3.8$^{\prime \prime}$ $\times$2.8$^{\prime \prime}$ (-78.5$^{\circ}$)\\
  G14.73-0.20 & 317.8 & 248.7 & 3.88 & 38.0\tablenotemark{\tiny  \textcolor{blue}{c}} & 1.1 & 4.1 & 3.5$^{\prime \prime}$ $\times$2.7$^{\prime \prime}$ (-78.1$^{\circ}$) & 4.4 & 3.7$^{\prime \prime}$ $\times$2.8$^{\prime \prime}$ (-78.8$^{\circ}$) & 2.5 & 4.4$^{\prime \prime}$ $\times$3.3$^{\prime \prime}$ (-77.1$^{\circ}$) \\
  G15.50-0.42 & 317.8 & 248.7 & 3.94 & 39.4\tablenotemark{\tiny  \textcolor{blue}{e}} & 1.1 & 4.0 & 3.5$^{\prime \prime}$ $\times$2.7$^{\prime \prime}$ (-78.3$^{\circ}$) & 4.5 & 3.6$^{\prime \prime}$ $\times$2.7$^{\prime \prime}$ (-78.7$^{\circ}$) & 4.3 & 3.7$^{\prime \prime}$ $\times$2.8$^{\prime \prime}$ (-79.1$^{\circ}$)\\
  G16.30-0.53 & 476.2 & 278.7 & 3.18 & 38.2\tablenotemark{\tiny  \textcolor{blue}{c}} & 1.2 & 4.0 & 3.5$^{\prime \prime}$ $\times$2.7$^{\prime \prime}$ (-79.1$^{\circ}$) & 4.2 & 3.8$^{\prime \prime}$ $\times$2.8$^{\prime \prime}$ (-79.2$^{\circ}$) & 3.6 & 5.6$^{\prime \prime}$ $\times$4.6$^{\prime \prime}$ (-74.9$^{\circ}$)\\
  G18.80-0.30 & 1184.6 & 382.7 & 3.88 & 65.4\tablenotemark{\tiny  \textcolor{blue}{e}} & 1.0 & 4.0 & 3.5$^{\prime \prime}$ $\times$2.7$^{\prime \prime}$ (-77.7$^{\circ}$) & 4.5 & 3.7$^{\prime \prime}$ $\times$2.8$^{\prime \prime}$ (-78.5$^{\circ}$) & 4.0 & 4.8$^{\prime \prime}$ $\times$3.7$^{\prime \prime}$ (-76.1$^{\circ}$)\\
  G22.53-0.19 & 883 & 320.4 & 4.73 & 76.4\tablenotemark{\tiny  \textcolor{blue}{c}} & 1.0 & 3.6 & 3.1$^{\prime \prime}$ $\times$2.2$^{\prime \prime}$ (-89.4$^{\circ}$) & 4.5 & 3.2$^{\prime \prime}$ $\times$2.1$^{\prime \prime}$ (-89.4$^{\circ}$) & 4.6 & 3.2$^{\prime \prime}$ $\times$2.1$^{\prime \prime}$ (-89.6$^{\circ}$)\\
  G28.27-0.17 & 2319.3 & 297.7 & 4.72 & 79.5\tablenotemark{\tiny  \textcolor{blue}{d}} & 1.1 & 3.4 & 3.1$^{\prime \prime}$ $\times$2.2$^{\prime \prime}$ (89.1$^{\circ}$) & 4.6 & 3.4$^{\prime \prime}$ $\times$2.2$^{\prime \prime}$ (89.7$^{\circ}$) & 5.0 & 3.8$^{\prime \prime}$ $\times$2.2$^{\prime \prime}$ (89.0$^{\circ}$)\\
  G28.52-0.25 & 443.4 & 183.7 & 4.47 & 87.3\tablenotemark{\tiny  \textcolor{blue}{d}} & 1.1 & 3.3 & 3.1$^{\prime \prime}$ $\times$2.2$^{\prime \prime}$ (89.0$^{\circ}$) & 4.2 & 3.4$^{\prime \prime}$ $\times$2.2$^{\prime \prime}$ (89.5$^{\circ}$) & 4.4 & 3.7$^{\prime \prime}$ $\times$2.3$^{\prime \prime}$ (89.7$^{\circ}$)\\
  G28.54-0.24 & 1233.5 & 511.0 & 4.43 & 86.4\tablenotemark{\tiny  \textcolor{blue}{d}} & 1.3 & 3.4 & 3.1$^{\prime \prime}$ $\times$2.2$^{\prime \prime}$ (89.1$^{\circ}$) & 4.7 & 3.3$^{\prime \prime}$ $\times$2.1$^{\prime \prime}$ (89.0$^{\circ}$) & 4.2 & 3.6$^{\prime \prime}$ $\times$2.3$^{\prime \prime}$ (89.4$^{\circ}$)\\
  \hline
  \hline
    \end{tabular}  
    \tablecomments{$^{(a)}$ Adopted from \citet{Yuan2017ApJS..231...11Y}; $^{(b)}$ Bayesian-corrected kinematic distance (Feng et al. in prep.); $^{(c)}$ \citet{Wienen2012A&A...544A.146W}, corrected with H$^{13}$CO$^+$ (1–0) from IRAM-30m \citep{Feng2020ApJ...901..145F}; $^{(d)}$ \citet{Csengeri2014A&A...565A..75C}, corrected with H$^{13}$CO$^+$ (1–0) from IRAM-30m \citep{Feng2020ApJ...901..145F}; $^{(e)}$ \citet{Shirley2013ApJS..209....2S}, corrected with H$^{13}$CO$^+$ (1–0) from IRAM-30m \citep{Feng2020ApJ...901..145F}; $^{(f)}$ Single-pointing observations with the Submillimeter Telescope (SMT) \citep{Yuan2017ApJS..231...11Y}, corrected with H$^{13}$CO$^+$ (1–0) from IRAM-30m \citep{Feng2020ApJ...901..145F};   
    $^{(g)}$ The HCO$^+$ abundance relative to H$_2$, estimated from H$^{13}$CO$^+$ observations with IRAM-30m and converted using the $^{12}$C/$^{13}$C isotopic ratio;
    $^{(h)}$ RMS value per channel (with a channel width of 61 kHz) before the primary beam correction. }
\end{table*}

\section{Observational Results}
\label{sec:result}

\subsection{SiO detection }
\label{Subsec:Detection}
SiO is originally formed in the gas phase around evolved stars and subsequently incorporated into silicate dust grains \citep[e.g.,][]{Gail1998FaDi..109..303G,Gail1999A&A...347..594G,Cherchneff2006A&A...456.1001C}. In star-forming regions, shocks associated with protostellar outflows can sputter or disrupt these grains, releasing Si (and/or SiO) back into the gas phase, where rapid gas-phase reactions efficiently enhance the SiO abundance, making it a powerful tracer of shocked gas \citep[e.g.,][]{Martin-Pintado1992A&A...254..315M,Schilke1997A&A...321..293S,Caselli1997A&A...322..296C,Gusdorf2008A&A...482..809G}.
Assuming optically thin conditions, we derive critical densities for the SiO~(2-1) of 5.4 $\times$ 10$^4$ cm$^{-3}$ at 20~K and 4.5 $\times$ 10$^4$ cm$^{-3}$ at 50~K, using the \emph{myRadex}\footnote{\href{https://github.com/fjdu/myRadex}{https://github.com/fjdu/myRadex}} package, which is built upon the RADEX radiative transfer code \citep{Van2007A&A...468..627V}.
In the MIAO sample, SiO~(2–1) emission is detected with the signal-to-noise ratio (S/N) $>$ 5 in all sixteen observed sources. In total, we detect 59 SiO~(2-1) emission features with diverse kinematic and morphological properties.

\begin{figure*}
\centering
\includegraphics[width=0.7\linewidth]{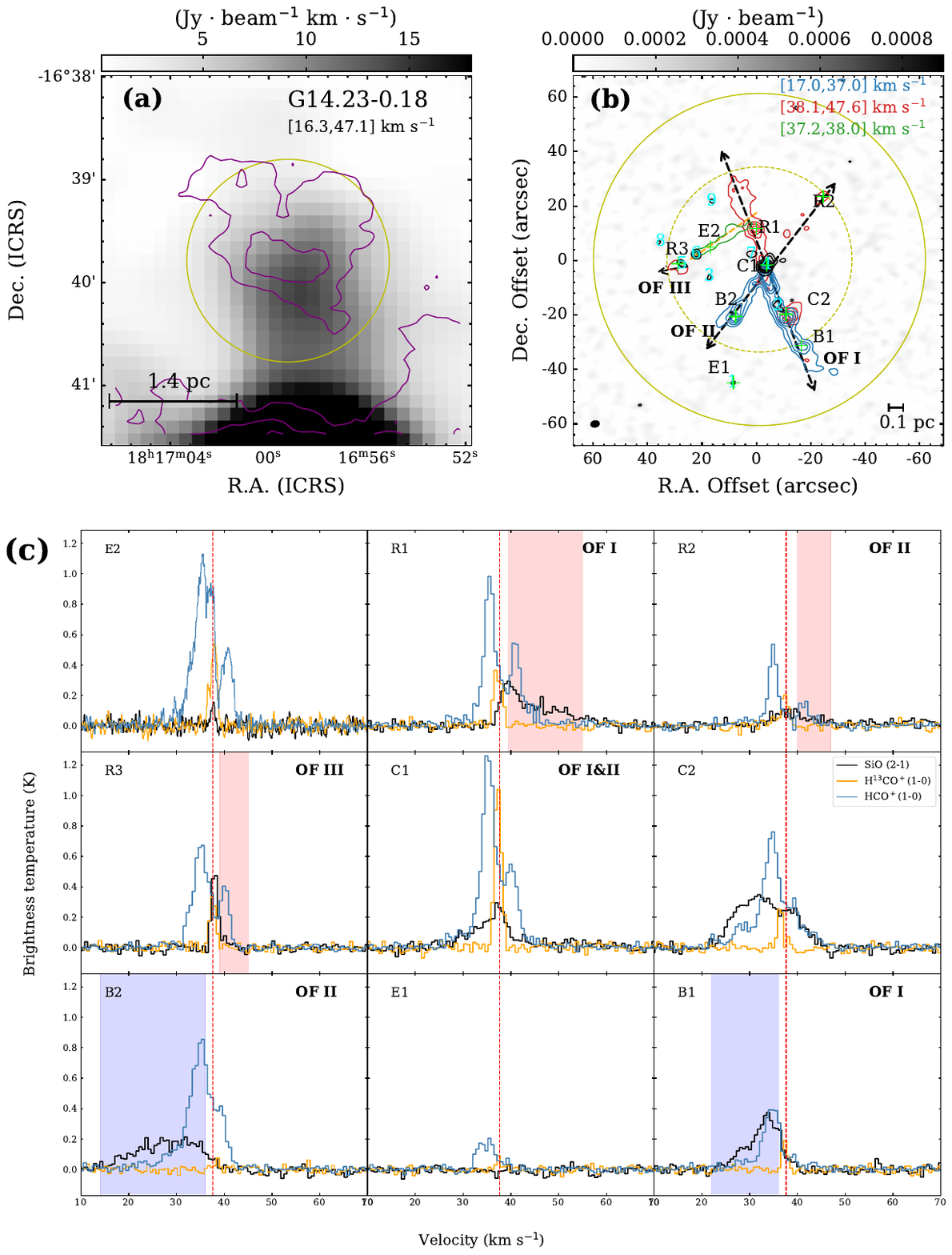}
\caption{Integrated intensity maps and line profiles of G14.23-0.18. (a) The greyscale map shows the SiO~(2–1) emission observed with the ALMA TP at an angular resolution of 74.5$^{\prime\prime}$ ($\sim$ 1 pc). Purple contours indicate the 870~$\mu$m continuum emission from APEX \citep{Schuller2009A&A...504..415S}, starting from 6$\sigma$ and increasing by 6$\sigma$ steps, with an angular resolution of 18.2$^{\prime\prime}$. The yellow circle represents the 0.1 primary-beam response of the ALMA 12 m observations, corresponding to 122$^{\prime\prime}$. (b): SiO~(2–1) emission (12m+7m+TP) overlaid on the 3.4~mm continuum grayscale map from the ALMA observation without primary-beam correction. The black contours trace the continuum emission, starting at 4$\sigma$ and increasing in steps of 2$\sigma$ (1$\sigma = 4.3 \times 10^{-5}$ Jy beam$^{-1}$). Blue and red contours represent the integrated intensities of the blue- and redshifted SiO~(2–1) emission, respectively, while the green contour marks the narrow-linewidth SiO component, with the velocity range labeled in the panel. The black dashed arrows indicate the outflow direction, while the orange dashed arrow marks the extended narrow-linewidth SiO component. Contours start at $4\sigma$ and increase in steps of $6\sigma$ ($1\sigma = 0.014$ Jy beam$^{-1}$ km s$^{-1}$ for the blue-shifted component, $1\sigma = 0.009$ Jy beam$^{-1}$ km~s$^{-1}$ for the red-shifted component, and $1\sigma = 0.009$ Jy beam$^{-1}$ km~s$^{-1}$ for the narrow component). The yellow solid circle is the same as that shown in panel (a), while the yellow dashed circle indicates the 0.5 primary-beam response of the ALMA 12 m observations (68$^{\prime\prime}$). The beam of the TP and the synthesized beam of the combined image are shown in panels (a) and (b), respectively. (c): Line profiles of SiO~(2–1), HCO$^+$(1–0), and H$^{13}$CO$^+$(1–0), shown in black, blue, and orange, respectively, are extracted from a single synthesized beam-averaged region at different positions labeled in panel (b). ``R" and ``B" label the redshifted and blueshifted components, ``C" represents the core, and ``E" indicates emission with no coincident core ($<$ 0.1 pc). ``OF I", ``OF II", and ``OF III" indicate the outflows we identified. All lines have a native velocity resolution of $\sim$ 0.21 km~s$^{-1}$, which is used for panels (E2) showing narrow linewidth components ($<1.5$ km~s$^{-1}$). For the other panels, the line profiles are smoothed to a velocity resolution of 0.63 km~s$^{-1}$. The vertical red dashed line indicates the systemic velocity of clump (37.6 km~s$^{-1}$, Table \ref{tab:clumps}). The blue and red windows indicate the velocity range integrated to estimate outflow parameters in Section \ref{subsec:Outflow_parameters}, excluding the velocity range associated with the central core. The complete figure set (15 images) is available in the online journal.}
\label{fig:mom0_and_spectra_1}
\end{figure*}

From the line profiles of the detected SiO (2–1) emission toward different regions, as well as their spectral distributions, compared with the dust continuum emission, we identify three types of features. As an illustration of these features, G14.23-0.18 is shown in Figure \ref{fig:mom0_and_spectra_1}, and the complete sample is presented in Figure Set 1.

(1) Sources showing broad red‑ or blue‑shifted line wings ($>$ 5 km~s$^{-1}$), some symmetry around the dust continuum peak and others monopolar, are classified as outflow candidates.
In detail, we used images before primary beam (PB) correction for identification, as they provide a uniform noise distribution. SiO~(2–1) emission was then searched within the primary beam, with a 10\% response cutoff applied to avoid low-sensitivity edges and to maximize the dynamic range of both the imaging and the evolutionary comparison between 70~$\mu$m-dark and bright clumps. We define an outflow candidate detection when all of the following criteria are met: (i) there is the presence of an asymmetric line profile with significant blue- or red-shifted line wings in the SiO~(2–1) emission; (ii) the velocity interval between the SiO~(2–1) emission peak and the most blue-shifted or red-shifted end of the line profile, down to the 1~$\sigma$ level, exceeds 5 km~s$^{-1}$; and (iii) the integrated intensity over the velocity range defined in criterion (ii) has a projected area larger than one synthesized beam. 
To identify the driving source, each outflow candidate is associated with the nearest dense core, 
labeled in Table \ref{tab:outflows} by the corresponding core ID or marked with a dagger ($\dagger$) when no core is detected. Cores are identified based on minimum significance $\geq$ 1$\sigma$ and spatial resolution (emission area above 3.5$\sigma$ exceeding the synthesized beam FWHM area); see Feng et al. (in preparation) for details.

To determine the systemic velocity ($V_{\rm sys}$), we use the H$^{13}$CO$^+$ (1-0) transition, as it traces relatively dense gas due to its high critical density ($\sim$ 5.5 $\times$ 10$^4$ cm$^{-3}$ at 20-50 K) and exhibits a single–peaked Gaussian profile toward most of the 3 mm dust continuum cores used in this work. 
Although the H$^{13}$CO$^+$~(1–0) line profile toward some dense cores shows multiple velocity components, the main component derived from multi-component fitting is consistent with the line core obtained from single-component fitting, with linewidth differences within 10\%. 
To cover all the outflows in each source and to increase the S/N for better identification, we integrated the SiO~(2-1) intensity over the velocity range for each source, as shown in Figure \ref{fig:mom0_and_spectra_1}, excluding velocities near $V_{\rm sys}$ to avoid contamination from quiescent dense core emission.
For example, the panel labeled “OF I” in Figure \ref{fig:mom0_and_spectra_1} shows one outflow candidate. Owing to the high velocity resolution, the red- and blue-shifted line profiles exhibit a characteristic asymmetric shape, with a sharp rise followed by a more gradual decay.
The morphology and kinematics of all these outflow candidates will be analyzed in Section \ref{Subsec:Morphology}.

(2) Broad Gaussian line profiles with linewidth exceeding 2 km~s$^{-1}$ show either spot-like structures on scales of a few synthesized beams or clumpy morphologies extending over $\sim$ 0.1 pc in the projection plane, and are not spatially associated with the 3~mm dust continuum emission. Although this feature is not evident in the example source G14.23-0.18, it is observed in other sources, such as the panel labeled B2 in G14.69-0.22 (Fig.~\ref{fig:mom0_and_spectra_1}).
We therefore do not classify this emission as an outflow candidate and discuss its possible origin in Section~\ref{Subsec:Origin_of_guassian_lineprofile}.

(3) Narrow linewidth components ($<$ 1.5 km s$^{-1}$), with some emission features spatially inconsistent with cores (e.g., E2 in G14.23-0.18, Fig. \ref{fig:mom0_and_spectra_1}). 
We discuss the possible origins of these narrow components in Section \ref{Subsec:Origin_of_narrow_linewidth}, and they are not included in the outflow analysis. 

Note that two pairs of clumps (G12.97-0.24 and G12.95-0.25; G28.52-0.25 and G28.54-0.24) have overlapping regions but were not observed as Nyquist-sampled mosaics. 
As a result, Core 8 in G12.95$-$0.25 corresponds to the same object as Core 1 in G12.97$-$0.24 (see Figure Set 1). No SiO emission is detected in the overlap regions. Therefore, this duplication does not affect the subsequent analysis or conclusions.

\figsetstart
\figsetnum{1}
\figsettitle{Integrated intensity maps and line profiles toward the observed sources.}
\label{fig:set_1}

\figsetgrpstart
\figsetgrpnum{1.1}
\figsetgrptitle{G14.23-0.18}
\figsetplot{f1_1.pdf}
\figsetgrpnote{The integrated intensity maps and line profiles of G14.23-0.18. The symbols and contour levels used are the same as in Figure \ref{fig:mom0_and_spectra_1}. The RMS $\sigma$ values are 0.014, 0.009, 0.009 Jy beam$^{-1}$ km~s$^{-1}$ for the blue-shifted, red-shifted, and narrow components, respectively. The vertical red dashed line indicates the systemic velocity of clump (37.6 km~s$^{-1}$).}
\figsetgrpend

\figsetgrpstart
\figsetgrpnum{1.2}
\figsetgrptitle{G11.10-0.11}
\figsetplot{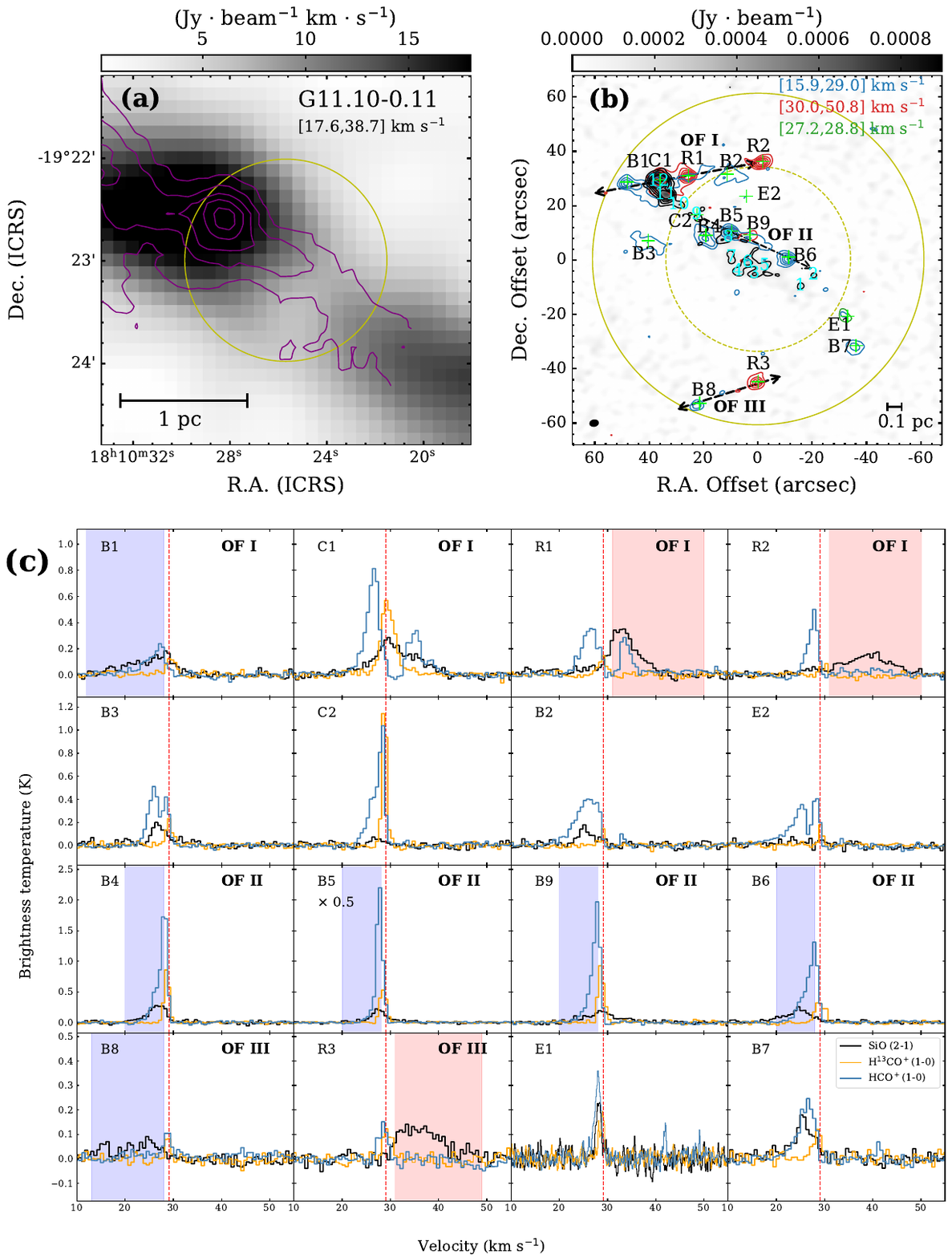}
\figsetgrpnote{The integrated intensity maps and line profiles of G11.10-0.11. The symbols and contour levels used are the same as in Figure \ref{fig:mom0_and_spectra_1}. The RMS $\sigma$ values are 0.008, 0.01, 0.004 Jy beam$^{-1}$ km~s$^{-1}$ for the blue-shifted, red-shifted, and narrow components, respectively. The vertical red dashed line indicates the systemic velocity of clump (29.1 km~s$^{-1}$).}
\figsetgrpend

\figsetgrpstart
\figsetgrpnum{1.3}
\figsetgrptitle{G11.38+0.81}
\figsetplot{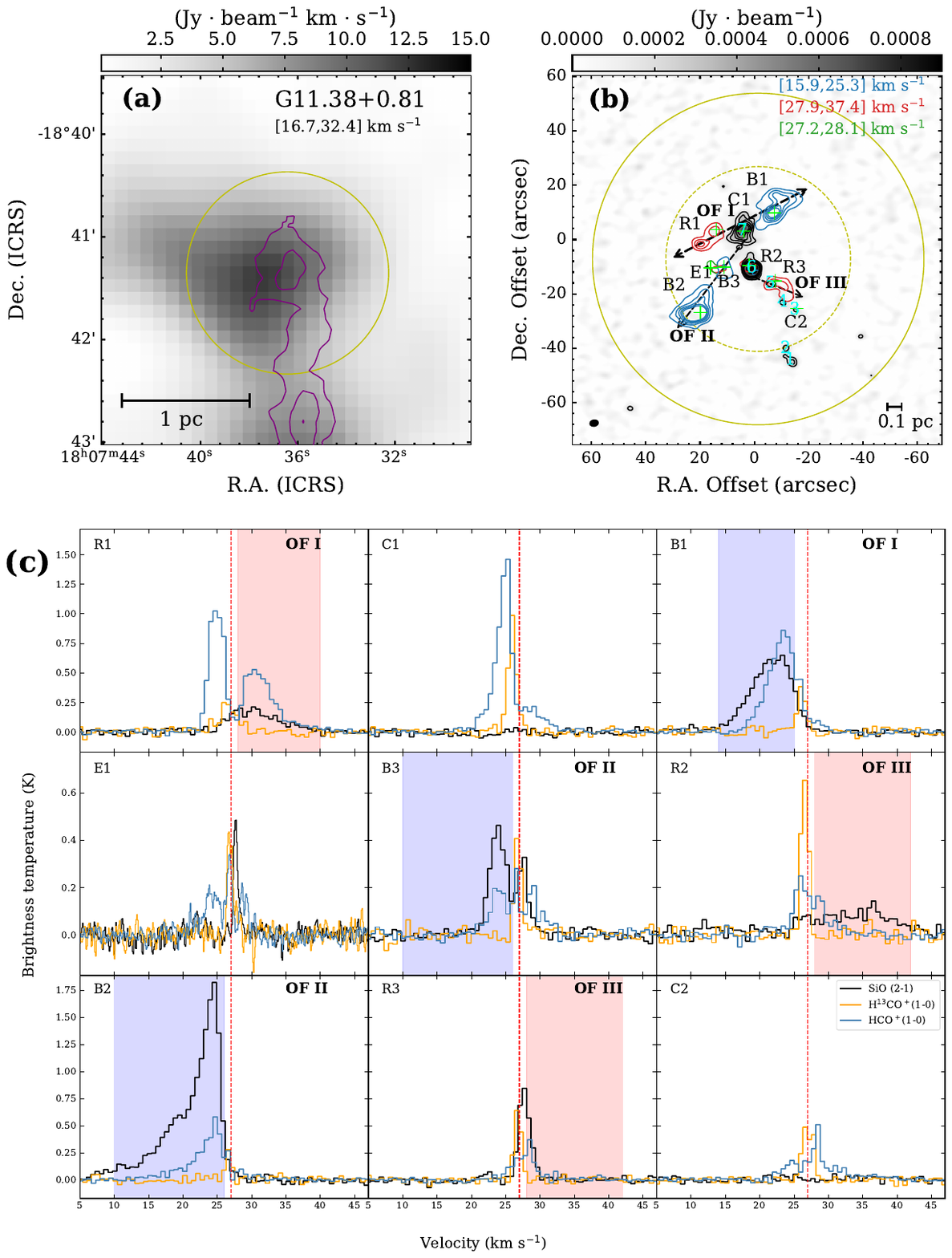}
\figsetgrpnote{The integrated intensity maps and line profiles of G11.38+0.81. The symbols and contour levels used are the same as in Figure \ref{fig:mom0_and_spectra_1}. The RMS $\sigma$ values are 0.01, 0.008, 0.008 Jy beam$^{-1}$ km~s$^{-1}$ for the blue-shifted, red-shifted, and narrow components, respectively. The vertical red dashed line indicates the systemic velocity of clump (27.0 km~s$^{-1}$).}
\figsetgrpend

\figsetgrpstart
\figsetgrpnum{1.4}
\figsetgrptitle{G12.95-0.25}
\figsetplot{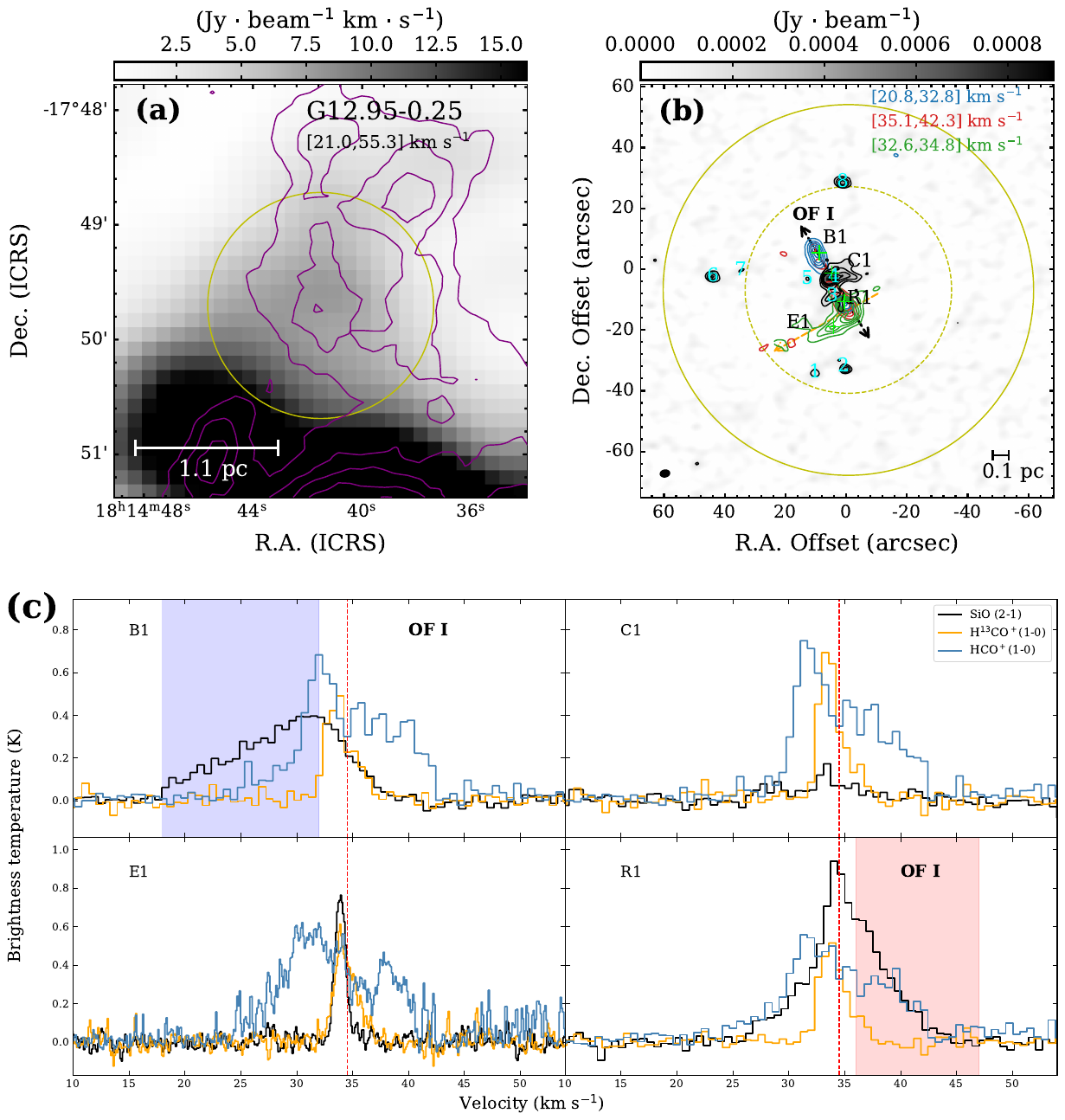}
\figsetgrpnote{The integrated intensity maps and line profiles of G12.95-0.25. The symbols and contour levels used are the same as in Figure \ref{fig:mom0_and_spectra_1}. The RMS $\sigma$ values are  0.009, 0.008, 0.004 Jy beam$^{-1}$ km~s$^{-1}$ for the blue- and red-shifted components, respectively. The vertical red dashed line indicates the systemic velocity of clump (34.5 km~s$^{-1}$).}
\figsetgrpend

\figsetgrpstart
\figsetgrpnum{1.5}
\figsetgrptitle{G12.97-0.24}
\figsetplot{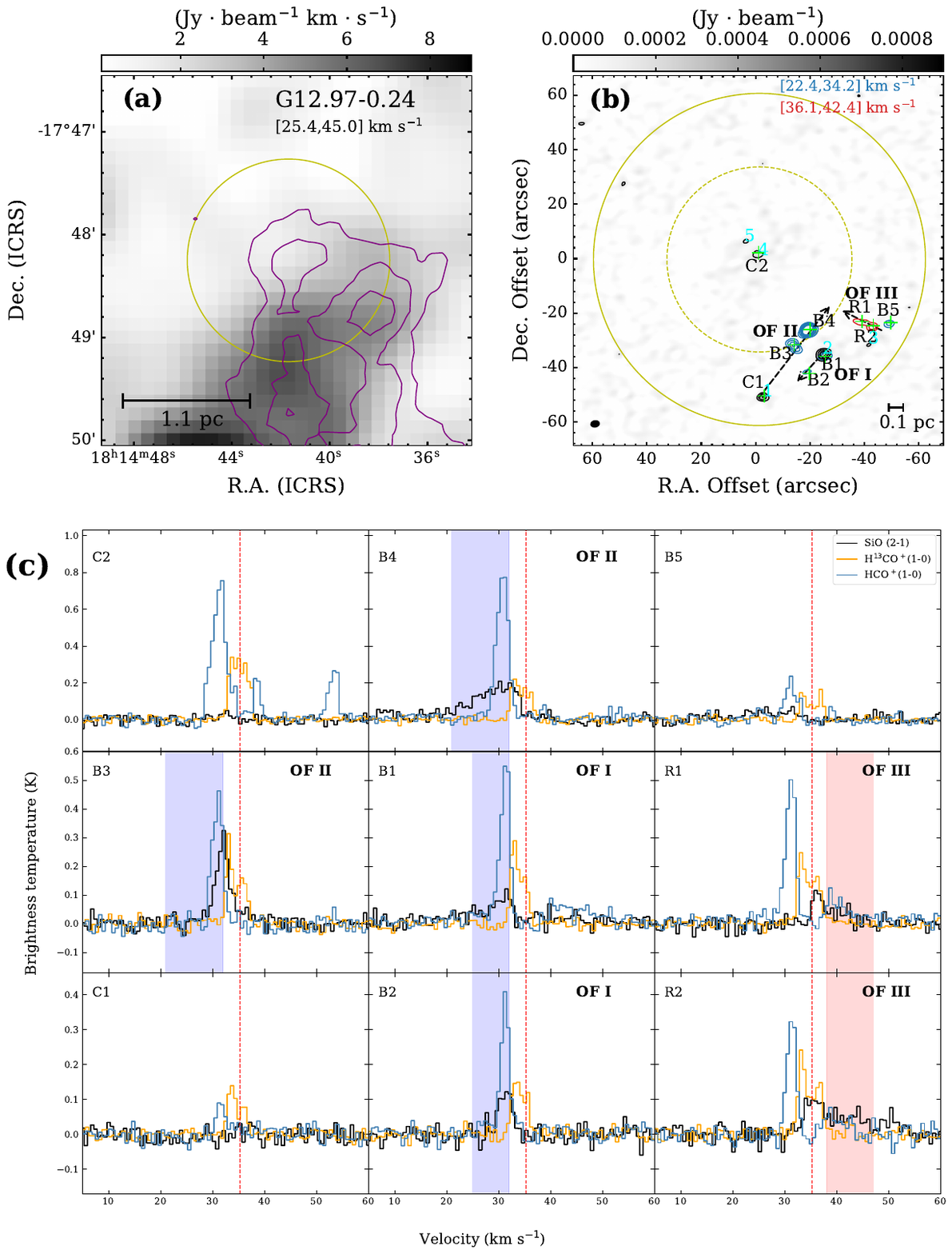}
\figsetgrpnote{The integrated intensity maps and line profiles of G12.97-0.24. The symbols and contour levels used are the same as in Figure \ref{fig:mom0_and_spectra_1}. The RMS $\sigma$ values are 0.01 and 0.007 Jy beam$^{-1}$ km~s$^{-1}$ for the blue- and red-shifted components, respectively. The vertical red dashed line indicates the systemic velocity of clump (35.3 km~s$^{-1}$).}
\figsetgrpend

\figsetgrpstart
\figsetgrpnum{1.6}
\figsetgrptitle{G14.18-0.23}
\figsetplot{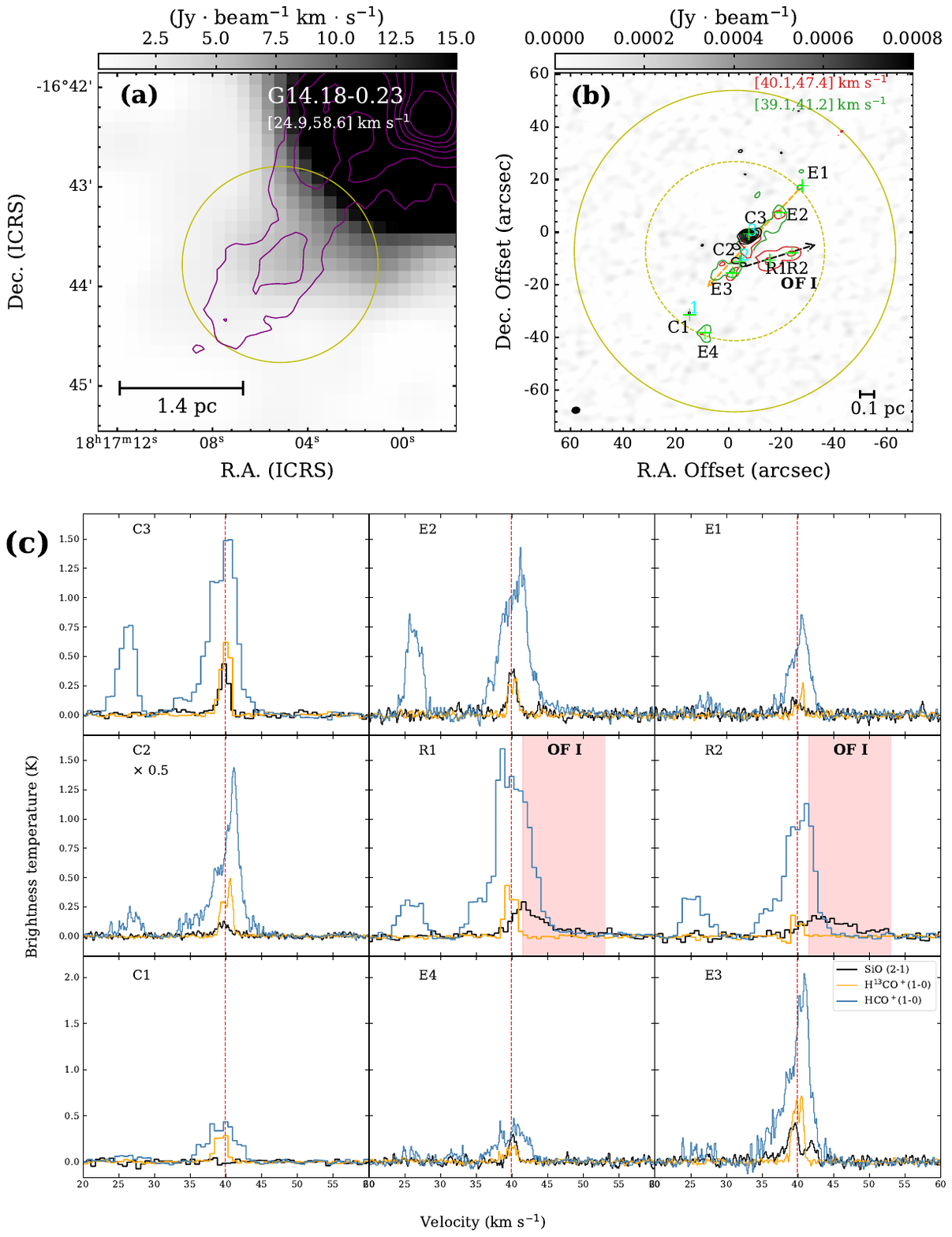}
\figsetgrpnote{The integrated intensity maps and line profiles of G14.18-0.23. The symbols and contour levels used are the same as in Figure \ref{fig:mom0_and_spectra_1}. The RMS $\sigma$ value is 0.007 Jy beam$^{-1}$ km~s$^{-1}$ for both the red-shifted and narrow components. The vertical red dashed line indicates the systemic velocity of clump (39.9 km~s$^{-1}$).}
\figsetgrpend

\figsetgrpstart
\figsetgrpnum{1.7}
\figsetgrptitle{G14.69-0.22}
\figsetplot{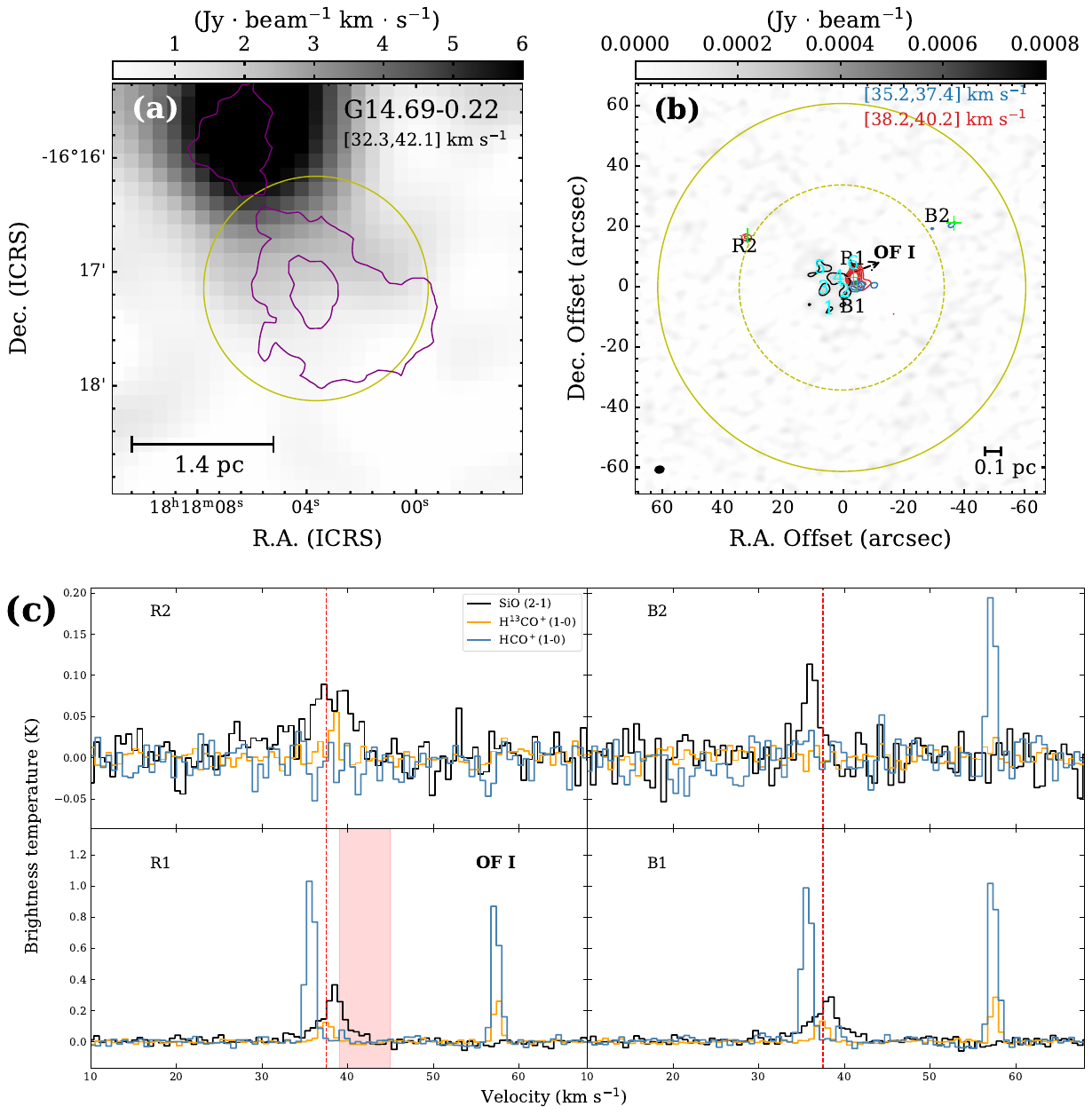}
\figsetgrpnote{The integrated intensity maps and line profiles of G14.69-0.22. The symbols and contour levels used are the same as in Figure \ref{fig:mom0_and_spectra_1}. The RMS $\sigma$ values are 0.004 Jy beam$^{-1}$ km~s$^{-1}$ for the red-shifted components. The vertical red dashed line indicates the systemic velocity of clump (37.5 km~s$^{-1}$).}
\figsetgrpend

\figsetgrpstart
\figsetgrpnum{1.8}
\figsetgrptitle{G14.73-0.20}
\figsetplot{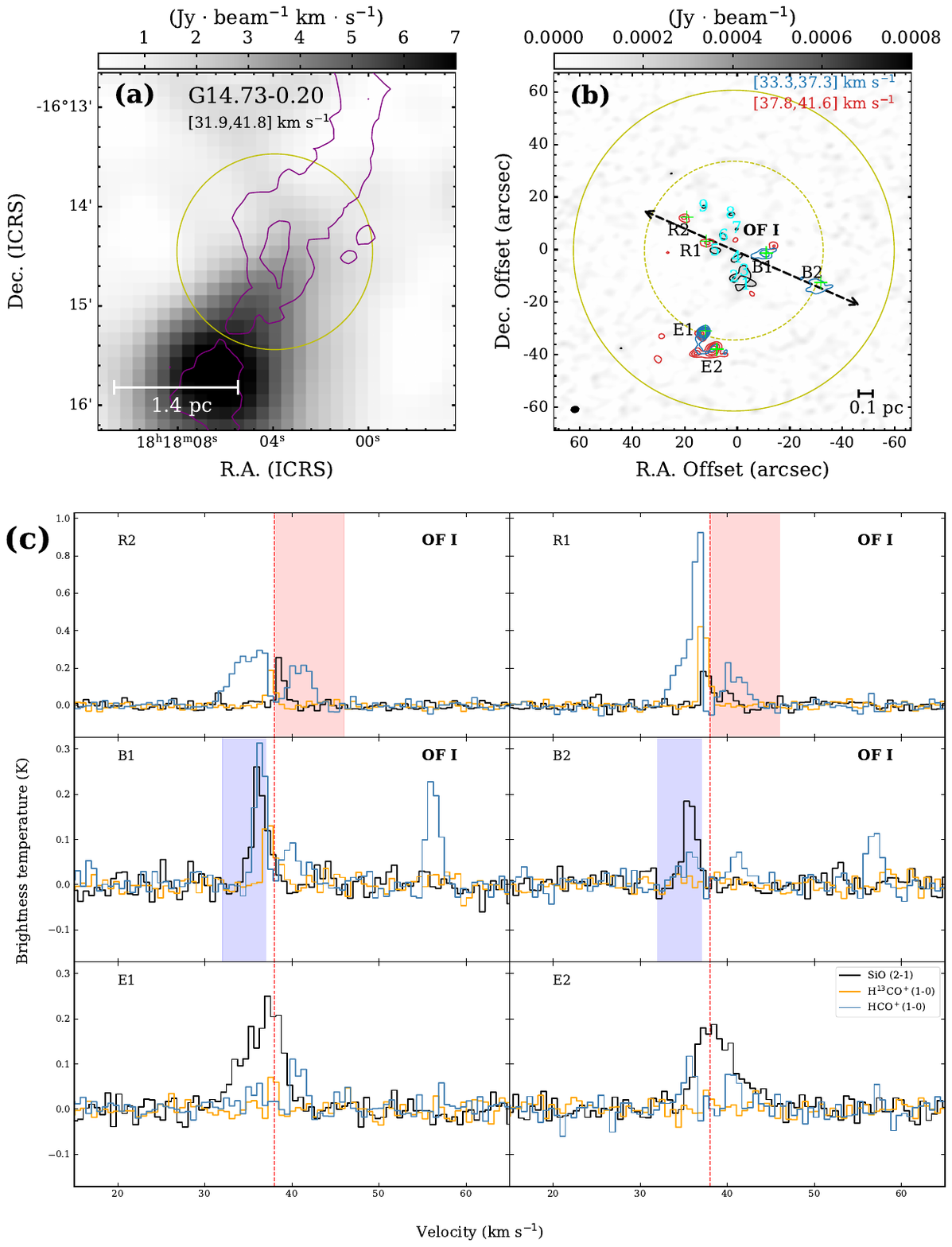}
\figsetgrpnote{The integrated intensity maps and line profiles of G14.73-0.20. The symbols and contour levels used are the same as in Figure \ref{fig:mom0_and_spectra_1}. The RMS $\sigma$ value is 0.005 Jy beam$^{-1}$ km~s$^{-1}$ for both the blue- and red-shifted components. The vertical red dashed line indicates the systemic velocity of clump (38.0 km~s$^{-1}$).}
\figsetgrpend

\figsetgrpstart
\figsetgrpnum{1.9}
\figsetgrptitle{G15.50-0.42}
\figsetplot{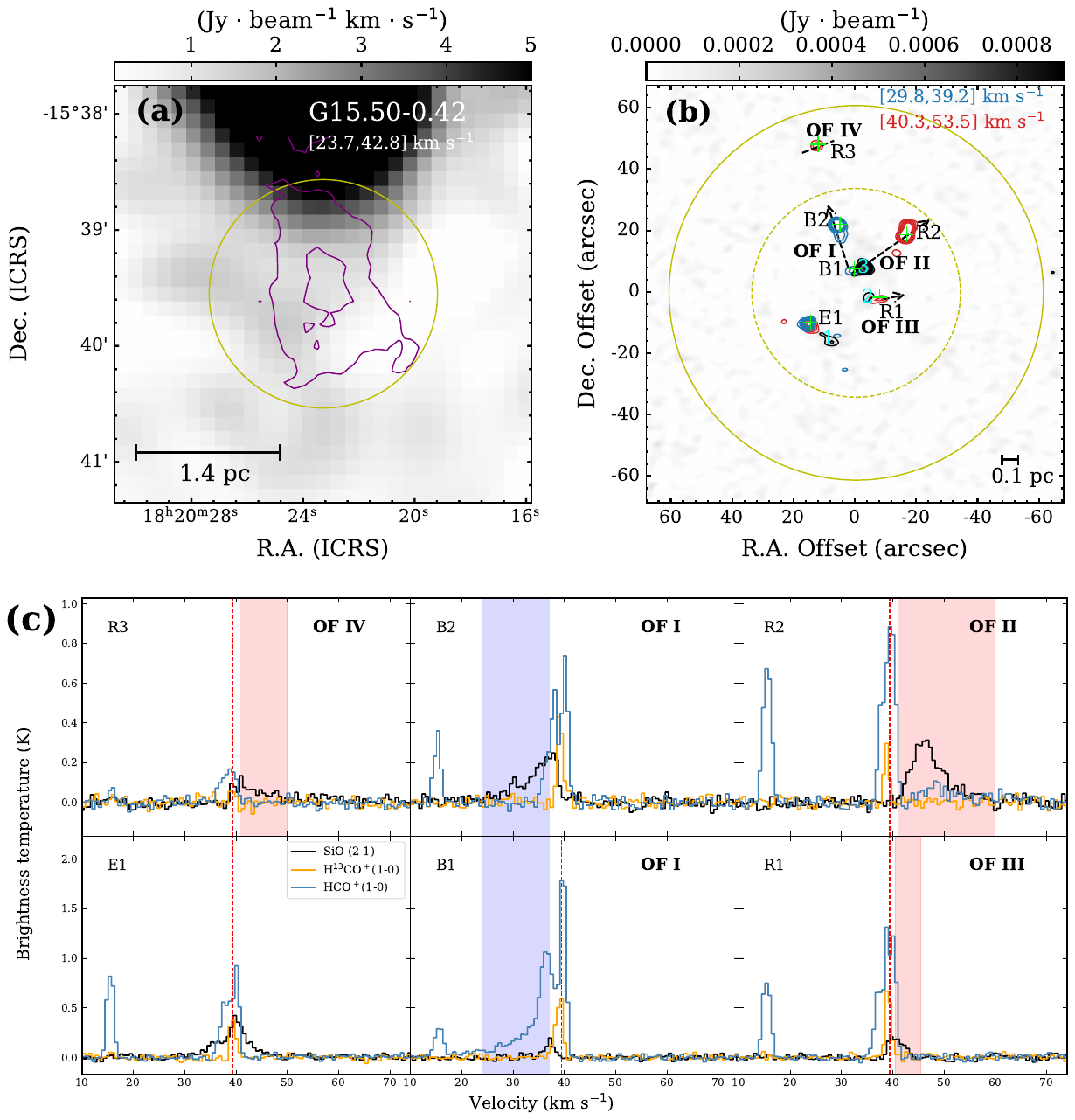}
\figsetgrpnote{The integrated intensity maps and line profiles of G15.50-0.42. The symbols and contour levels used are the same as in Figure \ref{fig:mom0_and_spectra_1}. The RMS $\sigma$ values are 0.008 and 0.01 Jy beam$^{-1}$ km~s$^{-1}$ for the blue- and red-shifted components, respectively. The vertical red dashed line indicates the systemic velocity of clump (39.4 km~s$^{-1}$).}
\figsetgrpend

\figsetgrpstart
\figsetgrpnum{1.10}
\figsetgrptitle{G16.30-0.53}
\figsetplot{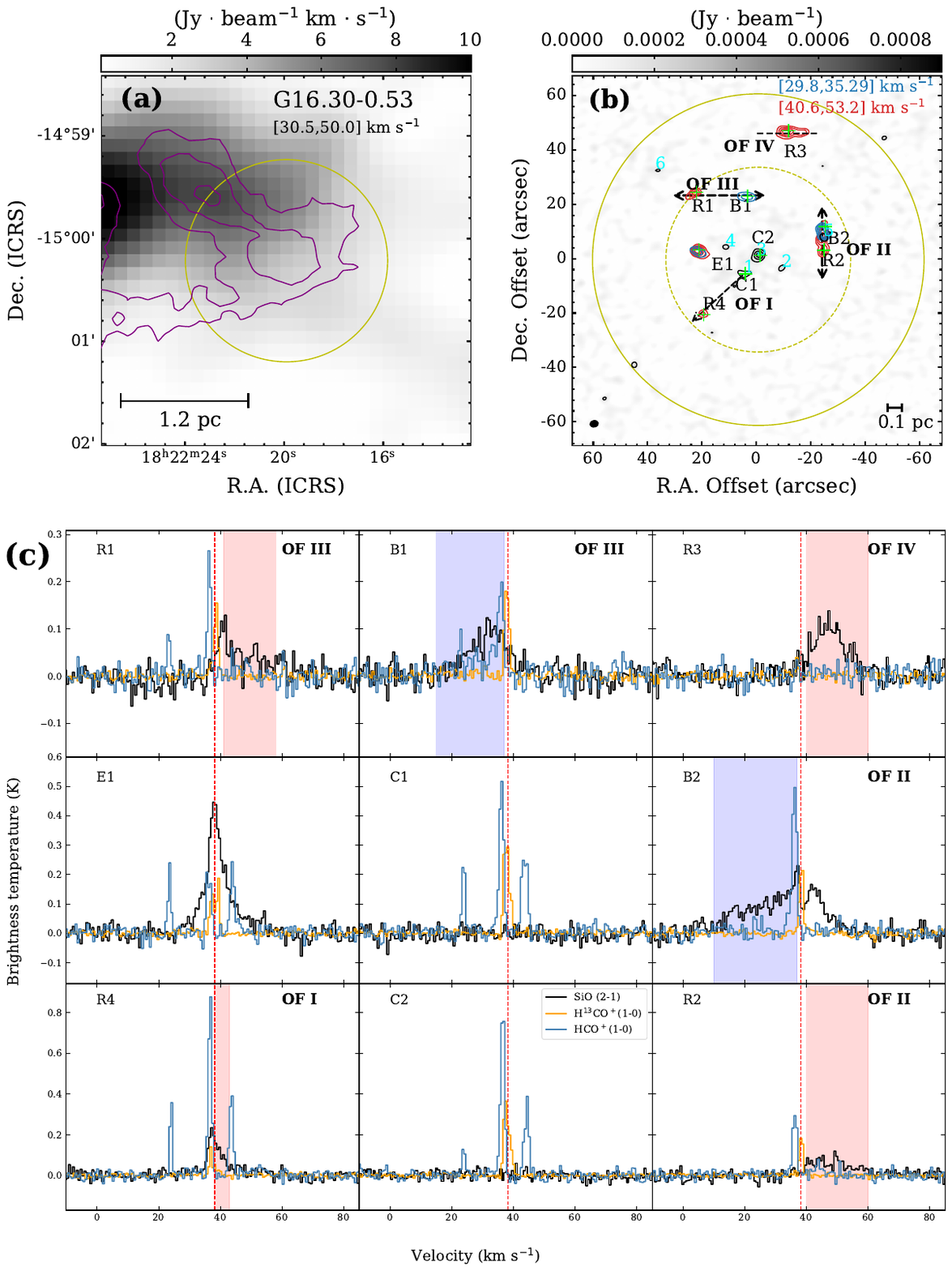}
\figsetgrpnote{The integrated intensity maps and line profiles of G16.30-0.53. The symbols and contour levels used are the same as in Figure \ref{fig:mom0_and_spectra_1}. The RMS $\sigma$ values are 0.008 and 0.01 Jy beam$^{-1}$ km~s$^{-1}$  for the blue- and red-shifted components, respectively. The vertical red dashed line indicates the systemic velocity of clump (38.2 km~s$^{-1}$).}
\figsetgrpend

\figsetgrpstart
\figsetgrpnum{1.11}
\figsetgrptitle{G18.80-0.30}
\figsetplot{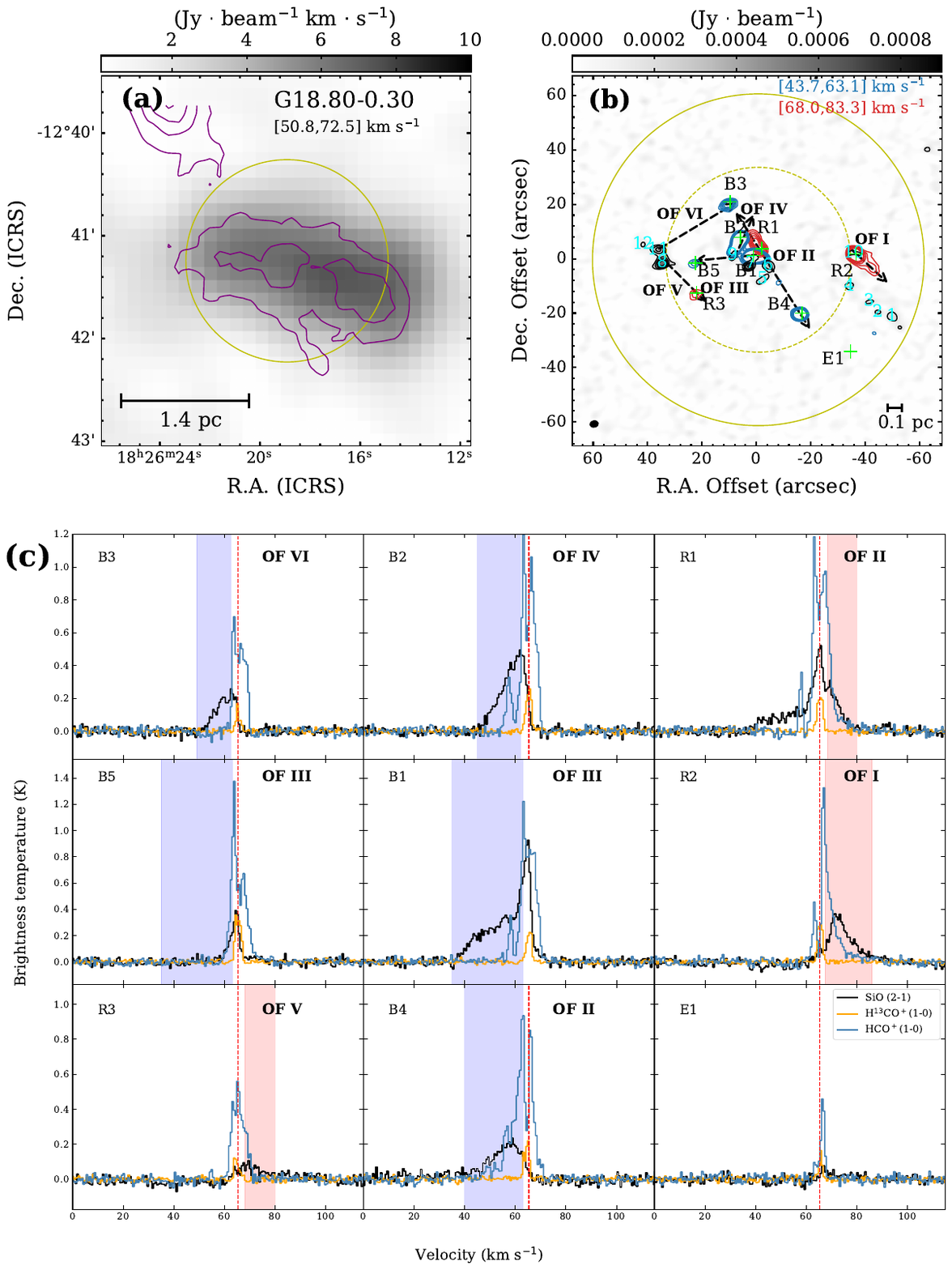}
\figsetgrpnote{The integrated intensity maps and line profiles of G18.80-0.30. The symbols and contour levels used are the same as in Figure \ref{fig:mom0_and_spectra_1}. The RMS $\sigma$ value is 0.012 Jy beam$^{-1}$ km~s$^{-1}$ for both the blue- and red-shifted components. The vertical red dashed line indicates the systemic velocity of clump (65.4 km~s$^{-1}$).}
\figsetgrpend

\figsetgrpstart
\figsetgrpnum{1.12}
\figsetgrptitle{G22.53-0.19}
\figsetplot{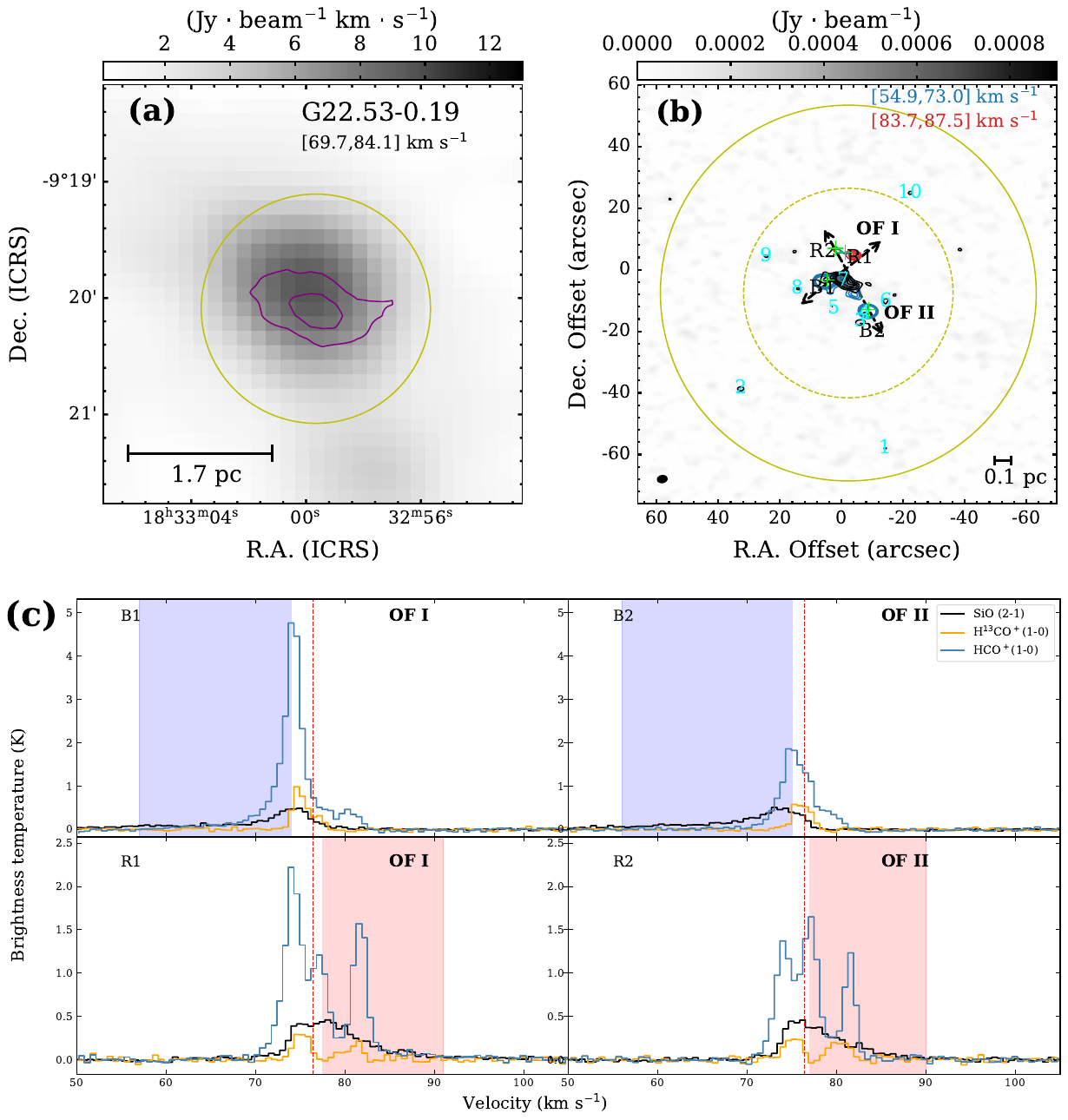}
\figsetgrpnote{The integrated intensity maps and line profiles of G22.53-0.19. The symbols and contour levels used are the same as in Figure \ref{fig:mom0_and_spectra_1}. The RMS $\sigma$ values are 0.009 Jy beam$^{-1}$ km~s$^{-1}$ for both the blue- and red-shifted components. The vertical red dashed line indicates the systemic velocity of clump (76.4 km~s$^{-1}$).}
\figsetgrpend

\figsetgrpstart
\figsetgrpnum{1.13}
\figsetgrptitle{G28.27-0.17}
\figsetplot{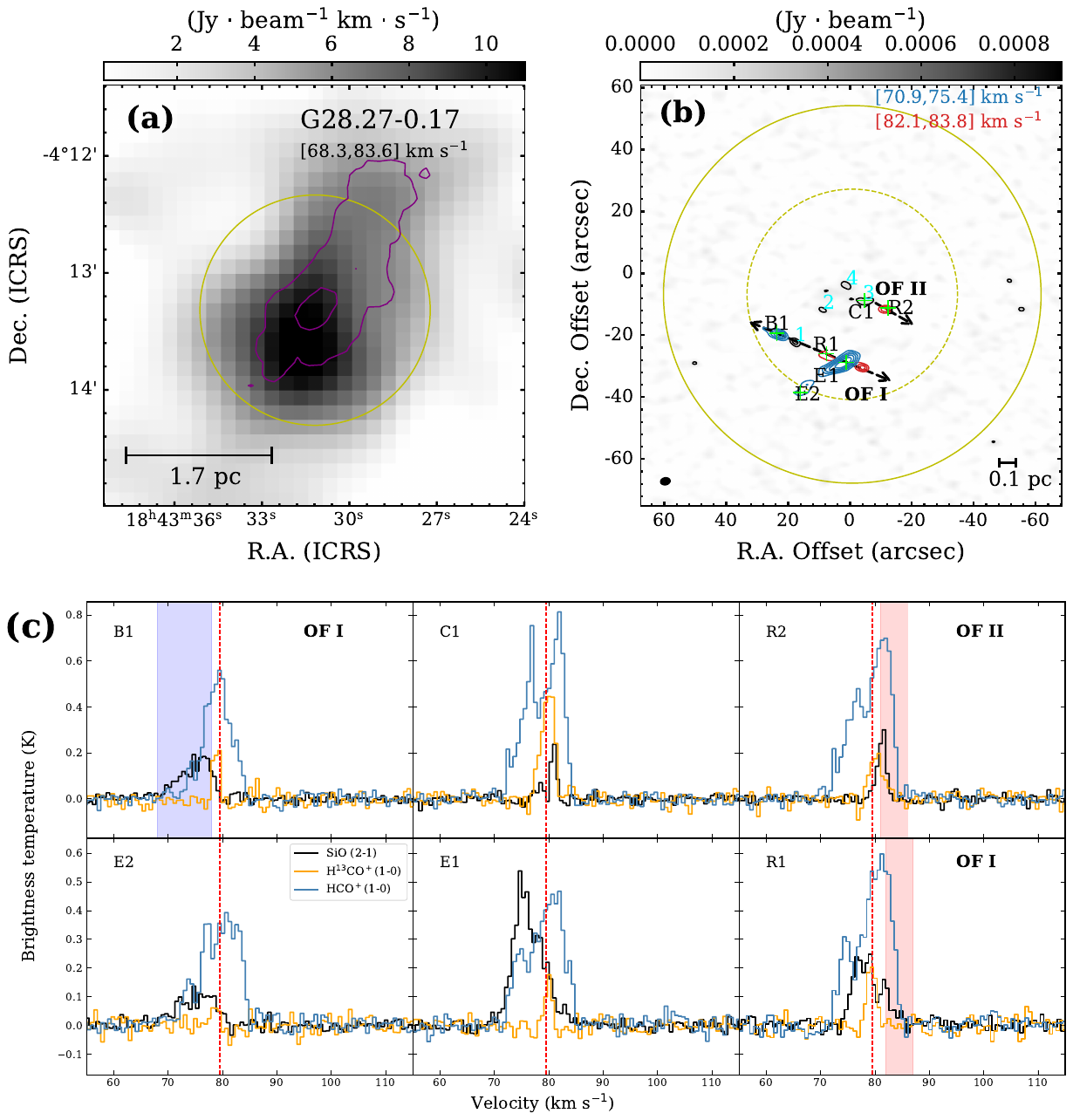}
\figsetgrpnote{The integrated intensity maps and line profiles of G28.27-0.17. The symbols and contour levels used are the same as in Figure \ref{fig:mom0_and_spectra_1}. The RMS $\sigma$ values are 0.007 and 0.04 Jy beam$^{-1}$ km~s$^{-1}$ for the blue- and red-shifted components, respectively. The vertical red dashed line indicates the systemic velocity of clump (79.5 km~s$^{-1}$).}
\figsetgrpend

\figsetgrpstart
\figsetgrpnum{1.14}
\figsetgrptitle{G28.52-0.25}
\figsetplot{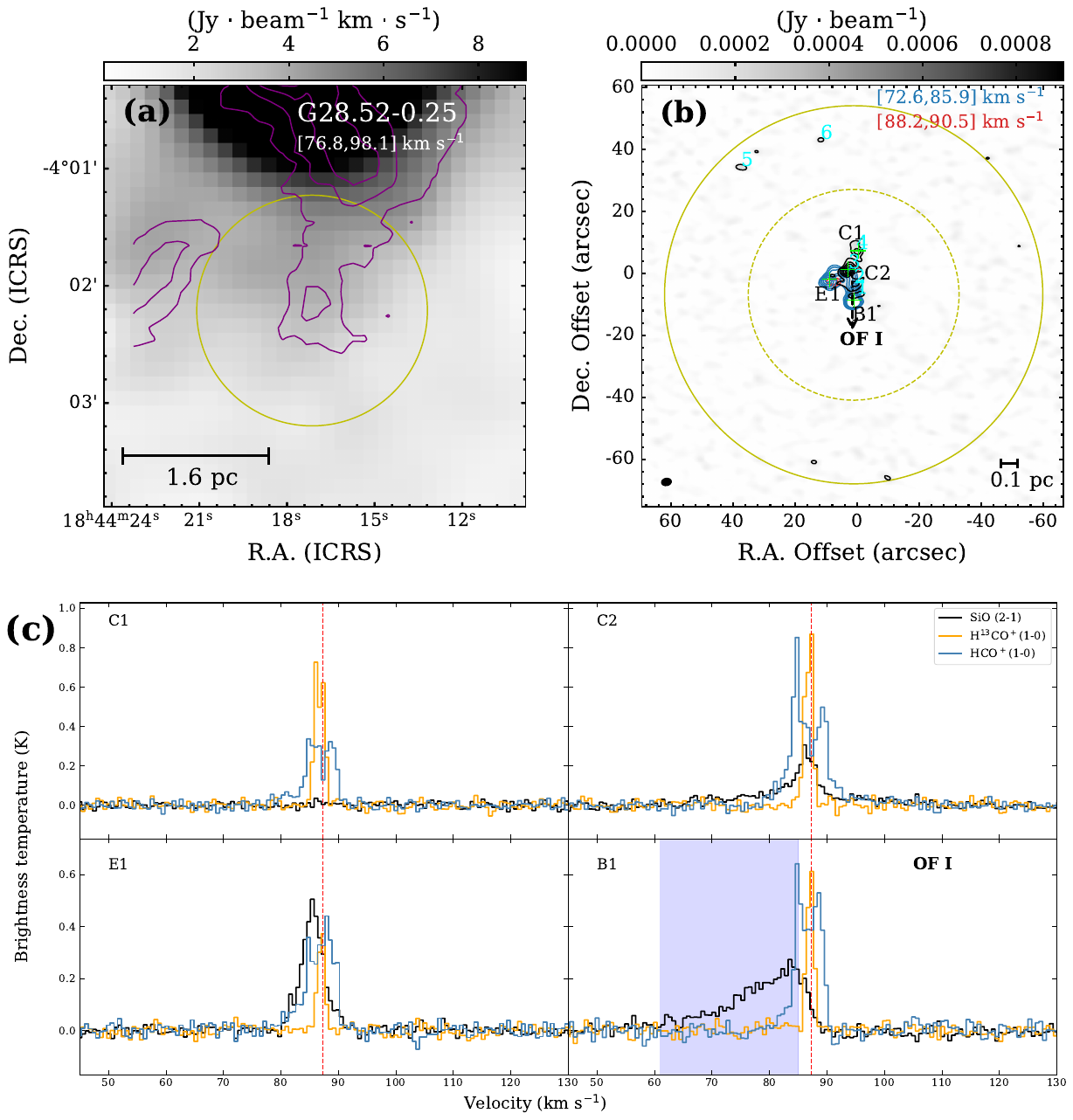}
\figsetgrpnote{The integrated intensity maps and line profiles of G28.52-0.25. The symbols and contour levels used are the same as in Figure \ref{fig:mom0_and_spectra_1}. The RMS $\sigma$ values are 0.009 and 0.007 Jy beam$^{-1}$ km~s$^{-1}$ for the blue- and red-shifted components, respectively. The vertical red dashed line indicates the systemic velocity of clump (87.3 km~s$^{-1}$).}
\figsetgrpend

\figsetgrpstart
\figsetgrpnum{1.15}
\figsetgrptitle{G28.54-0.24}
\figsetplot{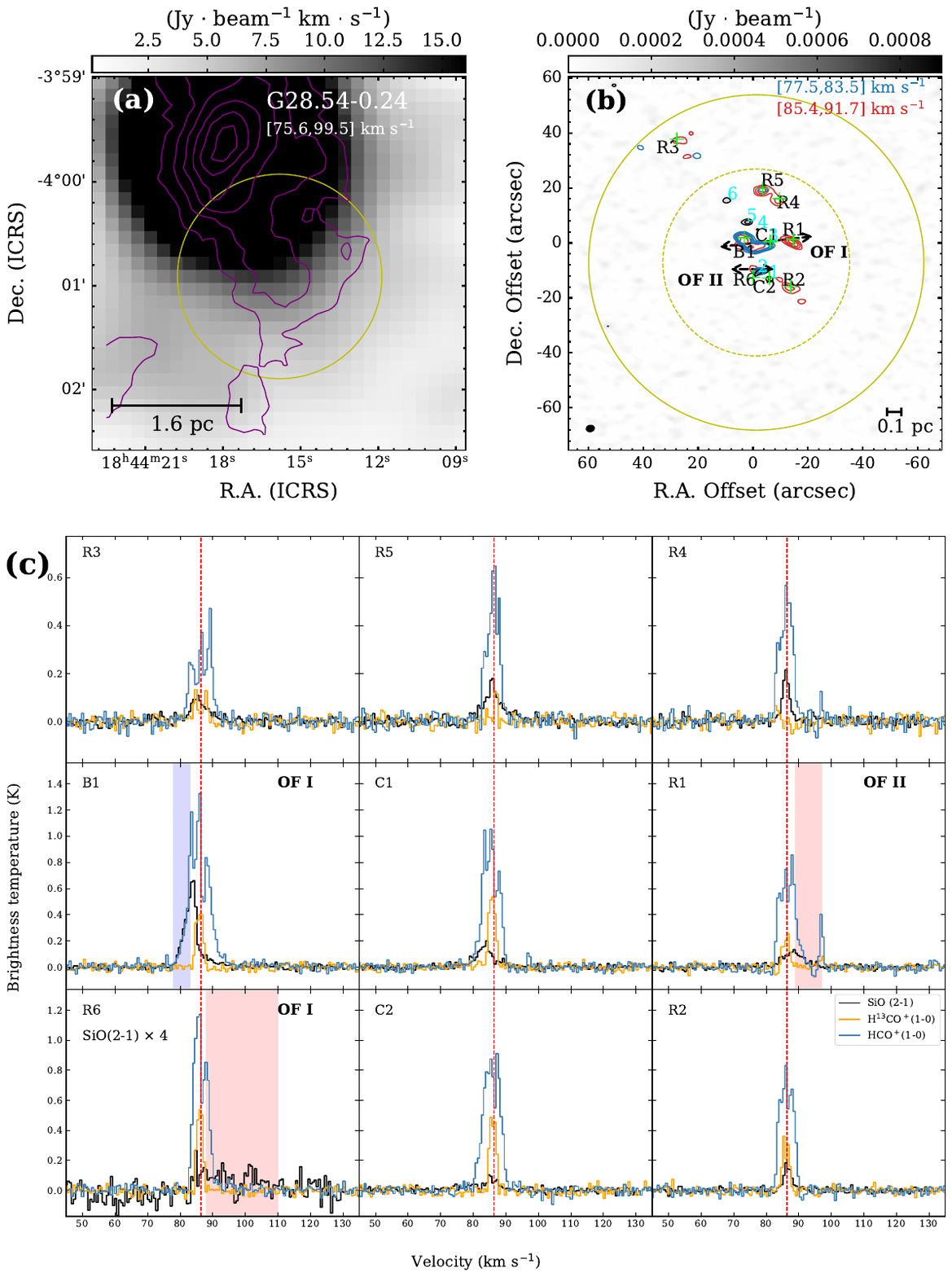}
\figsetgrpnote{The integrated intensity maps and line profiles of G28.54-0.24. The symbols and contour levels used are the same as in Figure \ref{fig:mom0_and_spectra_1}. The RMS $\sigma$ values are 0.006 and 0.007 Jy beam$^{-1}$ km~s$^{-1}$ for the blue- and red-shifted components, respectively. The vertical red dashed line indicates the systemic velocity of clump (86.4 km~s$^{-1}$).}
\figsetgrpend

\figsetend

\subsection{Outflow Morphology and Kinematics}
\label{Subsec:Morphology}
According to the definition described in Section~\ref{Subsec:Detection}, we identify 37 outflow candidates in 15 sources using the SiO (2-1) emission.
The remaining source, G34.78$-$0.57, shows no evidence of outflow activity and is therefore not included here (see Xie et al., submitted).
Among these outflow candidates, 30 are associated with dense cores.
Compared to the total number of dense cores identified in this sample, the resulting outflow detection rate is 28\%, which is higher than the 14\% reported for a sample of 70 $\mu$m dark High-mass clumps using SiO~(5-4) \citep{Li2020ApJ...903..119L}. 
Morphologically, the outflows can be categorized into two types: fourteen bipolar outflows with blue- and red-shifted lobes aligned along a common axis (e.g., OF I in G14.23-0.18, Figure~\ref{fig:mom0_and_spectra_1}) and twenty-three monopolar outflows, of which nine exhibit only blue-shifted lobes (e.g., OF II in G11.10-0.11, Fig. Set 1) and fourteen exhibit only red-shifted lobes (e.g., OF III in G14.23-0.18, Figure~\ref{fig:mom0_and_spectra_1}).
In Table \ref{tab:outflows}, features of the detected outflows are categorized to three types: type ``A'' represents the bipolar, ``B'' represents the monopolar, and ``C'' represents the knotty structure.
Note that these monopolar outflows may arise from inclination effects, and consequently the number of bipolar outflows in our sample should be regarded as a lower limit.

Moreover, among the identified outflow candidates, we also note several special cases: \\
(1) Unresolved outflows: In some regions, several outflows appear to originate from the same core. For example,
G14.23-0.18 displays an X-shaped outflow morphology (Fig. \ref{fig:mom0_and_spectra_1}). On the one hand, this structure could be interpreted as two distinct outflows, labeled I and II, both associated with Core1, which are likely unresolved at the current spatial resolution. On the other hand, the X-shaped morphology may arise from cavity walls produced by a wide-angle outflow launched from Core1, similar to those reported in previous studies \citep[e.g.,][]{Liu2025ApJ...979...17L}.
In this case, the opening angle is $\sim$30$^{\circ}$, broadly consistent with the classical opening‐angle evolution described by \citet{Arce2006ApJ...646.1070A}. 
Considering the spatial scale and the apparent angular separation between the lobes, as well as the short gas-phase lifetime of SiO in dense post-shock gas, where the pre-shock gas density of $n$ $\sim$ 10$^4$ - 10$^5$ cm$^{-3}$ is expected to be compressed by C-type shocks to $n_{\rm post}$ $\sim$ 10$^5$ - 10$^6$ cm$^{-3}$, leading to SiO freeze-out timescales of only $\sim$ 5 $\times$ 10$^2$ - 5 $\times$ 10$^3$ yr \citep[e.g.,][]{Martin-Pintado1992A&A...254..315M,
Schilke1997A&A...321..293S}, we interpret this structure as two distinct outflows, possibly originating from unresolved cores within the region.
Similar unresolved multiple sources are also shown in G11.38+0.81, G15.50-0.42, and G22.53–0.19 in our sample, and have been reported in previous studies \citep[e.g.,][]{Zhang2015ApJ...804..141Z,Li2024NatAs...8..472L}.

\begin{figure*}
    \includegraphics[width=\linewidth]{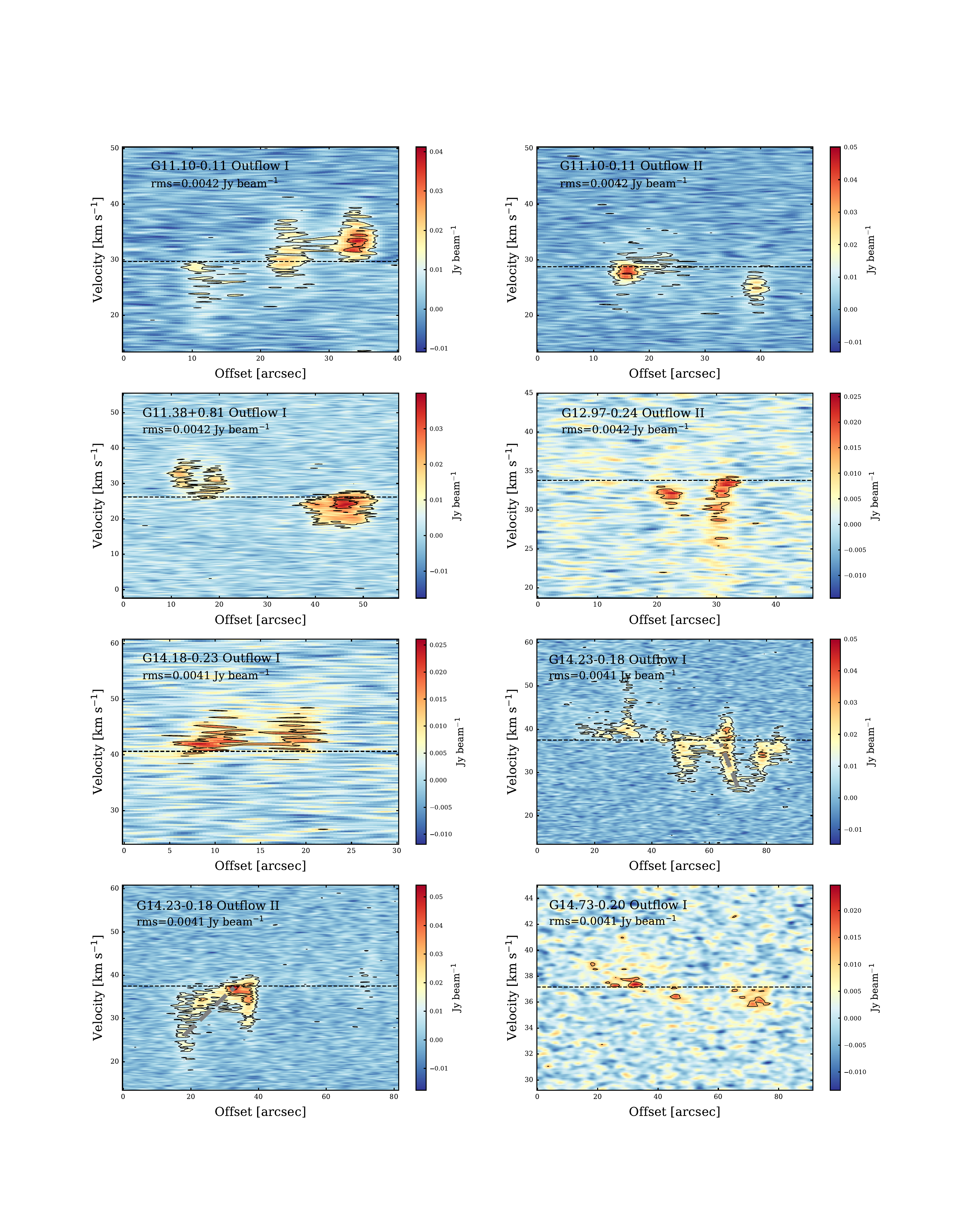}
    \caption{PV diagrams of SiO~(2–1) along the outflow direction axis (black arrows shown in the velocity integrated intensity maps (panel b) of Fig. \ref{fig:mom0_and_spectra_1}, extracted using a slice width of one synthesized beam (3.1$^{\prime\prime}$), indicating episodic injection.
Black contours represent the SiO~(2–1) emission, starting at 3$\sigma$ and increasing in steps of 4$\sigma$ (the rms $\sigma$ for each source is labeled in the respective panel). The increase in velocity with offset indicates the “Hubble wedge” structure (gray dashed lines).
The black horizontal dashed line marks the systemic velocity of the central core.}
    \label{fig:outflow_PV}
\end{figure*}

(2) Knotty structures. Eight outflows exhibit a series of knot-like features along their lobes in the integrated intensity maps (outflow categorized as type ``C'' in Table \ref{tab:outflows}).
To investigate the kinematic properties of the knotty structure, we examined their position–velocity (PV) diagrams (Figure~\ref{fig:outflow_PV}) using the ``pvextractor” tool \footnote{\url{http://pvextractor.readthedocs.org}}, which were extracted using a 3.1$^{\prime\prime}$-wide slice (comparable to the synthesized beam size) along the outflow axis, as indicated by the black dashed lines in Figure~\ref{fig:mom0_and_spectra_1}.
The PV diagrams reveal several discrete velocity peaks for these outflows.
Moreover, both outflow I and II in G14.23-0.18 exhibit “Hubble wedge” structures in their knots (indicated by gray dashed lines in Figure~\ref{fig:outflow_PV}), i.e., wedge-shaped features in the PV diagrams showing a Hubble-like relation where the velocity increases with distance from the driving source.  
Although such a structure is not observed in the remaining sources of our sample, it has been detected in our pilot source G34.74-0.12 \citep{Lin2025ApJ...990..229L} and in previous studies of 70 $\mu$m dark IRDCs \citep[e.g.,][]{Li2020ApJ...903..119L,Morii2021ApJ...923..147M}.
Such structures may be produced when episodically ejected high-velocity jets accelerate the surrounding gas as they propagate \citep{Bally2016ARA&A..54..491B, Cesaroni2018A&A...612A.103C,Guzman2024A&A...686A.143G}, or alternatively by multiple unresolved outflows.

(3) A narrow component is overlaid on a broad component. We note that this feature appears only in G18.80-0.30, where both the blueshifted lobe (B1) of Outflow III and the redshifted lobe (R1) of Outflow II exhibit such line profiles (Fig. Set 1).  
This feature is similar to the result reported by \citet{Feng2016ApJ...828..100F}, in which narrow SiO components were interpreted as arising from ion–neutral drift in the early magnetic precursor phase, while subsequent shock propagation and heating broaden the SiO profiles \citep{Jimenez-Serra2009ApJ...695..149J}. 
However, the presence of multiple outflows with different linewidths in this region may contaminate the results.
We will follow up with higher angular resolution observations to confirm whether this scenario is indeed the case.

(4) Outflow not associated with a core. 
There are four outflows (III in G11.10-0.11, IV in G15.50-0.42, III and IV in G16.30-0.53, see Fig. Set 1) that are not associated with any detected dense cores.
One possible explanation is that the driving dense cores lie below the detection limit of our current continuum observations, with a 3$\sigma$ core mass sensitivity of 0.5–2.2~$M_\odot$. For outflow IV in G15.50–0.42 and outflow IV in G16.30–0.53, only a single outflow lobe is detected, with the corresponding counter-lobe absent. In these cases, an alternative possibility is that the observed features represent secondary outflows driven by nearby detected cores.

\subsection{H$_2$ Abundance Estimation}
\label{subsec:SiO/HCO+}

To derive the outflow parameters, relatively precise estimates of molecular abundances with respect to H$_2$ are required.
However, SiO is known to be significantly enhanced in outflow and shocked regions \citep{Herbst1973ApJ...185..505H,Jorgensen2004A&A...416..603J}, rendering its abundance highly variable and uncertain in the interstellar medium. Although HCO$^+$ can also be enhanced in outflow regions, chemical modeling indicates that its abundance does not increase by several orders of magnitude, even in shocked environments \citep[e.g.,][]{Bachiller1997ApJ...487L..93B,Podio2014A&A...565A..64P}. On the other hand, HCO$^+$ traces not only outflow-related gas but also extended dense gas, as it is easily excited ($E_u$/k$_{\rm B}$ = 4.3 K) with the critical density of $\sim$ $5.9 \times 10^4$ cm$^{-3}$ at 20-50~K.

We adopt the mean H$^{13}$CO$^+$ abundance of 2.5 $\times$ 10$^{-12}$ measured with IRAM-30m for the MIAO–ALMA sample \citep{Feng2020ApJ...901..145F}, and convert it to HCO$^+$ abundance with respect to H$_2$ using the Galactocentric-distance-dependent $^{12}$C/$^{13}$C ratio \citep{Giannetti2014A&A...570A..65G}. The adopted HCO$^+$ abundances with respect to H$_2$ used for the estimation of outflow parameters (Section \ref{subsub:Derived_parameters}) for individual sources are listed in Table~\ref{tab:clumps}. 
Since H$^{13}$CO$^+$~(1–0) is generally optically thin, particularly in the outflow line wings, optical depth effects are largely mitigated, although the HCO$^+$ abundance may still be underestimated.

\section{Discussion}
\label{sec:discuss}

\subsection{Enhancement of SiO}
\label{subsec:SiO_enhancement}
SiO is known to be strongly enhanced in outflows and shocks.
To estimate the enhancement of SiO relative to HCO$^+$ in the outflow shock regions, we calculate their column-density ratio toward the line-wing emission under the assumptions of local thermodynamic equilibrium (LTE; excitation temperature equal to the kinetic temperature), optically thin emission, and a common spatial origin for both molecules.
To reduce contamination from low-velocity, non-shocked gas (i.e., quiescent dense-core material), we exclude velocity range close to $V_{\rm sys}$, which is determined from the H$^{13}$CO$^+$(1–0) emission (see Section~\ref{Subsec:Detection}).
For outflows located near the dense cores,
the integrated velocity ranges for both the SiO~(2–1) and HCO$^+$~(1–0) lines (as listed in Table~\ref{tab:outflows}) were then selected to exclude the velocity ranges where H$^{13}$CO$^+$~(1-0) shows $>$ 3 $\sigma$ emission toward the continuum cores.
For some outflows (e.g., OF II in G28.27-0.17; Fig. Set 1) that are not closely connected to the continuum core (i.e., located at a distance of one synthesized beam away, corresponding to $\sim$ 0.06 pc in projection), both HCO$^+$~(1–0) and SiO~(2–1) emission lines are assumed to be dominated by the outflowing gas.

Because precise gas temperatures are not available for the ALMA-MIAO sample, we assume that the outflow regions in these sources share the same gas kinetic temperature of 50 K as the pilot source G34.74-0.12, whose temperature was derived from fitting the H$_2$CO~(3$_{0,3}$-2$_{0,2}$), H$_2$CO~(3$_{2,2}$-2$_{2,1}$), and H$_2$CO~(3$_{2,1}$-2$_{2,0}$) transitions \citep{Lin2025ApJ...990..229L}. For simplicity, we assume the same excitation temperature for SiO~(2–1) and HCO$^+$~(1–0), equal to the gas kinetic temperature, since both transitions have similar critical densities ($\sim10^{4}$~cm$^{-3}$) and are expected to be excited in dense gas associated with shocked outflow regions.

In some regions, however, HCO$^+$ (1–0) emission is not detected toward weak and compact SiO-emitting features. 
For example, the non-detection of HCO$^+$ (1–0) emission in some regions may be attributed to limited sensitivity (OF I and III in G16.30-0.53, and OF VI in G18.80-0.30, see  Fig. Set 1), particularly for outflows located toward the edge of the primary beam (OF III in G11.10-0.11, OF IV in G16.30-0.53, and OF IV in G15.50-0.42, see Fig. Set 1).
For outflows in which HCO$^+$~(1–0) emission is not detected, we calculate lower limits on the relative SiO abundance ratio with respect to HCO$^+$.

\begin{figure}
    \includegraphics[width=\linewidth]{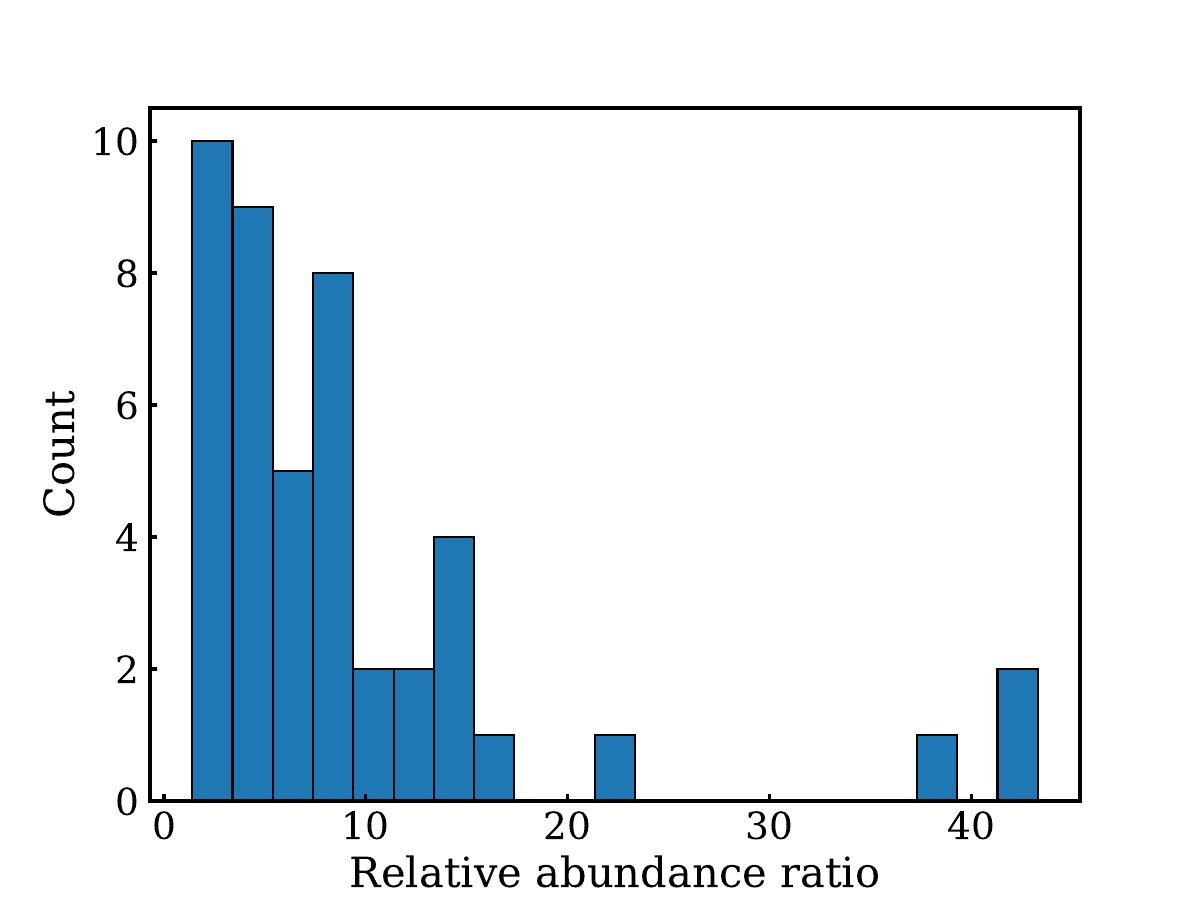}
    \caption{Histogram of the relative SiO abundance ratio with respect to HCO$^+$ for the outflow lobes detected in both SiO~(2-1) and HCO$^+$~(1-0) with S/N $>$ 3. The bin width of 2 was chosen to balance resolution and statistical significance.}
    \label{fig:abundance_ratio}
\end{figure}

We further derive the (upper limit of) relative SiO abundance with respect to H$_2$ for each outflow lobe detected in both SiO~(2–1) and HCO$^+$~(1–0) with S/N $>$ 3 in the ALMA observations, assuming optically thin emission for both lines. The abundance ratios of SiO relative to HCO$^+$ range from 1.4 to 43.3, with a mean value of 8.0 (Fig. \ref{fig:abundance_ratio}).
 The abundance ratios of SiO relative to HCO$^+$ are listed in Table \ref{tab:outflows}.
The relative SiO abundance ratios with respect to HCO$^+$ in our study are roughly one order of magnitude higher than those reported for outflows with $L/M \sim 2$–50 $L_{\odot}/M_{\odot}$ \citep[e.g.,][]{Sanchez-Monge2013A&A...557A..94S}. We note that differences in excitation assumptions, spatial filtering effects, and the treatment of optical depth between different studies may partly account for the discrepancy.
The variation in the SiO-to-HCO$^+$ abundance ratios could be linked to the evolutionary stage of the outflow, as previous studies have suggested that SiO predominantly traces the earliest, shock–dominated jet activity, whereas HCO$^+$ becomes a more effective tracer of the later, entrained outflow component as the system evolves \citep{Sakai2010ApJ...714.1658S,Sanchez-Monge2013A&A...557A..94S}. However, this interpretation should be treated with caution, as HCO$^+$~(1–0) emission may be optically thick even in the high-velocity line wings, which would lead to an underestimation of the HCO$^+$ column density and consequently an artificially enhanced SiO/HCO$^+$ ratio. Therefore, the relatively high SiO/HCO$^+$ ratios found in this work may not solely reflect the evolutionary stage of the outflows, but could also be influenced by optical depth effects.

\subsection{Outflow properties}
\label{subsec:Outflow_parameters}

\subsubsection{Observed outflow parameters}
\label{subsubsec:Observed outflow parameters}
For outflows detected in both SiO~(2-1) and HCO$^+$~(1-0), we use the SiO~(2-1) emission to determine the outflow area and velocity range, as described in Section~\ref{subsec:SiO_enhancement}, and use the HCO$^+$~(1–0) emission to calculate the integrated flux for each outflow lobe.
Note that three outflows (OF I in G11.10-0.11, OF I in G11.38+0.81, and OF I in G18.80-0.30) are associated with compact 70~$\mu$m bright ($>$ 300 MJy~sr$^{-1}$) sources in the Herschel images. The velocity-integrated HCO$^+$~(1-0) flux densities of these three outflows (OF I in G11.10-0.11, OF I in G11.38+0.81, and OF I in G18.80-0.30), which are located adjacent to 70~$\mu$m-bright regions, are typically about an order of magnitude higher than those of the outflows detected in 70~$\mu$m-dark regions in our sample (e.g., OF I in G12.97-0.24 and OF III in G14.23-0.18).
Our values are similar to those reported for outflows in the infrared dark clumps with $L/M$ $<$ 1 $L_{\odot}/M_{\odot}$ (e.g., \citet{Feng2016ApJ...828..100F,
Li2019ApJ...878...29L,Li2020ApJ...903..119L}), 
but lower than those reported in more evolved regions ($L/M$ $>$ 2 $L_{\odot}/M_{\odot}$) by one or two orders of magnitude \citep[e.g.,][]{Bontemps1996A&A...311..858B,Qin2008ApJ...677..353Q,Lu2021ApJ...909..177L,
Dutta2024AJ....167...72D,Towner2024ApJ...960...48T}. 

We measured the projected length of each red- and blue-shifted lobe of SiO outflows ($l_{\rm out}$) separately, defined as the projected distance from the driving source to the most distant SiO emission knot associated with that lobe.
These lobes have projected lengths ranging from 0.09~pc to 0.84~pc, with an average length of 0.34~pc.
The outflow timescales ($t_{\rm dyn}$) are then estimated using $l_{\rm out}$/$\Delta_v$, where $\Delta_v$ is the Full Width at Zero Intensity (FWZI)-defined velocity, which ranges from 4 to 28~km~s$^{-1}$, with a mean value of 14~km~s$^{-1}$. 
The linewidths of the outflows in our sample 
are smaller than those observed in more evolved regions such as Orion-KL, typically exceeding 50–100~km~s$^{-1}$ \citep{Zapata2012ApJ...754L..17Z,Wright2024ApJ...974..150W}.
The $t_{\rm dyn}$ in our study span from $3 \times 10^3$ to $2 \times 10^5$~yr, with an average of $4 \times 10^4$~yr and a median value of $2 \times 10^4$~yr.  
Note that this estimate does not include corrections for the outflow inclination.
These values are consistent with regions where $L/M$ $<$ 1 $L_{\odot}/M_{\odot}$ \citep{Feng2016ApJ...828..100F,
Li2020ApJ...903..119L}.
Although there are no significant differences in $t_{\rm dyn}$ compared to some earlier studies of regions with 1 $L_{\odot}/M_{\odot}$ $<$ $L/M$ $<$ 100 $L_{\odot}/M_{\odot}$ \citep[e.g.,][]{Beuther2002A&A...387..931B,Liu2021ApJ...921...96L}, 
this may be because our higher-sensitivity observations detect fainter SiO emission associated with the outflows, resulting in larger projected outflow lengths. In addition, our adopted velocity ranges exclude the low-velocity quiescent dense-core component, leading to smaller characteristic outflow velocities. Both effects contribute to larger dynamical timescales.

\subsubsection{Derived outflow parameters}
\label{subsub:Derived_parameters}
As mentioned in Section~\ref{subsec:SiO/HCO+}, the HCO$^+$ abundance derived in this work may be underestimated. Consequently, the outflow parameters estimated below, including masses ($M_{\rm out}$), momenta ($P_{\rm out}$), energies ($E_{\rm out}$), outflow rates ($\dot{M}_{\rm out}$), mechanical force ($F_{\rm out}$) and outflow energy rate ($L_{\rm out}$), should be regarded as reference values only.
For the detailed calculation formulas, see Appendix \ref{Appendix: B}. 
Assuming that the HCO$^+$~(1-0) line-wing emission is optically thin and in LTE, and adopting the HCO$^+$ abundance lower limits derived in Sect. \ref{subsec:SiO/HCO+}, all these parameters are calculated using the combined 12m, 7m, and TP datasets after applying primary beam correction, while inclination effects are not corrected for.

As listed in Table \ref{tab:outflows}, the derived outflow masses lie in the range of 0.01–3.7~$M_{\odot}$, with a mean value of 0.6~$M_{\odot}$ and a median of 0.3~$M_{\odot}$. The mass sensitivity is estimated to be 1 $\times$ 10$^{-4}$~$M_\odot$, based on the 3$\sigma$ threshold of the SiO~(2–1) emission and using the same method as that employed for deriving the outflow parameters.
The corresponding momenta and kinetic energies range from 0.06 to 56.6~$M_{\odot}$~km~s$^{-1}$, with a median value of 4.6~$M_{\odot}$~km~s$^{-1}$, and from $3.6 \times 10^{42}$ to $1.2 \times 10^{46}$~erg, with a median value of $6.0 \times 10^{44}$~erg, respectively.
Overall, our results are consistent with those reported for similar young clumps in the ASHES survey with $L/M$ $<$ 1 $L_{\odot}/M_{\odot}$ \citep{Li2020ApJ...903..119L,Lin2025ApJ...990..229L} and in G28.34 with $L/M$ $<$ 1 $L_{\odot}/M_{\odot}$ \citep[e.g.,][]{Feng2016ApJ...828..100F}.

The derived  outflow rate ($\dot{M}_{\mathrm{out}}$), mechanical force ($F_{\mathrm{out}}$), and outflow luminosity ($L_{\mathrm{out}}$) span $3.6 \times 10^{-7}$–$2.6 \times 10^{-4}$~$M_{\odot}$~yr$^{-1}$, $2.2 \times 10^{-6}$–$4.9 \times 10^{-3}$ $M_{\odot}$~km~s$^{-1}$~yr$^{-1}$, and $3.4 \times 10^{30}$–$2.9 \times 10^{34}$~erg~s$^{-1}$, respectively, with median values of $1.8 \times 10^{-5}$$M_{\odot}$~yr$^{-1}$, $2.4 \times 10^{-4}$$M_{\odot}$~km~s$^{-1}$~yr$^{-1}$, and $9.8 \times 10^{32}$~erg~s$^{-1}$.
Compared to the median values reported in the ASHES survey \citep[e.g.,][]{Li2020ApJ...903..119L} and G28.34 \citep{Feng2016ApJ...828..100F}, our estimates are consistent with their results.

We estimate the wind mass–loss rate ($\dot{M}_{\rm w}$) using $\dot{M}_{\rm w} = F_{\rm out}/v_{\rm w}$, where $F_{\rm out}$ is the outflow mechanical force and $v_{\rm w}$ is the characteristic wind velocity. A representative wind velocity of $v_{\rm w} = 100$ km s$^{-1}$ is adopted for typical low-mass stars \citep{Reipurth2001ARA&A..39..403R}. The mass accretion rate ($\dot{M}_{\rm acc}$) is then derived as $\dot{M}_{\rm acc} = k \dot{M}_{\rm w}$, assuming an accretion-to-wind mass–loss ratio of $k = 10$ based on a disk-wind model \citep{Hartmann2009apsf.book.....H}. This yields mass accretion rates in the range of 2.2 $\times$ 10$^{-7}$–4.9 $\times$ 10$^{-4}$~$M_{\odot}$ yr$^{-1}$, with a median value of 2.4 $\times$ 10$^{-4}$~$M_{\odot}$ yr$^{-1}$. The median value is broadly consistent with the accretion rates predicted by high-mass star formation models \citep[e.g.,][]{Staff2019ApJ...882..123S}, which typically require rates of $10^{-4}$–$10^{-3}$~$M_{\odot}$yr~$^{-1}$.
Compared with the accretion rates reported for the outflows in Orion-KL \citep[e.g.,][]{Wu2014ApJ...791..123W}, our measurements extend to substantially lower values, although the upper end of our distribution overlaps with the Orion-KL range.

Among all the outflows, OF II in G16.30-0.53, OF I in G28.52-0.25, and OF I in G28.54–0.24 exhibit the shortest dynamical timescales in our sample (ranging from $3.6 \times 10^3$ to $9.9 \times 10^3$ years), along with shorter lengths (0.09 to 0.24 pc), lower masses (0.03 to 0.21 $M_{\odot}$), and relatively modest kinetic energies (1.9 $\times$ 10$^{43}$ to 1.2 $\times$ 10$^{45}$ erg). Therefore, we classify them as young, low-mass outflows.

\subsubsection{Error Budget}
\label{subsec:uncertainty}
The above derived outflow parameters are subject to several sources of uncertainty, including: \\
(1) Molecular abundance uncertainties. The derived outflow parameters are sensitive to the adopted HCO$^+$ abundance. In calculating the outflow parameters, we derive the outflow parameters using the HCO$^+$~(1–0) abundance, which is obtained from the H$^{13}$CO$^+$~(1–0) abundance by adopting a source-dependent $^{12}$C/$^{13}$C ratio, as described in Section \ref{subsec:SiO/HCO+}. The adopted $^{12}$C/$^{13}$C ratio and H$^{13}$CO$^+$~(1--0) abundance may vary spatially within a clump and may therefore differ from their actual values in the outflowing gas. Therefore, these effects introduce systematic uncertainties in the adopted HCO$^+$ abundance and, consequently, in the derived outflow mass, momentum, and kinetic energy. \\
(2) Optical-depth effects. Although the use of H$^{13}$CO$^+$ significantly reduces optical depth effects, the HCO$^+$(1–0) emission still exhibits self-absorption features, suggesting that the optically thin assumption may not be fully valid, even in the outflow wings. Therefore, the derived HCO$^+$ column densities, as well as the outflow masses, momenta, and kinetic energies, may still be underestimated.\\
(3) Assumed gas temperature. We assume $T_{\rm ex} = $ $T_{\rm kin}$, as derived for G34.74–0.12 based on the H$_2$CO analysis. However, it remains uncertain whether H$_2$CO~(3-2) $K$-ladder lines, HCO$^+$~(1-0), and SiO~(2-1) trace the same physical regions, and whether the gas temperatures in the other sources are comparable to that of G34.74–0.12.
To assess the impact of this assumption, we repeat the analysis by adopting excitation temperatures of 15~K and 100~K. We find that the derived outflow mass, momentum, and kinetic energy decrease by a factor of $\sim$ 4 at 15~K and increase by a factor of $\sim$ 2 at 100~K, relative to the results obtained at 50~K. \\
(4) Outflow emission selection. The definition of the outflow emission in both velocity and area introduces additional systematic uncertainties. We select the outflow emission using velocity ranges that avoid emission dominated by quiescent dense-core gas near the systemic velocity. However, this velocity selection may exclude low-velocity outflowing gas, potentially leading to an underestimation of the derived outflow parameters in Section \ref{subsub:Derived_parameters}. In addition, the outflow area is defined using the 1$\sigma$ threshold of the SiO~(2–1) line wings, which introduces uncertainty due to sensitivity limitations. HCO$^+$(1–0) and SiO(2–1) may also trace different gas components, and their spatial mismatch may result in an underestimation of the HCO$^+$ integrated intensity and hence the derived outflow parameters. \\
(5) Unresolved outflow identification. As mentioned in Section \ref{Subsec:Morphology}, some outflows are likely to be unresolved.  
Consequently, the derived mass, momentum, and energy for them should be regarded as upper limits.
\\
(6) Inclination angle effects. No inclination correction is applied due to the lack of constraints on the outflow orientations. Consequently, the measured outflow lengths and velocities should be considered lower limits, due to projection effects onto the plane of the sky.   
The dynamical timescale, which depends on the inclination angle through a correction factor of $\cot(i)$, may therefore be overestimated or underestimated depending on the actual orientation of the outflow.
Assuming a mean inclination angle of 57.3$^{\circ}$, corresponding to a random distribution of outflow orientations, the inclination-corrected outflow velocity, momentum, energy, and mass-loss rate increase by factors ranging from 1.2 to 3.4  \citep{Dunham2014ApJ...783...29D,
Li2020ApJ...903..119L}.  \\
(7) Primary beam attenuation. Seven outflow candidates are located beyond the 50\% primary beam response (e.g., OF I in G11.10-0.11). Although we identify outflow structures in the uncorrected images with uniform noise and subsequently derive their fluxes and physical parameters from the primary-beam–corrected data, the outflow fluxes measured near the primary-beam edge are still likely to be underestimated due to reduced sensitivity, given that a relatively large primary-beam cutoff of 0.1 is adopted.

Additional sources of uncertainty include the kinematic distances to the clumps (6–12\%; Feng et al., in prep.), and the typical flux calibration uncertainty in ALMA Band 3 ($\sim$ 5\%).

\subsection{Origin of the broad Gaussian SiO emission}
\label{Subsec:Origin_of_guassian_lineprofile}
Broad SiO emission exhibits relatively Gaussian line profiles with full width at half maximum (FWHM) $>$ 2 km s$^{-1}$ and is not spatially associated with the 3 mm continuum cores (see Section \ref{Subsec:Morphology}), as seen in the regions labeled B2 in G14.69-0.22, E1 in G28.27-0.17, and R2 and R4 in G28.54-0.24.

To examine the large-scale cloud environment surrounding these sources, we analyze the $^{12}$CO~(1-0), $^{13}$CO~(1-0), and C$^{18}$O~(1-0) emission data obtained from the MWISP survey \citep{Su2019ApJS..240....9S}. We find clear evidence for CCC in G34.78-0.57, as indicated by two distinct velocity components (Xie et al., submitted.), and possible clues in G11.10-0.11, G11.38+0.81, G14.23-0.18, and G28.27-0.17. Moreover, \citet{Sanhueza2013ApJ...773..123S}, whose observations primarily targeted G28.23-0.19, also covered the G28.27-0.17 region and suggested the presence of ``subcloud–subcloud" collisions.
Therefore, the broad Gaussian SiO emission observed in G28.27–0.17 may be associated with either large-scale or small-scale CCCs, in which shock compression liberates silicon from dust grains and enhances the SiO abundance \citep[e.g.,][]{Cosentino2020MNRAS.499.1666C,Fukui2021PASJ...73S.405F}.
Alternatively, these broad Gaussian SiO emissions may arise from outflows lying close to the plane of the sky that do not exhibit prominent non-Gaussian line wings.
However, the current data do not provide sufficiently strong constraints to distinguish between these possibilities. Further observations with a larger field of view and higher spatial resolution are required to clarify the origin of the broad Gaussian SiO emission.

\subsection{Narrow-linewidth SiO emission}
\label{Subsec:Origin_of_narrow_linewidth}
As mentioned in Section \ref{Subsec:Morphology}, we found narrow-linewidth components without blue- or red-shifted line wings in the SiO (2-1) emission.
The identification of the narrow-linewidth components of SiO~(2–1) is based on the following criteria:
(i) S/N of SiO~(2-1) emission greater than 4;
(ii) spatial extent exceeding one synthesized beam size; and
(iii) a FWHM less than 1.5~km~s$^{-1}$.

At a velocity resolution of 0.21 km~s$^{-1}$, the FWHM linewidths of the narrow SiO~(2–1) linewidth components detected in G11.10-0.11 (E1), G11.38+0.81 (E1), G12.95-0.25 (E1), G14.18-0.23 (E1-E4), G14.23-0.18 (E2), and G34.78-0.57 are 1.2, 0.7, 1.1, 0.9, 1.0, and 0.6~km~s$^{-1}$, respectively (see Fig. Set 1 and Xie et al., submitted, for G34.78-0.57), which are several times larger than the thermal FWHM linewidth of SiO at 50~K ($\sim$ 0.23~km~s$^{-1}$). Their line profiles are well described by Gaussian functions and peak near the systemic velocities of the respective sources. An exception is E1 in G11.38+0.81, where the SiO~(2-1) peaks occur at the dip in the H$^{13}$CO$^+$~(1-0) and HCO$^+$~(1-0) emissions.

After integrating over the velocity range defined by the FWZI of these narrow-linewidth components, we find that their morphologies differ significantly. Only one source (G34.78-0.57) shows clumpy narrow-linewidth SiO emission spatially associated with the dense cores.

On the one hand, three narrow-linewidth components (E1 in G12.95–0.25, E1-E4 in G14.18–0.23, and E2 in G14.23–0.18) show slim, elongated morphologies, exceeding 0.33 pc in projected length.
In particular, the narrow component E1 in G12.95–0.25 partially overlaps with, and is oriented nearly perpendicular to, the redshifted outflow lobe traced by high-velocity SiO emission. In contrast, the narrow components E1-E4 in G14.18–0.23 and E2 in G14.23–0.18 show no clear spatial association with any known outflows.

\begin{figure*}
    \centering
    \includegraphics[width=\linewidth]{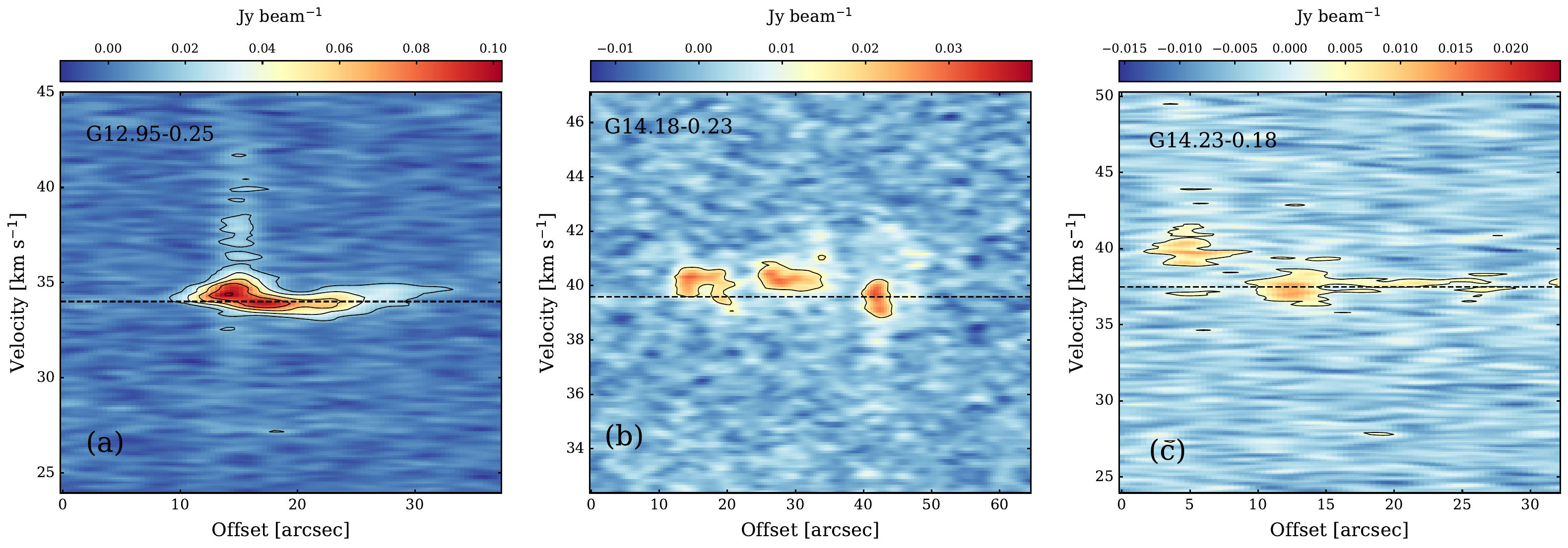}
    \caption{PV diagrams of SiO~(2–1) narrow-linewidth, spatially extended SiO~(2–1) components in G12.95-0.25, G14.18-0.23, and G14.23-0.18, extracted along the directions of the orange arrows shown in the velocity-integrated intensity maps (Fig. \ref{fig:mom0_and_spectra_1}).
    Black contours represent the narrow SiO~(2–1) component, starting at 3$\sigma$ and increasing in steps of 4$\sigma$, with 1$\sigma$ = 0.0041 Jy beam$^{-1}$. The black horizontal dashed line indicates the systemic velocity of the clump. 
    }
    \label{fig:PV}
\end{figure*}

To understand the origins of the extended narrow components, we extracted PV diagrams along their major axes in E1 in G12.95-0.25, E1-E4 in G14.18-0.23, and E2 in G14.23-0.18 (Fig. \ref{fig:PV}) to examine the velocity structure, using a slice width comparable to the synthesized beam size (3$^{\prime\prime}$.1).
The PV diagram of the SiO~(2-1) emission E1 in G12.95-0.25 reveals two distinct components: a broad velocity component associated with the red-shifted lobe of outflow I, and a narrow velocity component. The narrow component exhibits a relatively small velocity gradient. In contrast, no clear velocity gradients are observed in the spatially extended and narrow SiO components of G14.18–0.23 (E1-E4) and G14.23–0.18 (E2).
Therefore, the narrow SiO components detected in our sources may originate from the following scenarios.
(1) Considering the small velocity gradient shown in the narrow SiO~(2-1) component (E1) in G12.95-0.25, it may originate from a very young outflow. In this scenario, the shocked gas might not have undergone substantial acceleration yet, leading to narrow line widths \citep[e.g.,][]{Jimenez-Serra2004ApJ...603L..49J}.
(2) Although no clear evidence of CCCs is found for spatially extended and narrow SiO components of G14.18–0.23 (E1-E4) and G14.23–0.18 (E2) in the MWISP data, due to its linear resolution of $\sim$ 1~pc, we cannot rule out the possibility that the narrow SiO components in both regions may still originate from small-scale ``subcloud–subcloud" collisions \citep[e.g.,][]{Sanhueza2013ApJ...773..123S,Takahira2014ApJ...792...63T,Cosentino2020MNRAS.499.1666C,Yang2025A&A...694A..86Y}. 
(3) An alternative possibility is that SiO released by previous star-formation activity remains in the gas phase in relatively low-density regions, where the freeze-out timescale is long and the gas has dynamically relaxed, producing narrow, nearly Gaussian SiO line profiles without prominent high-velocity wings \citep[e.g.,][]{Lopez2016ApJ...822...85L}.

On the other hand, two narrow-linewidth SiO components (E1 in G11.10-0.11 and E1 in G11.38+0.81) exhibit compact structures with sizes comparable to only a few synthesized beams.
Because E1 in G11.10–0.11 and E1 in G11.38+0.81 are spatially located near the edges of the outflows, this suggests that the low-velocity SiO emission may originate from post-shock gas at the outflow cavities \citep[e.g.,][]{Podio2016A&A...593L...4P,Feng2022ApJ...933L..35F}, where material has been significantly decelerated due to interactions with the surrounding quiescent medium. 
Following the scenario of dissipated outflows proposed in previous studies  \citep[e.g.,][]{Lefloch1998ApJ...504L.109L,Codella1999A&A...343..585C}, we estimate the dissipation timescale of the shock as $t_{\rm dis} = \frac{8}{3} R_{\rho} \frac{L}{v_{\rm shock}}$,
where $L$ is the characteristic size of the SiO emission region, $v_{\rm shock}$ is the shock velocity, and $R_{\rho}$ denotes the ratio of the mass density of the shocked gas to that of the ambient medium, which is adopted as 10 \citep{Codella1999A&A...343..585C,Okoda2021ApJ...910...11O}.
Using an average shock velocity of 0.5~km~s$^{-1}$ measured from the SiO~(2-1) FWHM linewidth and observed shock lengths of 0.1–0.7~pc, the shock dissipation timescale is roughly estimated to be $\sim$ 10$^6$–10$^7$~yr, which is one to two orders of magnitude larger than the dynamical timescales of the outflows in these regions.
Alternatively, adopting a characteristic shock velocity of 10~km~s$^{-1}$, representative of both low-velocity shocks (e.g., \citealt{Nguyen2013ApJ...775...88N,Okoda2021ApJ...910...11O}) and the high-velocity outflow component in these regions, gives a shorter dissipation timescale of $\sim10^5$–$10^6$~yr. Nevertheless, this value remains approximately one order of magnitude longer than the outflow dynamical timescales, suggesting that the narrow component can persist before the shock is completely dissipated.
This long persistence is consistent with the scenario proposed by \citet{Lopez2016ApJ...822...85L}, in which SiO released by previous star-formation activity can remain in the gas phase for $\gtrsim 10^5$–$10^6$~yr in relatively low-density, dynamically relaxed environments.

For all the cases, we cannot rule out the possibility that projection effects may further affect the observed linewidths. When an outflow is oriented nearly parallel to the plane of the sky, its line-of-sight velocity component is significantly reduced, resulting in extremely narrow linewidth components.
In the context of outflows, we calculate the outflow parameters for these narrow components, using the same method described in Section \ref{subsec:Outflow_parameters}. The derived outflow masses of the narrow SiO components span 0.03–1.9~$M_\odot$, with corresponding momenta of 0.07–6.1~$M_\odot$~km~s$^{-1}$ and kinetic energies of $1.4 \times 10^{42}$ to $1.9 \times 10^{44}$~erg, consistent with those typically observed in outflows from low-mass protostars \citep[e.g.,][]{Wu2004A&A...426..503W,Dunham2014ApJ...783...29D}.

\subsection{Relationship between outflow and core mass}
\begin{figure*}
    \includegraphics[width=\linewidth]{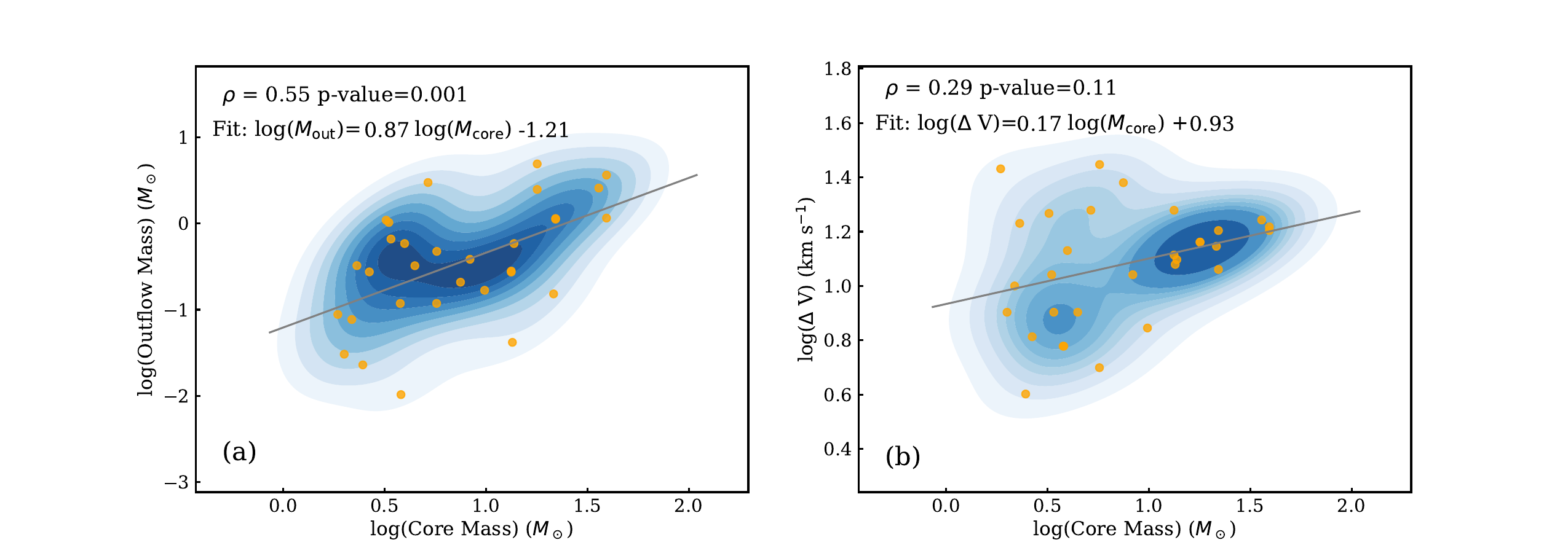}
    \caption{(a) Core mass versus outflow mass, with the bivariate Gaussian kernel density estimate (KDE) shown as blue contour lines. (b) Core mass versus outflow velocity, with a bivariate KDE. The core mass is adopted from Feng (in preparation).
    In both panels, each orange dot represents an individual outflow, while the grey lines show the linear fits. The Spearman's rank correlation coefficient ($\rho$), the corresponding p-value, and the best-fit parameters are indicated in each panel.}
    \label{fig:outflow_corr1}
\end{figure*}

The mass of an outflow likely depends on the properties of its central protostellar source. 
In Figure~\ref{fig:outflow_corr1}, we investigate the correlations between outflow masses ($M_{\mathrm{out}}$) and central core masses ($M_{\mathrm{core}}$), as well as between outflow velocity ($\Delta_v$) and $M_{\mathrm{core}}$, using $M_{\mathrm{out}}$ and $\Delta_v$ calculated as described in Section~\ref{subsec:Outflow_parameters} and  $M_{\mathrm{core}}$ adopted from Feng et al. (in preparation).  

Overall, we find that both the $M_{\mathrm{out}}$–$M_{\mathrm{core}}$ and $\Delta v$–$M_{\mathrm{core}}$ correlations exhibit slopes smaller than unity in log–log space, with best-fit slopes of 0.87 and 0.17, respectively. To assess the statistical significance of these trends, we performed a Spearman rank correlation analysis.
We find a statistically significant moderate correlation between $M_{\mathrm{out}}$ and $M_{\mathrm{core}}$ (Spearman's rank correlation coefficient $\rho$ = 0.54), indicating that more massive cores tend to drive more massive outflows. In contrast, the correlation between $\Delta_v$ and $M_{\mathrm{core}}$ is weak and not statistically significant ($\rho$ = 0.29), suggesting that the outflow velocity characteristic is not directly determined by the core mass and is likely influenced by additional factors such as inclination, evolutionary stage, and shock conditions.
This result is broadly consistent with previous studies of more evolved star-forming regions, which also find a clear link between outflow mass and core mass \citep[e.g.,][]{Beuther2002ASPC..267..341B,Zhang2005ApJ...625..864Z,Frank2014prpl.conf..451F,Maud2015MNRAS.453..645M,Lu2021ApJ...909..177L,Towner2024ApJ...960...48T,Xu2024AJ....167..285X}.
The positive $M_{\mathrm{out}}$–$M_{\mathrm{core}}$ correlation supports the idea that outflow strength primarily scales with the mass accretion rate.
Mass segregation has been found in massive protoclusters \citep[e.g.,][]{Xu2024ApJS..270....9X}.
As the core mass grows with time \citep{Xu2024ApJS..270....9X,Coletta2025A&A...696A.151C}, more massive cores will accrete more material, potentially leading to stronger outflows.
In contrast, the weak $\Delta v$–$M_{\mathrm{core}}$ relation implies that outflow kinematics are regulated by additional physical and geometrical effects beyond the core mass alone.

We note that the sources span a relatively narrow distance range (2.8–4.7 kpc). We examined the dependence of the derived correlations on source distance and find no significant evidence that distance systematically affects the observed trends. Although distance influences absolute detection limits, it impacts both quantities in a similar way and is therefore unlikely to dominate the observed correlations, although residual distance-related biases cannot be entirely excluded. 

In low-mass cores (2.5–6 $M_{\odot}$), several outflows (OF II in G11.10-0.11, OF I and II in G18.80-0.30, OF I in G14.18-0.23), with outflow masses ranging from 0.1 to 3 $M_{\odot}$, deviate from the main correlations (Fig. \ref{fig:outflow_corr1}a). 
Two outlier (OF I and II in G18.80-0.30) within this low-mass range exhibit velocities of 10-26 km$^{-1}$, and thus deviate from the correlations shown in both Fig. \ref{fig:outflow_corr1} (a) and (b). Other sources (OF II in G16.30-0.53, OF I and II in G22.53-0.19 and OF I in G28.52-0.25) contribute to the deviations in Fig. \ref{fig:outflow_corr1} (b). There are three possible reasons for these deviations, which may contribute to the uncertainties in the correlations discussed above. (1) Although the SiO~(2-1) and HCO$^+$~(1-0) emissions are recovered by combining interferometric and single-dish observations, the core masses are estimated using interferometric data alone, which may miss approximately 70–90\% of the total flux (Feng et al. in prep.). If the fraction of missing flux varies from core to core, the resulting comparison may be biased. In this case, bolometric continuum data would be required to properly correct the core masses. (2) The derived cores may not be fully resolved at the current angular resolution, leading to possible blending of multiple unresolved sources. (3) We associate each outflow with the nearest dense core, which may not always represent the true driving source and can introduce additional uncertainty in the correlation analysis.


\subsection{Relationship between outflow and $L/M$ ratio}
The $L/M$ ratio serves as a proxy for the evolutionary stage of star-forming clumps, with low values typically associated with young, cold clumps, and higher values indicating more evolved, luminous stages  \citep[e.g.,][]{Molinari2008A&A...481..345M,Molinari2016ApJ...826L...8M}. 

\begin{figure*}
    \includegraphics[width=1.0\linewidth]{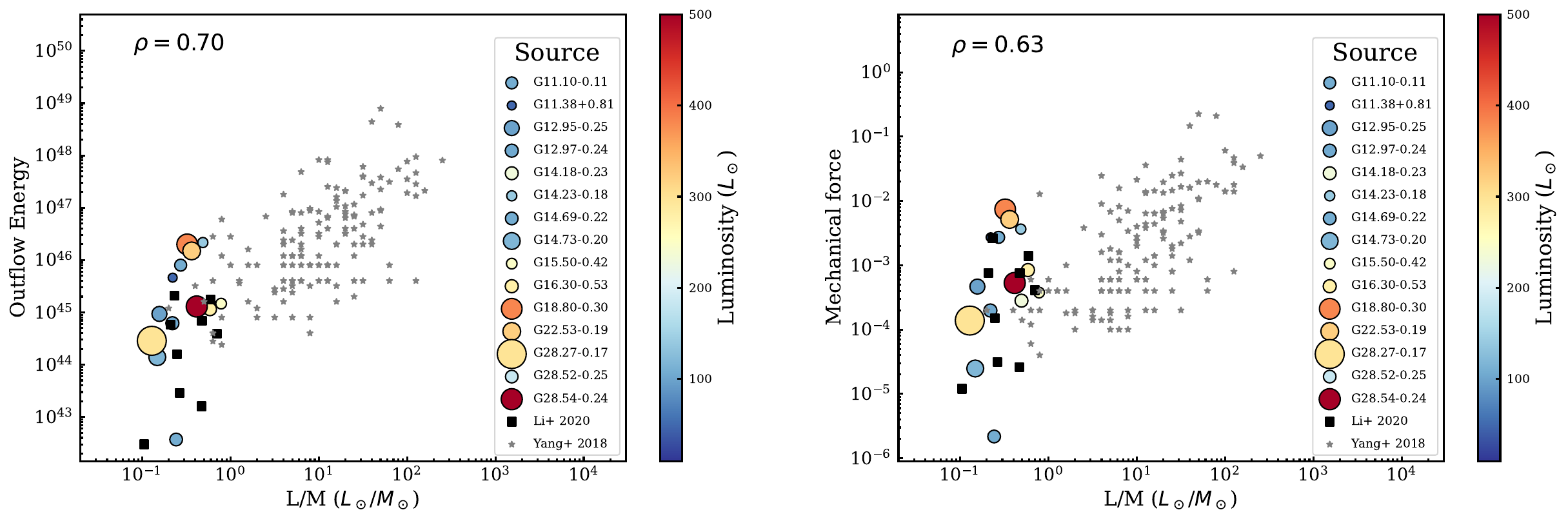}
    \caption{Outflow energy ($E_{\rm out}$) and outflow force ($F_{\rm out}$) are plotted against the $L/M$ ratio for each clump. Clump luminosities (color-coded) and masses (size-coded) for the sources in this work are adopted from \citet{Yuan2017ApJS..231...11Y}. The black squares indicate the sources from \cite{Li2020ApJ...903..119L}, and the gray stars indicate the sources from \cite{Yang2018ApJS..235....3Y}. The Spearman's rank correlation coefficients $\rho$ are provided in each panel.}
    \label{fig:corr_LM}
\end{figure*}

We sum the kinetic energy and mechanical force of all detected outflows within each clump and present them as a function of $L/M$ ratio in Fig. \ref{fig:corr_LM}.
To extend the dynamic range in $L/M$, we also include literature high-mass clumps from \citet{Yang2018ApJS..235....3Y} and \citet{Li2020ApJ...903..119L} in Fig.~\ref{fig:corr_LM}, allowing for a qualitative comparison of outflow trends across samples.  
When combined with literature samples, both the outflow energies and outflow forces show positive trends with the $L/M$ ratio, with Spearman’s rank correlation coefficients of 0.70 and 0.63, respectively. Note that these correlations are derived from the combined sample and are largely driven by the literature data. Therefore, the combined Spearman analysis is used primarily to assess the consistency of the observed trends across different samples. Overall, our results are broadly consistent with previously $L/M$–outflow correlations reported by \citet{Yang2018ApJS..235....3Y} and \citet{Li2020ApJ...903..119L}.
The positive correlations suggest that more evolved clumps, in which the central protostar has become more massive and luminous, tend to drive stronger outflows and exhibit increased mechanical feedback \citep[e.g.,][]{Maud2015MNRAS.453..645M,Bontemps1996A&A...311..858B}.

Note that the outflow energies and forces in this work are derived from HCO$^+$~(1–0) emission. The combined sample shown in Fig.~\ref{fig:corr_LM} includes results based on different molecular tracers and analysis methods, such as velocity integration ranges and inclination corrections, which may introduce additional scatter. 
In addition, the limited field of view of our single-pointing observations, which does not fully cover the entire clumps, may introduce additional uncertainties in these measurements. 
The current data do not allow a robust comparison with correlations involving turbulence energy or accretion rate. The follow-up observation will be carried in mosaic mode, covering 70 $\mu$m dark and bright clumps.

\section{Conclusions} 
\label{sec:summary}
In this paper, we study the shocks and outflows in extremely young, high-mass star-forming regions (70 $\mu$m dark, $L/M$ $<$ 1 $L_\odot$/$M_\odot$) by presenting ALMA observations of SiO~(2–1), HCO$^+$~(1-0), and H$^{13}$CO$^+$~(1-0) towards a sample of sixteen sources, as part of the MIAO survey. With a velocity resolution of 0.21~km~s$^{-1}$, we combine the data from 12 m, 7 m, and TP arrays to recover the extended structure, achieving a linear resolution of $\sim$ 0.06 pc. We find that: \\
(1) Using SiO~(2-1) emission, we identify thirty-seven outflow candidates across sixteen clumps.  Among these outflows, fourteen are classified as bipolar, nine as blue monopolar, and fourteen as red monopolar. In particular, eight outflows display knotty structures, possibly resulting from episodic accretion processes. These outflows, detected in the $L/M$ $<$ 1 $L_\odot$/$M_\odot$ regions, indicate that outflows have already begun at this early stage of high-mass star formation. \\
(2) To derive the outflow parameters, we integrated the HCO$^+$ (1–0) intensities over the line-wing velocity ranges, by excluding the emission dominated by low-velocity, quiescent dense-core gas near the systemic velocity, which is traced by the H$^{13}$CO$^+$ (1–0) line. 
We find that the outflow mass exhibits a moderately positive correlation with the mass of the central dense core, whereas the outflow velocity shows only a weak correlation.\\
(3) Benefiting from the high sensitivity of our observations and the use of the SiO~(2–1) transition, which is more easily excited than higher-$J$ transitions, our adopted methodology yields slightly narrower velocity ranges and longer projected outflow lengths compared to previous studies. As a result, we derive outflow dynamical timescales spanning $10^{3}$–$10^{5}$~yr, with a median value of $\sim 4 \times 10^{4}$~yr.\\
(4) Six narrow SiO~(2-1) components (FWHM = 0.6–1.4~km~s$^{-1}$) are detected, potentially tracing recently shocked gas, cloud–cloud collision interfaces, remnant SiO from past star formation, or outflows with elongation along the plane of the sky. 
Moreover, we detect three regions of broad Gaussian SiO emission ($>$ 2~km~s$^{-1}$) that are not associated with any 3~mm continuum cores. Although their origins remain uncertain, follow-up observations will be required to constrain their nature.

\begin{acknowledgments}
S.L. and S.F. acknowledge support from the National Key R\&D program of China grant (2025YFE0108200), National Science Foundation of China (12373023,12133008), the starting grant at Xiamen University, and the presidential excellence fund at Xiamen University. N.S. is grateful for support by the MEXT/JSPS Grant-in-Aid from the Ministry of Education, Culture, Sports, Science, and Technology of Japan (20H05845 and 25H00676) and the Chinese MOST and RIKEN Joint Research Project. PS was partially supported by a Grant-in-Aid for Scientific Research (KAKENHI Numbers JP26H02066, JP23H01221, and JP26K00748) of JSPS.
This paper makes use of the ALMA data ADS/JAO.ALMA\#2018.1.00101.S (PI: S.~Feng) and 2024.1.01696.S (PI: S.~Lin). ALMA is a partnership of ESO (representing its member states), NSF (USA) and NINS (Japan), together with NRC (Canada), MOST and ASIAA (Taiwan), and KASI (Republic of Korea), in cooperation with the Republic of Chile. The Joint ALMA Observatory is operated by ESO, AUI/NRAO and NAOJ. 
We thank the anonymous referee for the constructive comments and suggestions that improved this manuscript.
\end{acknowledgments}

\begin{contribution}
S.L. carried out the data reduction and analysis under the supervision of S.F.. S.F. conceived the project, coordinated the research activities, and served as the corresponding author. J.W. provided critical guidance and scientific insights.
All authors discussed the results and contributed to the final manuscript.


\end{contribution}

%
\facilities{ALMA}

\software{CASA \citep{McMullin2007ASPC..376..127M}, APLpy \citep{aplpy2012,aplpy2019}, \href{http://www.astropy.org}{Astropy} (\citealt{astropy:2013,astropy:2018,astropy:2022}), Matplotlib \citep{Hunter2007CSE.....9...90H}
          }


\appendix
\setcounter{section}{0}
\renewcommand{\thetable}{A\arabic{section}}
\setcounter{table}{0}
\renewcommand{\thetable}{A\arabic{table}}
\setcounter{figure}{0}
\renewcommand{\thefigure}{A\arabic{figure}}



\section{Outflow Parameters Estimation}
\label{Appendix: B}
Assuming that the HCO$^+$ (1–0) line-wing emission is optically thin and in LTE, and adopting the HCO$^+$ abundance derived in Sect. \ref{subsec:SiO/HCO+}, we derive the basic physical parameters of the outflows, including the outflow mass ($M_{\rm out}$), momentum ($P_{\rm out}$), energy ($E_{\rm out}$), and outflow rate ($\dot{M}_{\rm out}$) (Table \ref{tab:outflows}). 
The $M_{\rm out}$ is calculated from:
\begin{equation}
    M_{\rm out} = d^2X_{\rm mol}^{-1}\overline{m}_{\rm H_2} \int_{\Omega} N_{\rm mol}(\Omega)d\Omega, 
\end{equation}
where $d$ is the source distance, $\Omega$ is the total solid angle, $X_{\rm mol}$ is the abundance of the molecule with respect to H$_2$, $\overline{m}_{\rm H_2}$ is the mean mass per H$_2$ molecule, and $N_{\rm mol}$ is the column density of the molecule. The $P_{\rm out}$ and $E_{\rm out}$ are estimated from:
\begin{equation}
    P_{\rm out} = M_{\rm out} \Delta_v,
\end{equation}
\begin{equation}
    E_{\rm out} = \frac{1}{2}M_{\rm out} \Delta_v^2,
\end{equation}
where $\Delta_v$ is the velocity range defined by the FWZI.

The dynamical age and outflow rate are estimated from: 
\begin{equation}
    t_{\rm dyn} = \frac{l_{\rm out}}{\Delta_v},
\end{equation}
where $l_{\rm out}$ is the outflow physical length.

The outflow rate ($\dot{M}_{\rm out}$), mechanical force ($F_{\rm out}$), and outflow luminosity ($L_{\rm out}$) are estimated from:
\begin{equation}
    \dot{M}_{\rm out} = \frac{M_{\rm out}}{t_{\rm dyn}} ,
\end{equation}
\begin{equation}
    F_{\rm out} = P_{\rm out} / t_{\rm dyn},
\end{equation}
\begin{equation}
    L_{\rm out} = E_{\rm out} / t_{\rm dyn}.
\end{equation}

Table \ref{tab:outflows} presents the derived parameters of the outflow candidates.

\begin{longrotatetable}
\setlength{\tabcolsep}{1.9pt}       
\begin{deluxetable*}{lcccccccc|ccccccccccc}
\tabletypesize{\footnotesize}
\tablecaption{Outflow parameters}.\label{tab:outflows}
\tablewidth{-3pt}
\tablehead{
& \multicolumn{8}{c}{Measured from observations} & \multicolumn{8}{c}{Derived parameters}\\
Source &  Outflow  & $\chi_{\rm [SiO/HCO^+]}$ & Core ID\tablenotemark{\tiny  \textcolor{blue}{a}} & $V_{\rm sys}$\tablenotemark{\tiny  \textcolor{blue}{b}} & lobe  & $\Delta v$\tablenotemark{\tiny  \textcolor{blue}{c}} & Integral flux\tablenotemark{\tiny  \textcolor{blue}{d}} & $l_{\rm out}$\tablenotemark{\tiny  \textcolor{blue}{e}}  & $M_{\rm out}$ & $P_{\rm out}$ & $E_{\rm out}$  & $t_{\rm dyn}$ & $\dot{M}_{\rm out}$ & $F_{\rm out}$ & $L_{\rm out}$  & label\tablenotemark{\tiny  \textcolor{blue}{f}} \\
\colhead{} &  & \colhead{} & \colhead{} & [km~s$^{-1}$] & & [km~s$^{-1}$]  & [Jy km~s$^{-1}$] & [pc] & [$M_{\odot}$] & [$M_{\odot}$~km~s$^{-1}$] & [erg]  & [yr] & [$M_{\odot}$~yr$^{-1}$] & [$M_{\odot}$~km~s$^{-1}$~yr$^{-1}$] & [erg~s$^{-1}$] & }
\startdata
G11.10-0.11&I&4.3&12&29.7&blue&[12,28]& 0.6 & 0.21&1.59&25.44&4.05e+45&1.29e+04&1.24e-04&1.98e-03&9.97e+33 &A,C*\\
&I&10.1&12&29.7&red&[31,50] & 1.3 & 0.57&0.98&18.63&3.52e+45&2.93e+04&3.35e-05&6.36e-04&3.81e+33 & A,C*\\
&II&2.0&9&28.7&blue&[20,28]& 0.6 & 0.41&0.66&5.28&4.20e+44&5.00e+04&1.32e-05&1.06e-04&2.66e+32 & B,C\\
&III&$>38.2$&$\dagger$&-&blue&[13,28]& 0.06 &-&-&-&-&-&-&-&-& A\\
&III&$>40.2$&$\dagger$&-&red&[31,49]& 0.23 &-&-&-&-&-&-&-&-& A\\
\hline
G11.38+0.81&I&8.0&7&26.2&blue&[14,25]& 1.9&0.33 &0.75&8.30&9.08e+44&2.97e+04&2.54e-05&2.80e-04&9.70e+32 & A,C*\\
&I&4.2&7&26.2&red&[28,40]& 0.5&0.27 &0.39&4.63&5.53e+44&2.16e+04&1.78e-05&2.14e-04&8.10e+32& A,C*\\
&II&14.7&7&26.2&blue&[10,26]&3.2 &0.42 &1.12&17.92&2.85e+45&2.56e+04&4.37e-05&6.99e-04&3.52e+33&B\\
&III&8.8&6&26.6&red&
[28,42]& 0.5&0.27 &0.15&2.13&2.97e+44&1.91e+04&7.96e-06&1.11e-04&4.92e+32&B\\
\hline
G12.97-0.24&I&3.7&2&34.2&blue&[25,32]&0.09 &0.16 &0.17&1.17&8.18e+43&2.23e+04&7.53e-06&5.27e-05&1.16e+32 &B\\
&II&4.4&1&33.8&blue&[21,32]&0.43 &0.45 &0.38&4.22&4.62e+44&4.02e+04&9.55e-06&1.05e-04&3.64e+32&B,C\\
&III&10.4&3&33.8&red&[38,48]& 0.07&0.20 &0.08&0.77&7.65e+43&1.92e+04&4.01e-06&4.01e-05&1.26e+32&B\\
\hline
G12.95-0.25&I&8.1&4&33.6&blue&[18,32]& 0.83&0.20 &0.31&4.34&6.05e+44&1.40e+04&2.22e-05&3.11e-04&1.37e+33&A\\
&I&5.6&4&33.6&red&[36,47]&0.71 &0.22 &0.27&3.01&3.29e+44&1.94e+04&1.41e-05&1.55e-04&5.38e+32&A\\
\hline
G14.18-0.23&I&2.5&2&40.6&red&[42,53]&0.5 &0.45 &1.03&11.28&1.23e+45&4.01e+04&2.56e-05&2.81e-04&9.75e+32 &B,C\\
\hline
G14.23-0.18&I&5.7&4&37.5&blue&[22,36]& 3.2&0.84 &3.74&52.30&7.28e+45&5.85e+04&6.39e-05&8.95e-04&3.95e+33&A,C\\
&I&4.5&4&37.5&red&[40,55]& 1.0&0.71 &1.16&17.40&2.60e+45&4.66e+04&2.49e-05&3.73e-04&1.76e+33&A,C\\
&II&3.2&4&37.5&blue&[14,36]&1.5 &0.51 &2.43&53.44&1.17e+46&2.26e+04&1.07e-04&2.36e-03&1.64e+34&A,C\\
&II&9.2&4&37.5&red&[40,47]& 0.06&0.66 &0.05&0.33&2.30e+43&9.25e+04&5.10e-07&3.57e-06&7.87e+30&A,C\\
&III&2.4&5&37.9&red&[39,45]& 0.08&0.23 &0.12&0.71&4.23e+43&3.69e+04&3.20e-06&1.92e-05&3.63e+31&B\\
\hline
G14.69-0.22&I&43.3&4&37.4&red&[39,45]& 0.14&0.18 &0.01&0.06&3.71e+42&2.86e+04&3.62e-07&2.17e-06&4.11e+30&B\\
\hline
G14.73-0.20&I&12.4&4&37.3&blue&[32,37]&0.16 &0.64 &0.09&0.45&2.26e+43&1.25e+05&7.27e-07&3.64e-06&5.73e+30&A,C\\
&I&1.5&4&37.3&red&[38,46]& 0.06&0.57 &0.18&1.47&1.17e+44&6.92e+04&2.65e-06&2.12e-05&5.34e+31&A,C\\
\hline
G15.50-0.42&I&5.8&3&39.7&blue&[24,37]&0.33 &0.36 &0.27&3.55&4.59e+44&2.69e+04&1.02e-05&1.32e-04&5.41e+32&B\\
&II&14.5&3&39.7&red&[41,60]&0.47 &0.43 &0.28&5.33&1.01e+45&2.21e+04&1.27e-05&2.41e-04&1.44e+33&B\\
&III&8.9&2&39.4&red&[41,45]& 0.05&0.14 &0.02&0.09&3.64e+42&3.43e+04&6.68e-07&2.67e-06&3.37e+30&B\\
&IV&$>116.3$&$\dagger$&-&red&[41,50]& 0.05&- &-&-&-&-&-&-&-&A\\
\hline
G16.30-0.53&I&$>88.5$&1&-&red&[38,43]& 0.04&- &-&-&-&-&-&-&-&B\\
&II&8.5&5&38.9&blue&[10,37]& 0.08&0.10 &0.09&2.38&6.38e+44&3.70e+03&2.38e-05&6.43e-04&5.47e+33&A\\
&II&$>41.4$&5&-&red&[40,60]& 0.09&-&-&-&-&-&-&-&-&A\\
&III&16.9&$\dagger$&-&blue&[15,37]& 0.15&- & 0.11 &2.34 &5.13e+44&-&-&-&-&A\\
&III&$>40.1$&$\dagger$&-&red&[41,58]& 0.26&- &-&-&-&-&-&-&-&A\\
&IV&$>$67.8&$\dagger$&-&red&[40,60]& 0.27&- &-&-&-&-&-&-&-&B\\
\hline
G18.80-0.30&I&4.0&10&66.3&red&[67,86]& 0.87&0.22 &2.98&56.62&1.07e+46&1.16e+04&2.58e-04&4.90e-03&2.94e+34&B\\
&II&4.8&6&65.1&blue&[40,63]& 0.46&0.44 &0.74&16.97&3.88e+45&1.86e+04&3.96e-05&9.11e-04&6.60e+33&A*\\
&II&6.4&6&65.1&red&[68,82]& 0.47&0.28 &0.35&4.94&6.88e+44&1.93e+04&1.83e-05&2.56e-04&1.13e+33&(1)\\
&III&14.0&7&65.3&blue&[36,63]& 1.3&0.45 &0.47&13.27&3.69e+45&1.56e+04&3.04e-05&8.50e-04&7.50e+33&B\\
&IV&14.1&9&65.4&blue&[45,62]& 1.0&0.21 &0.32&5.51&9.31e+44&1.20e+04&2.70e-05&4.58e-04&2.46e+33&B\\
&V&12.3&8&65.4&red&[68,80]& 0.77&0.36 &0.04&0.50&5.99e+43&2.94e+04&1.42e-06&1.70e-05&6.44e+31&B\\
&VI &$>$64.5&11&65.4&blue&[49,63]& 0.27&-&-&-&-&-&-&-&-&B\\
\hline
G22.53-0.19&I&6.3&7&75.3&blue&[55,74]& 0.74&0.19 &0.71&13.49&2.55e+45&1.00e+04&7.09e-05&1.35e-03&8.07e+33&A\\
&I&1.6&7&75.3&red&[77,91]& 0.05&0.25 &0.44&6.20&8.64e+44&1.73e+04&2.57e-05&3.60e-04&1.59e+33&A\\
&II&1.6&7&75.3&blue&[56,75]& 0.66&0.36 &2.85&54.09&1.02e+46&1.83e+04&1.55e-04&2.95e-03&1.77e+34&A\\
&II&3.8&7&75.3&red&[77,90]&0.12 &0.27 &0.78&10.17&1.31e+45&2.03e+04&3.85e-05&5.00e-04&2.05e+33&A\\
\hline
G28.27-0.17&I&8.8&1&79.5&blue&[67,78]& 0.11&0.20 &0.18&2.03&2.22e+44&1.79e+04&1.03e-05&1.14e-04&3.94e+32&A\\
&I&3.2&1&79.5&red&[82,87]& 0.54&0.52 &0.14&0.69&3.45e+43&1.01e+05&1.38e-06&6.88e-06&1.08e+31&A\\
&II&2.1&3&79.9&red&[81,86]& 0.26&0.17 &0.12&0.59&2.94e+43&3.38e+04&3.50e-06&1.75e-05&2.76e+31&B\\
\hline
G28.52-0.25&I&21.6&3&87.5&blue&[61,85]& 0.85&0.24 &0.21&5.00&1.19e+45&9.88e+03&2.11e-05&5.06e-04&3.82e+33&B\\
\hline
G28.54-0.24&I&7.6&3&86.3&blue&[78,83]& 0.69 &0.27 &0.34&1.68&8.38e+43&5.25e+04&6.41e-06&3.21e-05&5.05e+31&A\\
&I&3.2&3&86.3&red&[88,110]& 0.15& 0.26&0.25&5.45&1.19e+45&1.14e+04&2.17e-05&4.77e-04&3.30e+33&A\\
&II&1.4&2&86.9&red&[89,97]& 0.02 &0.09 &0.03&0.24&1.94e+43&1.13e+04&2.69e-06&2.15e-05&5.43e+31&B\\
\enddata
\tablecomments{$^{(a)}$ Core IDs are adopted from Feng et al. (in preparation), where the symbol ``$\dagger$” denotes no cores coincident within 0.2 pc; $^{(b)}$ Systemic velocity of central cores, measured from H$^{13}$CO$^+$~(1-0) of this observation; $^{(c)}$ velocity ranges of blue- and redshifted components for parameter measurements are defined to exclude contamination from the central core emission; $^{(d)}$ HCO$^+$ (1–0) integrated flux density;
$^{(e)}$ Defined as the region where the emission exceeds $4\sigma$; $^{(f)}$ The features of the detected outflows: `A' represents the bipolar, `B' represents the monopolar, and `C' represents the knotty structure. `*' represents the outflow detected in the 70~$\mu$m bright regions.} 

\end{deluxetable*}
\end{longrotatetable}
\vspace{-32pt}
\noindent


\bibliography{ref}{}

\begin{thebibliography}{}
\expandafter\ifx\csname natexlab\endcsname\relax\def\natexlab#1{#1}\fi
\providecommand{\url}[1]{\href{#1}{#1}}
\providecommand{\dodoi}[1]{doi:~\href{http://doi.org/#1}{\nolinkurl{#1}}}
\providecommand{\doeprint}[1]{\href{http://ascl.net/#1}{\nolinkurl{http://ascl.net/#1}}}
\providecommand{\doarXiv}[1]{\href{https://arxiv.org/abs/#1}{\nolinkurl{https://arxiv.org/abs/#1}}}

\bibitem[{H.~G. {Arce} \& A.~A. {Goodman}(2001){Arce} \& {Goodman}}]{Arce2001ApJ...551L.171A}
{Arce}, H.~G., \& {Goodman}, A.~A. 2001, \bibinfo{title}{{The Mass-Velocity and Position-Velocity Relations in Episodic Outflows},} \apjl, 551, L171, \dodoi{10.1086/320031}

\bibitem[{H.~G. {Arce} \& A.~I. {Sargent}(2006){Arce} \& {Sargent}}]{Arce2006ApJ...646.1070A}
{Arce}, H.~G., \& {Sargent}, A.~I. 2006, \bibinfo{title}{{The Evolution of Outflow-Envelope Interactions in Low-Mass Protostars},} \apj, 646, 1070, \dodoi{10.1086/505104}

\bibitem[{H.~G. {Arce} {et~al.}(2007){Arce}, {Shepherd}, {Gueth}, {Lee}, {Bachiller}, {Rosen}, \& {Beuther}}]{Arce2007prpl.conf..245A}
{Arce}, H.~G., {Shepherd}, D., {Gueth}, F., {et~al.} 2007, in Protostars and Planets V, ed. B.~{Reipurth}, D.~{Jewitt}, \& K.~{Keil}, 245, \dodoi{10.48550/arXiv.astro-ph/0603071}

\bibitem[{ {Astropy Collaboration} {et~al.}(2013){Astropy Collaboration}, {Robitaille}, {Tollerud}, {Greenfield}, {Droettboom}, {Bray}, {Aldcroft}, {Davis}, {Ginsburg}, {Price-Whelan}, {Kerzendorf}, {Conley}, {Crighton}, {Barbary}, {Muna}, {Ferguson}, {Grollier}, {Parikh}, {Nair}, {Unther}, {Deil}, {Woillez}, {Conseil}, {Kramer}, {Turner}, {Singer}, {Fox}, {Weaver}, {Zabalza}, {Edwards}, {Azalee Bostroem}, {Burke}, {Casey}, {Crawford}, {Dencheva}, {Ely}, {Jenness}, {Labrie}, {Lim}, {Pierfederici}, {Pontzen}, {Ptak}, {Refsdal}, {Servillat}, \& {Streicher}}]{astropy:2013}
{Astropy Collaboration}, {Robitaille}, T.~P., {Tollerud}, E.~J., {et~al.} 2013, \bibinfo{title}{{Astropy: A community Python package for astronomy},} \aap, 558, A33, \dodoi{10.1051/0004-6361/201322068}

\bibitem[{ {Astropy Collaboration} {et~al.}(2018){Astropy Collaboration}, {Price-Whelan}, {Sip{\H{o}}cz}, {G{\"u}nther}, {Lim}, {Crawford}, {Conseil}, {Shupe}, {Craig}, {Dencheva}, {Ginsburg}, {Vand erPlas}, {Bradley}, {P{\'e}rez-Su{\'a}rez}, {de Val-Borro}, {Aldcroft}, {Cruz}, {Robitaille}, {Tollerud}, {Ardelean}, {Babej}, {Bach}, {Bachetti}, {Bakanov}, {Bamford}, {Barentsen}, {Barmby}, {Baumbach}, {Berry}, {Biscani}, {Boquien}, {Bostroem}, {Bouma}, {Brammer}, {Bray}, {Breytenbach}, {Buddelmeijer}, {Burke}, {Calderone}, {Cano Rodr{\'\i}guez}, {Cara}, {Cardoso}, {Cheedella}, {Copin}, {Corrales}, {Crichton}, {D'Avella}, {Deil}, {Depagne}, {Dietrich}, {Donath}, {Droettboom}, {Earl}, {Erben}, {Fabbro}, {Ferreira}, {Finethy}, {Fox}, {Garrison}, {Gibbons}, {Goldstein}, {Gommers}, {Greco}, {Greenfield}, {Groener}, {Grollier}, {Hagen}, {Hirst}, {Homeier}, {Horton}, {Hosseinzadeh}, {Hu}, {Hunkeler}, {Ivezi{\'c}}, {Jain}, {Jenness}, {Kanarek}, {Kendrew}, {Kern}, {Kerzendorf}, {Khvalko}, {King}, {Kirkby}, {Kulkarni},
  {Kumar}, {Lee}, {Lenz}, {Littlefair}, {Ma}, {Macleod}, {Mastropietro}, {McCully}, {Montagnac}, {Morris}, {Mueller}, {Mumford}, {Muna}, {Murphy}, {Nelson}, {Nguyen}, {Ninan}, {N{\"o}the}, {Ogaz}, {Oh}, {Parejko}, {Parley}, {Pascual}, {Patil}, {Patil}, {Plunkett}, {Prochaska}, {Rastogi}, {Reddy Janga}, {Sabater}, {Sakurikar}, {Seifert}, {Sherbert}, {Sherwood-Taylor}, {Shih}, {Sick}, {Silbiger}, {Singanamalla}, {Singer}, {Sladen}, {Sooley}, {Sornarajah}, {Streicher}, {Teuben}, {Thomas}, {Tremblay}, {Turner}, {Terr{\'o}n}, {van Kerkwijk}, {de la Vega}, {Watkins}, {Weaver}, {Whitmore}, {Woillez}, {Zabalza}, \& {Astropy Contributors}}]{astropy:2018}
{Astropy Collaboration}, {Price-Whelan}, A.~M., {Sip{\H{o}}cz}, B.~M., {et~al.} 2018, \bibinfo{title}{{The Astropy Project: Building an Open-science Project and Status of the v2.0 Core Package},} \aj, 156, 123, \dodoi{10.3847/1538-3881/aabc4f}

\bibitem[{ {Astropy Collaboration} {et~al.}(2022){Astropy Collaboration}, {Price-Whelan}, {Lim}, {Earl}, {Starkman}, {Bradley}, {Shupe}, {Patil}, {Corrales}, {Brasseur}, {N{"o}the}, {Donath}, {Tollerud}, {Morris}, {Ginsburg}, {Vaher}, {Weaver}, {Tocknell}, {Jamieson}, {van Kerkwijk}, {Robitaille}, {Merry}, {Bachetti}, {G{"u}nther}, {Aldcroft}, {Alvarado-Montes}, {Archibald}, {B{'o}di}, {Bapat}, {Barentsen}, {Baz{'a}n}, {Biswas}, {Boquien}, {Burke}, {Cara}, {Cara}, {Conroy}, {Conseil}, {Craig}, {Cross}, {Cruz}, {D'Eugenio}, {Dencheva}, {Devillepoix}, {Dietrich}, {Eigenbrot}, {Erben}, {Ferreira}, {Foreman-Mackey}, {Fox}, {Freij}, {Garg}, {Geda}, {Glattly}, {Gondhalekar}, {Gordon}, {Grant}, {Greenfield}, {Groener}, {Guest}, {Gurovich}, {Handberg}, {Hart}, {Hatfield-Dodds}, {Homeier}, {Hosseinzadeh}, {Jenness}, {Jones}, {Joseph}, {Kalmbach}, {Karamehmetoglu}, {Ka{l}uszy{'n}ski}, {Kelley}, {Kern}, {Kerzendorf}, {Koch}, {Kulumani}, {Lee}, {Ly}, {Ma}, {MacBride}, {Maljaars}, {Muna}, {Murphy}, {Norman}, {O'Steen},
  {Oman}, {Pacifici}, {Pascual}, {Pascual-Granado}, {Patil}, {Perren}, {Pickering}, {Rastogi}, {Roulston}, {Ryan}, {Rykoff}, {Sabater}, {Sakurikar}, {Salgado}, {Sanghi}, {Saunders}, {Savchenko}, {Schwardt}, {Seifert-Eckert}, {Shih}, {Jain}, {Shukla}, {Sick}, {Simpson}, {Singanamalla}, {Singer}, {Singhal}, {Sinha}, {Sip{H{o}}cz}, {Spitler}, {Stansby}, {Streicher}, {{{S}}umak}, {Swinbank}, {Taranu}, {Tewary}, {Tremblay}, {Val-Borro}, {Van Kooten}, {Vasovi{'c}}, {Verma}, {de Miranda Cardoso}, {Williams}, {Wilson}, {Winkel}, {Wood-Vasey}, {Xue}, {Yoachim}, {Zhang}, {Zonca}, \& {Astropy Project Contributors}}]{astropy:2022}
{Astropy Collaboration}, {Price-Whelan}, A.~M., {Lim}, P.~L., {et~al.} 2022, \bibinfo{title}{{The Astropy Project: Sustaining and Growing a Community-oriented Open-source Project and the Latest Major Release (v5.0) of the Core Package},} \apj, 935, 167, \dodoi{10.3847/1538-4357/ac7c74}

\bibitem[{R. {Bachiller} \& M. {P{\'e}rez Guti{\'e}rrez}(1997){Bachiller} \& {P{\'e}rez Guti{\'e}rrez}}]{Bachiller1997ApJ...487L..93B}
{Bachiller}, R., \& {P{\'e}rez Guti{\'e}rrez}, M. 1997, \bibinfo{title}{{Shock Chemistry in the Young Bipolar Outflow L1157},} \apjl, 487, L93, \dodoi{10.1086/310877}

\bibitem[{J. {Bally}(2016){Bally}}]{Bally2016ARA&A..54..491B}
{Bally}, J. 2016, \bibinfo{title}{{Protostellar Outflows},} \araa, 54, 491, \dodoi{10.1146/annurev-astro-081915-023341}

\bibitem[{H. {Beuther} {et~al.}(2014){Beuther}, {Klessen}, {Dullemond}, \& {Henning}}]{PPVI2014prpl.conf.....B}
{Beuther}, H., {Klessen}, R.~S., {Dullemond}, C.~P., \& {Henning}, T., eds. 2014, {Protostars and Planets VI}, \dodoi{10.2458/azu_uapress_9780816531240}

\bibitem[{H. {Beuther} {et~al.}(2025){Beuther}, {Kuiper}, \& {Tafalla}}]{Beuther2025ARA&A..63....1B}
{Beuther}, H., {Kuiper}, R., \& {Tafalla}, M. 2025, \bibinfo{title}{{Star Formation from Low to High Mass: A Comparative View},} \araa, 63, 1, \dodoi{10.1146/annurev-astro-013125-122023}

\bibitem[{H. {Beuther} {et~al.}(2002{\natexlab{a}}){Beuther}, {Schilke}, {Gueth}, {McCaughrean}, {Andersen}, {Sridharan}, \& {Menten}}]{Beuther2002A&A...387..931B}
{Beuther}, H., {Schilke}, P., {Gueth}, F., {et~al.} 2002{\natexlab{a}}, \bibinfo{title}{{IRAS 05358+3543: Multiple outflows at the earliest stages of massive star formation},} \aap, 387, 931, \dodoi{10.1051/0004-6361:20020319}

\bibitem[{H. {Beuther} {et~al.}(2002{\natexlab{b}}){Beuther}, {Schilke}, {Menten}, {Walmsley}, {Sridharan}, \& {Wyrowski}}]{Beuther2002ASPC..267..341B}
{Beuther}, H., {Schilke}, P., {Menten}, K.~M., {et~al.} 2002{\natexlab{b}}, in Astronomical Society of the Pacific Conference Series, Vol. 267, Hot Star Workshop III: The Earliest Phases of Massive Star Birth, ed. P.~{Crowther}, 341, \dodoi{10.48550/arXiv.astro-ph/0112508}

\bibitem[{H. {Beuther} {et~al.}(2002{\natexlab{c}}){Beuther}, {Schilke}, {Sridharan}, {Menten}, {Walmsley}, \& {Wyrowski}}]{Beuther2002A&A...383..892B}
{Beuther}, H., {Schilke}, P., {Sridharan}, T.~K., {et~al.} 2002{\natexlab{c}}, \bibinfo{title}{{Massive molecular outflows},} \aap, 383, 892, \dodoi{10.1051/0004-6361:20011808}

\bibitem[{S. {Bontemps} {et~al.}(1996){Bontemps}, {Andre}, {Terebey}, \& {Cabrit}}]{Bontemps1996A&A...311..858B}
{Bontemps}, S., {Andre}, P., {Terebey}, S., \& {Cabrit}, S. 1996, \bibinfo{title}{{Evolution of outflow activity around low-mass embedded young stellar objects},} \aap, 311, 858

\bibitem[{L.~A. {Busch} {et~al.}(2020){Busch}, {Belloche}, {Cabrit}, {Hennebelle}, \& {Commer{\c{c}}on}}]{Busch2020A&A...633A.126B}
{Busch}, L.~A., {Belloche}, A., {Cabrit}, S., {Hennebelle}, P., \& {Commer{\c{c}}on}, B. 2020, \bibinfo{title}{{The dynamically young outflow of the Class 0 protostar Cha-MMS1},} \aap, 633, A126, \dodoi{10.1051/0004-6361/201936432}

\bibitem[{ {CASA Team} {et~al.}(2022){CASA Team}, {Bean}, {Bhatnagar}, {Castro}, {Donovan Meyer}, {Emonts}, {Garcia}, {Garwood}, {Golap}, {Gonzalez Villalba}, {Harris}, {Hayashi}, {Hoskins}, {Hsieh}, {Jagannathan}, {Kawasaki}, {Keimpema}, {Kettenis}, {Lopez}, {Marvil}, {Masters}, {McNichols}, {Mehringer}, {Miel}, {Moellenbrock}, {Montesino}, {Nakazato}, {Ott}, {Petry}, {Pokorny}, {Raba}, {Rau}, {Schiebel}, {Schweighart}, {Sekhar}, {Shimada}, {Small}, {Steeb}, {Sugimoto}, {Suoranta}, {Tsutsumi}, {van Bemmel}, {Verkouter}, {Wells}, {Xiong}, {Szomoru}, {Griffith}, {Glendenning}, \& {Kern}}]{CASA2022PASP..134k4501C}
{CASA Team}, {Bean}, B., {Bhatnagar}, S., {et~al.} 2022, \bibinfo{title}{{CASA, the Common Astronomy Software Applications for Radio Astronomy},} \pasp, 134, 114501, \dodoi{10.1088/1538-3873/ac9642}

\bibitem[{P. {Caselli} {et~al.}(1997){Caselli}, {Hartquist}, \& {Havnes}}]{Caselli1997A&A...322..296C}
{Caselli}, P., {Hartquist}, T.~W., \& {Havnes}, O. 1997, \bibinfo{title}{{Grain-grain collisions and sputtering in oblique C-type shocks.},} \aap, 322, 296

\bibitem[{R. {Cesaroni} {et~al.}(2018){Cesaroni}, {Moscadelli}, {Neri}, {Sanna}, {Caratti o Garatti}, {Eisloffel}, {Stecklum}, {Ray}, \& {Walmsley}}]{Cesaroni2018A&A...612A.103C}
{Cesaroni}, R., {Moscadelli}, L., {Neri}, R., {et~al.} 2018, \bibinfo{title}{{Radio outburst from a massive (proto)star. When accretion turns into ejection},} \aap, 612, A103, \dodoi{10.1051/0004-6361/201732238}

\bibitem[{Y. {Cheng} {et~al.}(2019){Cheng}, {Qiu}, {Zhang}, {Wyrowski}, {Menten}, \& {G{\"u}sten}}]{Cheng2019ApJ...877..112C}
{Cheng}, Y., {Qiu}, K., {Zhang}, Q., {et~al.} 2019, \bibinfo{title}{{Multiline Observations of Molecular Bullets from a High-mass Protostar},} \apj, 877, 112, \dodoi{10.3847/1538-4357/ab15d4}

\bibitem[{I. {Cherchneff}(2006){Cherchneff}}]{Cherchneff2006A&A...456.1001C}
{Cherchneff}, I. 2006, \bibinfo{title}{{A chemical study of the inner winds of asymptotic giant branch stars},} \aap, 456, 1001, \dodoi{10.1051/0004-6361:20064827}

\bibitem[{C. {Codella} {et~al.}(1999){Codella}, {Bachiller}, \& {Reipurth}}]{Codella1999A&A...343..585C}
{Codella}, C., {Bachiller}, R., \& {Reipurth}, B. 1999, \bibinfo{title}{{Low and high velocity SiO emission around young stellar objects},} \aap, 343, 585

\bibitem[{A. {Coletta} {et~al.}(2025){Coletta}, {Molinari}, {Schisano}, {Traficante}, {Elia}, {Benedettini}, {Mininni}, {Soler}, {S{\'a}nchez-Monge}, {Schilke}, {Battersby}, {Fuller}, {Beuther}, {Zhang}, {Beltr{\'a}n}, {Jones}, {Klessen}, {Walch}, {Fontani}, {Avison}, {Brogan}, {Clarke}, {Hatchfield}, {Hennebelle}, {Ho}, {Hunter}, {Johnston}, {Klaassen}, {Koch}, {Kuiper}, {Lis}, {Liu}, {Lumsden}, {Maruccia}, {M{\"o}ller}, {Moscadelli}, {Nucara}, {Rigby}, {Rygl}, {Sanhueza}, {van der Tak}, {Wells}, {Wyrowski}, {De Angelis}, {Liu}, {Ahmadi}, {Bronfman}, {Liu}, {Su}, {Tang}, {Testi}, \& {Zinnecker}}]{Coletta2025A&A...696A.151C}
{Coletta}, A., {Molinari}, S., {Schisano}, E., {et~al.} 2025, \bibinfo{title}{{ALMAGAL: III. Compact source catalog: Fragmentation statistics and physical evolution of the core population},} \aap, 696, A151, \dodoi{10.1051/0004-6361/202452706}

\bibitem[{G. {Cosentino} {et~al.}(2020){Cosentino}, {Jim{\'e}nez-Serra}, {Henshaw}, {Caselli}, {Viti}, {Barnes}, {Tan}, {Fontani}, \& {Wu}}]{Cosentino2020MNRAS.499.1666C}
{Cosentino}, G., {Jim{\'e}nez-Serra}, I., {Henshaw}, J.~D., {et~al.} 2020, \bibinfo{title}{{SiO emission as a probe of cloud-cloud collisions in infrared dark clouds},} \mnras, 499, 1666, \dodoi{10.1093/mnras/staa2942}

\bibitem[{T. {Csengeri} {et~al.}(2014){Csengeri}, {Urquhart}, {Schuller}, {Motte}, {Bontemps}, {Wyrowski}, {Menten}, {Bronfman}, {Beuther}, {Henning}, {Testi}, {Zavagno}, \& {Walmsley}}]{Csengeri2014A&A...565A..75C}
{Csengeri}, T., {Urquhart}, J.~S., {Schuller}, F., {et~al.} 2014, \bibinfo{title}{{The ATLASGAL survey: a catalog of dust condensations in the Galactic plane},} \aap, 565, A75, \dodoi{10.1051/0004-6361/201322434}

\bibitem[{B.~T. {Draine} {et~al.}(1983){Draine}, {Roberge}, \& {Dalgarno}}]{Draine1983ApJ...264..485D}
{Draine}, B.~T., {Roberge}, W.~G., \& {Dalgarno}, A. 1983, \bibinfo{title}{{Magnetohydrodynamic shock waves in molecular clouds.},} \apj, 264, 485, \dodoi{10.1086/160617}

\bibitem[{M.~M. {Dunham} {et~al.}(2014){Dunham}, {Arce}, {Mardones}, {Lee}, {Matthews}, {Stutz}, \& {Williams}}]{Dunham2014ApJ...783...29D}
{Dunham}, M.~M., {Arce}, H.~G., {Mardones}, D., {et~al.} 2014, \bibinfo{title}{{Molecular Outflows Driven by Low-mass Protostars. I. Correcting for Underestimates When Measuring Outflow Masses and Dynamical Properties},} \apj, 783, 29, \dodoi{10.1088/0004-637X/783/1/29}

\bibitem[{S. {Dutta} {et~al.}(2024){Dutta}, {Lee}, {Johnstone}, {Lee}, {Hirano}, {Di Francesco}, {Moraghan}, {Liu}, {Sahu}, {Liu}, {Tatematsu}, {Goldsmith}, {Lee}, {Li}, {Eden}, {Juvela}, {Bronfman}, {Hsu}, {Kim}, {Kwon}, {Sanhueza}, {Liu}, {L{\'o}pez-V{\'a}zquez}, {Luo}, \& {Yi}}]{Dutta2024AJ....167...72D}
{Dutta}, S., {Lee}, C.-F., {Johnstone}, D., {et~al.} 2024, \bibinfo{title}{{ALMA Survey of Orion Planck Galactic Cold Clumps (ALMASOP): Molecular Jets and Episodic Accretion in Protostars},} \aj, 167, 72, \dodoi{10.3847/1538-3881/ad152b}

\bibitem[{S. {Feng} {et~al.}(2016){Feng}, {Beuther}, {Zhang}, {Liu}, {Zhang}, {Wang}, \& {Qiu}}]{Feng2016ApJ...828..100F}
{Feng}, S., {Beuther}, H., {Zhang}, Q., {et~al.} 2016, \bibinfo{title}{{Outflow Detection in a 70 {\ensuremath{\mu}}m Dark High-Mass Core},} \apj, 828, 100, \dodoi{10.3847/0004-637X/828/2/100}

\bibitem[{S. {Feng} {et~al.}(2019){Feng}, {Caselli}, {Wang}, {Lin}, {Beuther}, \& {Sipil{\"a}}}]{Feng2019ApJ...883..202F}
{Feng}, S., {Caselli}, P., {Wang}, K., {et~al.} 2019, \bibinfo{title}{{The Chemical Structure of Young High-mass Star-forming Clumps. I. Deuteration},} \apj, 883, 202, \dodoi{10.3847/1538-4357/ab3a42}

\bibitem[{S. {Feng} {et~al.}(2022){Feng}, {Liu}, {Caselli}, {Burkhardt}, {Du}, {Bachiller}, {Codella}, \& {Ceccarelli}}]{Feng2022ApJ...933L..35F}
{Feng}, S., {Liu}, H.~B., {Caselli}, P., {et~al.} 2022, \bibinfo{title}{{A Detailed Temperature Map of the Archetypal Protostellar Shocks in L1157},} \apjl, 933, L35, \dodoi{10.3847/2041-8213/ac75d7}

\bibitem[{S. {Feng} {et~al.}(2020){Feng}, {Li}, {Caselli}, {Du}, {Lin}, {Sipil{\"a}}, {Beuther}, {Sanhueza}, {Tatematsu}, {Liu}, {Zhang}, {Wang}, {Hogge}, {Jimenez-Serra}, {Lu}, {Liu}, {Wang}, {Zhang}, {Zahorecz}, {Li}, {Liu}, \& {Yuan}}]{Feng2020ApJ...901..145F}
{Feng}, S., {Li}, D., {Caselli}, P., {et~al.} 2020, \bibinfo{title}{{The Chemical Structure of Young High-mass Star-forming Clumps. II. Parsec-scale CO Depletion and Deuterium Fraction of HCO$^{+}$},} \apj, 901, 145, \dodoi{10.3847/1538-4357/abada3}

\bibitem[{M. {Fern{\'a}ndez-L{\'o}pez} {et~al.}(2013){Fern{\'a}ndez-L{\'o}pez}, {Girart}, {Curiel}, {Zapata}, {Fonfr{\'\i}a}, \& {Qiu}}]{FernandezLopez2013ApJ...778...72F}
{Fern{\'a}ndez-L{\'o}pez}, M., {Girart}, J.~M., {Curiel}, S., {et~al.} 2013, \bibinfo{title}{{Multiple Monopolar Outflows Driven by Massive Protostars in IRAS 18162-2048},} \apj, 778, 72, \dodoi{10.1088/0004-637X/778/1/72}

\bibitem[{M. {Fern{\'a}ndez-L{\'o}pez} {et~al.}(2021){Fern{\'a}ndez-L{\'o}pez}, {Sanhueza}, {Zapata}, {Stephens}, {Hull}, {Zhang}, {Girart}, {Koch}, {Cort{\'e}s}, {Silva}, {Tatematsu}, {Nakamura}, {Guzm{\'a}n}, {Nguyen Luong}, {Guzm{\'a}n Ccolque}, {Tang}, \& {Chen}}]{Fernandez2021ApJ...913...29F}
{Fern{\'a}ndez-L{\'o}pez}, M., {Sanhueza}, P., {Zapata}, L.~A., {et~al.} 2021, \bibinfo{title}{{Magnetic Fields in Massive Star-forming Regions (MagMaR). I. Linear Polarized Imaging of the Ultracompact H II Region G5.89-0.39},} \apj, 913, 29, \dodoi{10.3847/1538-4357/abf2b6}

\bibitem[{D.~R. {Flower} \& G. {Pineau Des For{\^e}ts}(2010){Flower} \& {Pineau Des For{\^e}ts}}]{Flower2010MNRAS.406.1745F}
{Flower}, D.~R., \& {Pineau Des For{\^e}ts}, G. 2010, \bibinfo{title}{{Excitation and emission of H2, CO and H2O molecules in interstellar shock waves},} \mnras, 406, 1745, \dodoi{10.1111/j.1365-2966.2010.16834.x}

\bibitem[{A. {Frank} {et~al.}(2014){Frank}, {Ray}, {Cabrit}, {Hartigan}, {Arce}, {Bacciotti}, {Bally}, {Benisty}, {Eisl{\"o}ffel}, {G{\"u}del}, {Lebedev}, {Nisini}, \& {Raga}}]{Frank2014prpl.conf..451F}
{Frank}, A., {Ray}, T.~P., {Cabrit}, S., {et~al.} 2014, in Protostars and Planets VI, ed. H.~{Beuther}, R.~S. {Klessen}, C.~P. {Dullemond}, \& T.~{Henning}, 451--474, \dodoi{10.2458/azu_uapress_9780816531240-ch020}

\bibitem[{Y. {Fukui} {et~al.}(2021){Fukui}, {Inoue}, {Hayakawa}, \& {Torii}}]{Fukui2021PASJ...73S.405F}
{Fukui}, Y., {Inoue}, T., {Hayakawa}, T., \& {Torii}, K. 2021, \bibinfo{title}{{Rapid and efficient mass collection by a supersonic cloud-cloud collision as a major mechanism of high-mass star formation},} \pasj, 73, S405, \dodoi{10.1093/pasj/psaa079}

\bibitem[{Y. {Fukui} {et~al.}(2014){Fukui}, {Ohama}, {Hanaoka}, {Furukawa}, {Torii}, {Dawson}, {Mizuno}, {Hasegawa}, {Fukuda}, {Soga}, {Moribe}, {Kuroda}, {Hayakawa}, {Kawamura}, {Kuwahara}, {Yamamoto}, {Okuda}, {Onishi}, {Maezawa}, \& {Mizuno}}]{Fukui2014ApJ...780...36F}
{Fukui}, Y., {Ohama}, A., {Hanaoka}, N., {et~al.} 2014, \bibinfo{title}{{Molecular Clouds toward the Super Star Cluster NGC 3603 Possible Evidence for a Cloud-Cloud Collision in Triggering the Cluster Formation},} \apj, 780, 36, \dodoi{10.1088/0004-637X/780/1/36}

\bibitem[{H.-P. {Gail} \& E. {Sedlmayr}(1998){Gail} \& {Sedlmayr}}]{Gail1998FaDi..109..303G}
{Gail}, H.-P., \& {Sedlmayr}, E. 1998, \bibinfo{title}{{Inorganic dust formation in astrophysical environments},} Faraday Discussions, 109, 303, \dodoi{10.1039/a709290c}

\bibitem[{H.-P. {Gail} \& E. {Sedlmayr}(1999){Gail} \& {Sedlmayr}}]{Gail1999A&A...347..594G}
{Gail}, H.-P., \& {Sedlmayr}, E. 1999, \bibinfo{title}{{Mineral formation in stellar winds. I. Condensation sequence of silicate and iron grains in stationary oxygen rich outflows},} \aap, 347, 594

\bibitem[{A. {Giannetti} {et~al.}(2014){Giannetti}, {Wyrowski}, {Brand}, {Csengeri}, {Fontani}, {Walmsley}, {Nguyen Luong}, {Beuther}, {Schuller}, {G{\"u}sten}, \& {Menten}}]{Giannetti2014A&A...570A..65G}
{Giannetti}, A., {Wyrowski}, F., {Brand}, J., {et~al.} 2014, \bibinfo{title}{{ATLASGAL-selected massive clumps in the inner Galaxy. I. CO depletion and isotopic ratios},} \aap, 570, A65, \dodoi{10.1051/0004-6361/201423692}

\bibitem[{B. {Godard} {et~al.}(2009){Godard}, {Falgarone}, \& {Pineau Des For{\^e}ts}}]{Godard2009A&A...495..847G}
{Godard}, B., {Falgarone}, E., \& {Pineau Des For{\^e}ts}, G. 2009, \bibinfo{title}{{Models of turbulent dissipation regions in the diffuse interstellar medium},} \aap, 495, 847, \dodoi{10.1051/0004-6361:200810803}

\bibitem[{A. {Gusdorf} {et~al.}(2008){Gusdorf}, {Cabrit}, {Flower}, \& {Pineau Des For{\^e}ts}}]{Gusdorf2008A&A...482..809G}
{Gusdorf}, A., {Cabrit}, S., {Flower}, D.~R., \& {Pineau Des For{\^e}ts}, G. 2008, \bibinfo{title}{{SiO line emission from C-type shock waves: interstellar jets and outflows},} \aap, 482, 809, \dodoi{10.1051/0004-6361:20078900}

\bibitem[{E. {Guzm{\'a}n Ccolque} {et~al.}(2024){Guzm{\'a}n Ccolque}, {Fern{\'a}ndez L{\'o}pez}, {Vazzano}, {de Gregorio}, {Plunkett}, \& {Santamar{\'\i}a-Miranda}}]{Guzman2024A&A...686A.143G}
{Guzm{\'a}n Ccolque}, E., {Fern{\'a}ndez L{\'o}pez}, M., {Vazzano}, M.~M., {et~al.} 2024, \bibinfo{title}{{Episodicity in accretion-ejection processes associated with IRAS 15398-3359},} \aap, 686, A143, \dodoi{10.1051/0004-6361/202348816}

\bibitem[{L. {Hartmann}(2009){Hartmann}}]{Hartmann2009apsf.book.....H}
{Hartmann}, L. 2009, {Accretion Processes in Star Formation: Second Edition}

\bibitem[{E. {Herbst} \& W. {Klemperer}(1973){Herbst} \& {Klemperer}}]{Herbst1973ApJ...185..505H}
{Herbst}, E., \& {Klemperer}, W. 1973, \bibinfo{title}{{The Formation and Depletion of Molecules in Dense Interstellar Clouds},} \apj, 185, 505, \dodoi{10.1086/152436}

\bibitem[{T. {Hirota} {et~al.}(2017){Hirota}, {Machida}, {Matsushita}, {Motogi}, {Matsumoto}, {Kim}, {Burns}, \& {Honma}}]{Hirota2017NatAs...1E.146H}
{Hirota}, T., {Machida}, M.~N., {Matsushita}, Y., {et~al.} 2017, \bibinfo{title}{{Disk-driven rotating bipolar outflow in Orion Source I},} Nature Astronomy, 1, 0146, \dodoi{10.1038/s41550-017-0146}

\bibitem[{J.~D. {Hunter}(2007){Hunter}}]{Hunter2007CSE.....9...90H}
{Hunter}, J.~D. 2007, \bibinfo{title}{{Matplotlib: A 2D Graphics Environment},} Computing in Science and Engineering, 9, 90, \dodoi{10.1109/MCSE.2007.55}

\bibitem[{N. {Issac} {et~al.}(2025){Issac}, {Lu}, {Liu}, {Zapata}, {Liu}, {Tej}, {Zhang}, {Jiao}, \& {Zhang}}]{Issac2025AJ....169..324I}
{Issac}, N., {Lu}, X., {Liu}, T., {et~al.} 2025, \bibinfo{title}{{Detection of an Explosive Outflow in G34.26+0.15},} \aj, 169, 324, \dodoi{10.3847/1538-3881/adcfa0}

\bibitem[{N. {Izumi} {et~al.}(2024){Izumi}, {Sanhueza}, {Koch}, {Lu}, {Li}, {Sabatini}, {Olguin}, {Zhang}, {Nakamura}, {Tatematsu}, {Morii}, {Sakai}, \& {Tafoya}}]{Izumi2024ApJ...963..163I}
{Izumi}, N., {Sanhueza}, P., {Koch}, P.~M., {et~al.} 2024, \bibinfo{title}{{The ALMA Survey of 70 {\ensuremath{\mu}}m Dark High-mass Clumps in Early Stages (ASHES). X. Hot Gas Reveals Deeply Embedded Star Formation},} \apj, 963, 163, \dodoi{10.3847/1538-4357/ad18c6}

\bibitem[{K.-S. {Jhan} {et~al.}(2022){Jhan}, {Lee}, {Johnstone}, {Liu}, {Liu}, {Hirano}, {Tatematsu}, {Dutta}, {Moraghan}, {Shang}, {Lee}, {Li}, {Liu}, {Hsu}, {Kwon}, {Sahu}, {Liu}, {Kim}, {Luo}, {Qin}, {Sanhueza}, {Bronfman}, {Qizhou}, {Eden}, {Traficante}, {Lee}, \& {Almasop Team}}]{Jhan2022ApJ...931L...5J}
{Jhan}, K.-S., {Lee}, C.-F., {Johnstone}, D., {et~al.} 2022, \bibinfo{title}{{ALMA Survey of Orion Planck Galactic Cold Clumps (ALMASOP): Deriving Inclination Angle and Velocity of the Protostellar Jets from Their SiO Knots},} \apjl, 931, L5, \dodoi{10.3847/2041-8213/ac6a53}

\bibitem[{I. {Jim{\'e}nez-Serra} {et~al.}(2009){Jim{\'e}nez-Serra}, {Mart{\'\i}n-Pintado}, {Caselli}, {Viti}, \& {Rodr{\'\i}guez-Franco}}]{Jimenez-Serra2009ApJ...695..149J}
{Jim{\'e}nez-Serra}, I., {Mart{\'\i}n-Pintado}, J., {Caselli}, P., {Viti}, S., \& {Rodr{\'\i}guez-Franco}, A. 2009, \bibinfo{title}{{The Evolution of Molecular Line Profiles Induced by the Propagation of C-Shock Waves},} \apj, 695, 149, \dodoi{10.1088/0004-637X/695/1/149}

\bibitem[{I. {Jim{\'e}nez-Serra} {et~al.}(2004){Jim{\'e}nez-Serra}, {Mart{\'\i}n-Pintado}, {Rodr{\'\i}guez-Franco}, \& {Marcelino}}]{Jimenez-Serra2004ApJ...603L..49J}
{Jim{\'e}nez-Serra}, I., {Mart{\'\i}n-Pintado}, J., {Rodr{\'\i}guez-Franco}, A., \& {Marcelino}, N. 2004, \bibinfo{title}{{Tracing the Shock Precursors in the L1448-mm/IRS 3 Outflows},} \apjl, 603, L49, \dodoi{10.1086/382784}

\bibitem[{J.~K. {J{\o}rgensen} {et~al.}(2004){J{\o}rgensen}, {Sch{\"o}ier}, \& {van Dishoeck}}]{Jorgensen2004A&A...416..603J}
{J{\o}rgensen}, J.~K., {Sch{\"o}ier}, F.~L., \& {van Dishoeck}, E.~F. 2004, \bibinfo{title}{{Molecular inventories and chemical evolution of low-mass protostellar envelopes},} \aap, 416, 603, \dodoi{10.1051/0004-6361:20034440}

\bibitem[{S. {Kong} {et~al.}(2019){Kong}, {Arce}, {Maureira}, {Caselli}, {Tan}, \& {Fontani}}]{Kong2019ApJ...874..104K}
{Kong}, S., {Arce}, H.~G., {Maureira}, M.~J., {et~al.} 2019, \bibinfo{title}{{Widespread Molecular Outflows in the Infrared Dark Cloud G28.37+0.07: Indications of Orthogonal Outflow-filament Alignment},} \apj, 874, 104, \dodoi{10.3847/1538-4357/ab07b9}

\bibitem[{C.-F. {Lee} {et~al.}(2007){Lee}, {Ho}, {Palau}, {Hirano}, {Bourke}, {Shang}, \& {Zhang}}]{Lee2007ApJ...670.1188L}
{Lee}, C.-F., {Ho}, P. T.~P., {Palau}, A., {et~al.} 2007, \bibinfo{title}{{Submillimeter Arcsecond-Resolution Mapping of the Highly Collimated Protostellar Jet HH 211},} \apj, 670, 1188, \dodoi{10.1086/522333}

\bibitem[{B. {Lefloch} {et~al.}(1998){Lefloch}, {Castets}, {Cernicharo}, \& {Loinard}}]{Lefloch1998ApJ...504L.109L}
{Lefloch}, B., {Castets}, A., {Cernicharo}, J., \& {Loinard}, L. 1998, \bibinfo{title}{{Widespread SiO Emission in NGC 1333},} \apjl, 504, L109, \dodoi{10.1086/311581}

\bibitem[{B. {Lefloch} {et~al.}(2012){Lefloch}, {Cabrit}, {Busquet}, {Codella}, {Ceccarelli}, {Cernicharo}, {Pardo}, {Benedettini}, {Lis}, \& {Nisini}}]{Lefloch2012ApJ...757L..25L}
{Lefloch}, B., {Cabrit}, S., {Busquet}, G., {et~al.} 2012, \bibinfo{title}{{The CHESS Survey of the L1157-B1 Shock Region: CO Spectral Signatures of Jet-driven Bow Shocks},} \apjl, 757, L25, \dodoi{10.1088/2041-8205/757/2/L25}

\bibitem[{P. {Lesaffre} {et~al.}(2013){Lesaffre}, {Pineau des For{\^e}ts}, {Godard}, {Guillard}, {Boulanger}, \& {Falgarone}}]{Lesaffre2013A&A...550A.106L}
{Lesaffre}, P., {Pineau des For{\^e}ts}, G., {Godard}, B., {et~al.} 2013, \bibinfo{title}{{Low-velocity shocks: signatures of turbulent dissipation in diffuse irradiated gas},} \aap, 550, A106, \dodoi{10.1051/0004-6361/201219928}

\bibitem[{S. {Li} {et~al.}(2019{\natexlab{a}}){Li}, {Zhang}, {Pillai}, {Stephens}, {Wang}, \& {Li}}]{Li2019ApJ...886..130L}
{Li}, S., {Zhang}, Q., {Pillai}, T., {et~al.} 2019{\natexlab{a}}, \bibinfo{title}{{Formation of Massive Protostellar Clusters{\textemdash}Observations of Massive 70 {\ensuremath{\mu}}m Dark Molecular Clouds},} \apj, 886, 130, \dodoi{10.3847/1538-4357/ab464e}

\bibitem[{S. {Li} {et~al.}(2019{\natexlab{b}}){Li}, {Wang}, {Fang}, {Zhang}, {Li}, {Zhang}, {Li}, {Zhu}, \& {Zeng}}]{Li2019ApJ...878...29L}
{Li}, S., {Wang}, J., {Fang}, M., {et~al.} 2019{\natexlab{b}}, \bibinfo{title}{{A SiO J = 5 {\textrightarrow} 4 Survey Toward Massive Star Formation Regions},} \apj, 878, 29, \dodoi{10.3847/1538-4357/ab1e4c}

\bibitem[{S. {Li} {et~al.}(2020){Li}, {Sanhueza}, {Zhang}, {Nakamura}, {Lu}, {Wang}, {Liu}, {Tatematsu}, {Jackson}, {Silva}, {Guzm{\'a}n}, {Sakai}, {Izumi}, {Tafoya}, {Li}, {Contreras}, {Morii}, \& {Kim}}]{Li2020ApJ...903..119L}
{Li}, S., {Sanhueza}, P., {Zhang}, Q., {et~al.} 2020, \bibinfo{title}{{The ALMA Survey of 70 {\ensuremath{\mu}}m Dark High-mass Clumps in Early Stages (ASHES). II. Molecular Outflows in the Extreme Early Stages of Protocluster Formation},} \apj, 903, 119, \dodoi{10.3847/1538-4357/abb81f}

\bibitem[{S. {Li} {et~al.}(2024){Li}, {Sanhueza}, {Beuther}, {Chen}, {Kuiper}, {Olguin}, {Pudritz}, {Stephens}, {Zhang}, {Nakamura}, {Lu}, {Kuruwita}, {Sakai}, {Henning}, {Taniguchi}, \& {Li}}]{Li2024NatAs...8..472L}
{Li}, S., {Sanhueza}, P., {Beuther}, H., {et~al.} 2024, \bibinfo{title}{{Observations of high-order multiplicity in a high-mass stellar protocluster},} Nature Astronomy, 8, 472, \dodoi{10.1038/s41550-023-02181-9}

\bibitem[{S. {Lin} {et~al.}(2025){Lin}, {Feng}, {Sanhueza}, {Wang}, {Zhang}, {Zhang}, {Xu}, {Wang}, {Morii}, {Liu}, {Liu}, {Wang}, {Li}, {Tafoya}, {Baan}, {Li}, \& {Sabatini}}]{Lin2025ApJ...990..229L}
{Lin}, S., {Feng}, S., {Sanhueza}, P., {et~al.} 2025, \bibinfo{title}{{The ALMA Survey of 70 {\ensuremath{\mu}}m Dark High-mass Clumps in Early Stages (ASHES). XII. Unanchored Forked Stream in the Propagating Path of a Protostellar Outflow},} \apj, 990, 229, \dodoi{10.3847/1538-4357/adf208}

\bibitem[{C.-F. {Liu} {et~al.}(2025){Liu}, {Shang}, {Johnstone}, {Ai}, {Lee}, {Krasnopolsky}, {Hirano}, {Dutta}, {Hsu}, {L{\'o}pez-V{\'a}zquez}, {Liu}, {Liu}, {Tatematsu}, {Zhang}, {Rawlings}, {Eden}, {Ren}, {Sanhueza}, {Kwon}, {Lee}, {Kuan}, {Bandopadhyay}, {V{\"a}is{\"a}l{\"a}}, {Lee}, \& {Das}}]{Liu2025ApJ...979...17L}
{Liu}, C.-F., {Shang}, H., {Johnstone}, D., {et~al.} 2025, \bibinfo{title}{{ALMA Survey of Orion Planck Galactic Cold Clumps (ALMASOP): Nested Morphological and Kinematic Structures of Outflows Revealed in SiO and CO Emission},} \apj, 979, 17, \dodoi{10.3847/1538-4357/ad9275}

\bibitem[{M. {Liu} {et~al.}(2021){Liu}, {Tan}, {Marvil}, {Kong}, {Rosero}, {Caselli}, \& {Cosentino}}]{Liu2021ApJ...921...96L}
{Liu}, M., {Tan}, J.~C., {Marvil}, J., {et~al.} 2021, \bibinfo{title}{{SiO Outflows as Tracers of Massive Star Formation in Infrared Dark Clouds},} \apj, 921, 96, \dodoi{10.3847/1538-4357/ac0829}

\bibitem[{A. {L{\'o}pez-Sepulcre} {et~al.}(2016){L{\'o}pez-Sepulcre}, {Watanabe}, {Sakai}, {Furuya}, {Saruwatari}, \& {Yamamoto}}]{Lopez2016ApJ...822...85L}
{L{\'o}pez-Sepulcre}, A., {Watanabe}, Y., {Sakai}, N., {et~al.} 2016, \bibinfo{title}{{The Role of SiO as a Tracer of Past Star-formation Events: The Case of the High-mass Protocluster NGC 2264-C},} \apj, 822, 85, \dodoi{10.3847/0004-637X/822/2/85}

\bibitem[{X. {Lu} {et~al.}(2021){Lu}, {Li}, {Ginsburg}, {Longmore}, {Kruijssen}, {Walker}, {Feng}, {Zhang}, {Battersby}, {Pillai}, {Mills}, {Kauffmann}, {Cheng}, \& {Inutsuka}}]{Lu2021ApJ...909..177L}
{Lu}, X., {Li}, S., {Ginsburg}, A., {et~al.} 2021, \bibinfo{title}{{ALMA Observations of Massive Clouds in the Central Molecular Zone: Ubiquitous Protostellar Outflows},} \apj, 909, 177, \dodoi{10.3847/1538-4357/abde3c}

\bibitem[{S.~V. {Makin} \& D. {Froebrich}(2018){Makin} \& {Froebrich}}]{Makin2018ApJS..234....8M}
{Makin}, S.~V., \& {Froebrich}, D. 2018, \bibinfo{title}{{YSO Jets in the Galactic Plane from UWISH2. IV. Jets and Outflows in Cygnus-X},} \apjs, 234, 8, \dodoi{10.3847/1538-4365/aa8862}

\bibitem[{J. {Martin-Pintado} {et~al.}(1992){Martin-Pintado}, {Bachiller}, \& {Fuente}}]{Martin-Pintado1992A&A...254..315M}
{Martin-Pintado}, J., {Bachiller}, R., \& {Fuente}, A. 1992, \bibinfo{title}{{SiO emission as a tracer of shocked gas in molecular outflows.},} \aap, 254, 315

\bibitem[{L.~T. {Maud} {et~al.}(2015){Maud}, {Moore}, {Lumsden}, {Mottram}, {Urquhart}, \& {Hoare}}]{Maud2015MNRAS.453..645M}
{Maud}, L.~T., {Moore}, T.~J.~T., {Lumsden}, S.~L., {et~al.} 2015, \bibinfo{title}{{A distance-limited sample of massive molecular outflows},} \mnras, 453, 645, \dodoi{10.1093/mnras/stv1635}

\bibitem[{J.~P. {McMullin} {et~al.}(2007){McMullin}, {Waters}, {Schiebel}, {Young}, \& {Golap}}]{McMullin2007ASPC..376..127M}
{McMullin}, J.~P., {Waters}, B., {Schiebel}, D., {Young}, W., \& {Golap}, K. 2007, in Astronomical Society of the Pacific Conference Series, Vol. 376, Astronomical Data Analysis Software and Systems XVI, ed. R.~A. {Shaw}, F.~{Hill}, \& D.~J. {Bell}, 127

\bibitem[{H. {Mikami} {et~al.}(1992){Mikami}, {Umemoto}, {Yamamoto}, \& {Saito}}]{Mikami1992ApJ...392L..87M}
{Mikami}, H., {Umemoto}, T., {Yamamoto}, S., \& {Saito}, S. 1992, \bibinfo{title}{{Detection of SiO Emission in the L1157 Dark Cloud},} \apjl, 392, L87, \dodoi{10.1086/186432}

\bibitem[{S. {Molinari} {et~al.}(2016){Molinari}, {Merello}, {Elia}, {Cesaroni}, {Testi}, \& {Robitaille}}]{Molinari2016ApJ...826L...8M}
{Molinari}, S., {Merello}, M., {Elia}, D., {et~al.} 2016, \bibinfo{title}{{Calibration of Evolutionary Diagnostics in High-mass Star Formation},} \apjl, 826, L8, \dodoi{10.3847/2041-8205/826/1/L8}

\bibitem[{S. {Molinari} {et~al.}(2008){Molinari}, {Pezzuto}, {Cesaroni}, {Brand}, {Faustini}, \& {Testi}}]{Molinari2008A&A...481..345M}
{Molinari}, S., {Pezzuto}, S., {Cesaroni}, R., {et~al.} 2008, \bibinfo{title}{{The evolution of the spectral energy distribution in massive young stellar objects},} \aap, 481, 345, \dodoi{10.1051/0004-6361:20078661}

\bibitem[{K. {Morii} {et~al.}(2021){Morii}, {Sanhueza}, {Nakamura}, {Jackson}, {Li}, {Beuther}, {Zhang}, {Feng}, {Tafoya}, {Guzm{\'a}n}, {Izumi}, {Sakai}, {Lu}, {Tatematsu}, {Ohashi}, {Silva}, {Olguin}, \& {Contreras}}]{Morii2021ApJ...923..147M}
{Morii}, K., {Sanhueza}, P., {Nakamura}, F., {et~al.} 2021, \bibinfo{title}{{The ALMA Survey of 70 {\ensuremath{\mu}}m Dark High-mass Clumps in Early Stages (ASHES). IV. Star Formation Signatures in G023.477},} \apj, 923, 147, \dodoi{10.3847/1538-4357/ac2365}

\bibitem[{K. {Morii} {et~al.}(2023){Morii}, {Sanhueza}, {Nakamura}, {Zhang}, {Sabatini}, {Beuther}, {Lu}, {Li}, {Garay}, {Jackson}, {Olguin}, {Tafoya}, {Tatematsu}, {Izumi}, {Sakai}, \& {Silva}}]{Morii2023ApJ...950..148M}
{Morii}, K., {Sanhueza}, P., {Nakamura}, F., {et~al.} 2023, \bibinfo{title}{{The ALMA Survey of 70 {\ensuremath{\mu}}m Dark High-mass Clumps in Early Stages (ASHES). IX. Physical Properties and Spatial Distribution of Cores in IRDCs},} \apj, 950, 148, \dodoi{10.3847/1538-4357/acccea}

\bibitem[{K. {Morii} {et~al.}(2024){Morii}, {Sanhueza}, {Zhang}, {Nakamura}, {Li}, {Sabatini}, {Olguin}, {Beuther}, {Tafoya}, {Izumi}, {Tatematsu}, \& {Sakai}}]{Morii2024ApJ...966..171M}
{Morii}, K., {Sanhueza}, P., {Zhang}, Q., {et~al.} 2024, \bibinfo{title}{{The ALMA Survey of 70 {\ensuremath{\mu}}m Dark High-mass Clumps in Early Stages (ASHES). XI. Statistical Study of Early Fragmentation},} \apj, 966, 171, \dodoi{10.3847/1538-4357/ad32d0}

\bibitem[{F. {Motte} {et~al.}(2018){Motte}, {Bontemps}, \& {Louvet}}]{Motte2018ARA&A..56...41M}
{Motte}, F., {Bontemps}, S., \& {Louvet}, F. 2018, \bibinfo{title}{{High-Mass Star and Massive Cluster Formation in the Milky Way},} \araa, 56, 41, \dodoi{10.1146/annurev-astro-091916-055235}

\bibitem[{Q. {Nguyen-Lu'o'ng} {et~al.}(2013){Nguyen-Lu'o'ng}, {Motte}, {Carlhoff}, {Louvet}, {Lesaffre}, {Schilke}, {Hill}, {Hennemann}, {Gusdorf}, {Didelon}, {Schneider}, {Bontemps}, {Duarte-Cabral}, {Menten}, {Martin}, {Wyrowski}, {Bendo}, {Roussel}, {Bernard}, {Bronfman}, {Henning}, {Kramer}, \& {Heitsch}}]{Nguyen2013ApJ...775...88N}
{Nguyen-Lu'o'ng}, Q., {Motte}, F., {Carlhoff}, P., {et~al.} 2013, \bibinfo{title}{{Low-velocity Shocks Traced by Extended SiO Emission along the W43 Ridges: Witnessing the Formation of Young Massive Clusters},} \apj, 775, 88, \dodoi{10.1088/0004-637X/775/2/88}

\bibitem[{T. {Nony} {et~al.}(2020){Nony}, {Motte}, {Louvet}, {Plunkett}, {Gusdorf}, {Fechtenbaum}, {Pouteau}, {Lefloch}, {Bontemps}, {Molet}, \& {Robitaille}}]{Nony2020A&A...636A..38N}
{Nony}, T., {Motte}, F., {Louvet}, F., {et~al.} 2020, \bibinfo{title}{{Episodic accretion constrained by a rich cluster of outflows},} \aap, 636, A38, \dodoi{10.1051/0004-6361/201937046}

\bibitem[{Y. {Okoda} {et~al.}(2021){Okoda}, {Oya}, {Francis}, {Johnstone}, {Inutsuka}, {Ceccarelli}, {Codella}, {Chandler}, {Sakai}, {Aikawa}, {Alves}, {Balucani}, {Bianchi}, {Bouvier}, {Caselli}, {Caux}, {Charnley}, {Choudhury}, {De Simone}, {Dulieu}, {Dur{\'a}n}, {Evans}, {Favre}, {Fedele}, {Feng}, {Fontani}, {Hama}, {Hanawa}, {Herbst}, {Hirota}, {Imai}, {Isella}, {J{\'\i}menez-Serra}, {Kahane}, {Lefloch}, {Loinard}, {L{\'o}pez-Sepulcre}, {Maud}, {Maureira}, {Menard}, {Mercimek}, {Miotello}, {Moellenbrock}, {Mori}, {Murillo}, {Nakatani}, {Nomura}, {Oba}, {O'Donoghue}, {Ohashi}, {Ospina-Zamudio}, {Pineda}, {Podio}, {Rimola}, {Sakai}, {Segura-Cox}, {Shirley}, {Svoboda}, {Taquet}, {Testi}, {Vastel}, {Viti}, {Watanabe}, {Watanabe}, {Witzel}, {Xue}, {Zhang}, {Zhao}, \& {Yamamoto}}]{Okoda2021ApJ...910...11O}
{Okoda}, Y., {Oya}, Y., {Francis}, L., {et~al.} 2021, \bibinfo{title}{{FAUST. II. Discovery of a Secondary Outflow in IRAS 15398-3359: Variability in Outflow Direction during the Earliest Stage of Star Formation?},} \apj, 910, 11, \dodoi{10.3847/1538-4357/abddb1}

\bibitem[{T. {Pillai} {et~al.}(2019){Pillai}, {Kauffmann}, {Zhang}, {Sanhueza}, {Leurini}, {Wang}, {Sridharan}, \& {K{\"o}nig}}]{Pillai2019A&A...622A..54P}
{Pillai}, T., {Kauffmann}, J., {Zhang}, Q., {et~al.} 2019, \bibinfo{title}{{Massive and low-mass protostars in massive ``starless'' cores},} \aap, 622, A54, \dodoi{10.1051/0004-6361/201732570}

\bibitem[{L. {Podio} {et~al.}(2014){Podio}, {Lefloch}, {Ceccarelli}, {Codella}, \& {Bachiller}}]{Podio2014A&A...565A..64P}
{Podio}, L., {Lefloch}, B., {Ceccarelli}, C., {Codella}, C., \& {Bachiller}, R. 2014, \bibinfo{title}{{Molecular ions in the protostellar shock L1157-B1},} \aap, 565, A64, \dodoi{10.1051/0004-6361/201322928}

\bibitem[{L. {Podio} {et~al.}(2016){Podio}, {Codella}, {Gueth}, {Cabrit}, {Maury}, {Tabone}, {Lef{\`e}vre}, {Anderl}, {Andr{\'e}}, {Belloche}, {Bontemps}, {Hennebelle}, {Lefloch}, {Maret}, \& {Testi}}]{Podio2016A&A...593L...4P}
{Podio}, L., {Codella}, C., {Gueth}, F., {et~al.} 2016, \bibinfo{title}{{First image of the L1157 molecular jet by the CALYPSO IRAM-PdBI survey},} \aap, 593, L4, \dodoi{10.1051/0004-6361/201628876}

\bibitem[{S.-L. {Qin} {et~al.}(2008){Qin}, {Zhao}, {Moran}, {Marrone}, {Patel}, {Wang}, {Liu}, \& {Kuan}}]{Qin2008ApJ...677..353Q}
{Qin}, S.-L., {Zhao}, J.-H., {Moran}, J.~M., {et~al.} 2008, \bibinfo{title}{{Infall and Outflow of Molecular Gas in Sgr B2},} \apj, 677, 353, \dodoi{10.1086/529067}

\bibitem[{K. {Qiu} {et~al.}(2008){Qiu}, {Zhang}, {Megeath}, {Gutermuth}, {Beuther}, {Shepherd}, {Sridharan}, {Testi}, \& {De Pree}}]{Qiu2008ApJ...685.1005Q}
{Qiu}, K., {Zhang}, Q., {Megeath}, S.~T., {et~al.} 2008, \bibinfo{title}{{Spitzer IRAC and MIPS Imaging of Clusters and Outflows in Nine High-Mass Star Forming Regions},} \apj, 685, 1005, \dodoi{10.1086/591044}

\bibitem[{B. {Reipurth} \& J. {Bally}(2001){Reipurth} \& {Bally}}]{Reipurth2001ARA&A..39..403R}
{Reipurth}, B., \& {Bally}, J. 2001, \bibinfo{title}{{Herbig-Haro Flows: Probes of Early Stellar Evolution},} \araa, 39, 403, \dodoi{10.1146/annurev.astro.39.1.403}

\bibitem[{T. Robitaille(2019)Robitaille}]{aplpy2019}
Robitaille, T. 2019, \bibinfo{title}{{APLpy v2.0: The Astronomical Plotting Library in Python},} \dodoi{10.5281/zenodo.2567476}

\bibitem[{T. {Robitaille} \& E. {Bressert}(2012){Robitaille} \& {Bressert}}]{aplpy2012}
{Robitaille}, T., \& {Bressert}, E. 2012, \bibinfo{title}{{APLpy: Astronomical Plotting Library in Python},}, Astrophysics Source Code Library \doeprint{1208.017}

\bibitem[{T. {Sakai} {et~al.}(2010){Sakai}, {Sakai}, {Hirota}, \& {Yamamoto}}]{Sakai2010ApJ...714.1658S}
{Sakai}, T., {Sakai}, N., {Hirota}, T., \& {Yamamoto}, S. 2010, \bibinfo{title}{{A Survey of Molecular Lines Toward Massive Clumps in Early Evolutionary Stages of High-mass Star Formation},} \apj, 714, 1658, \dodoi{10.1088/0004-637X/714/2/1658}

\bibitem[{{\'A}. {S{\'a}nchez-Monge} {et~al.}(2013){S{\'a}nchez-Monge}, {L{\'o}pez-Sepulcre}, {Cesaroni}, {Walmsley}, {Codella}, {Beltr{\'a}n}, {Pestalozzi}, \& {Molinari}}]{Sanchez-Monge2013A&A...557A..94S}
{S{\'a}nchez-Monge}, {\'A}., {L{\'o}pez-Sepulcre}, A., {Cesaroni}, R., {et~al.} 2013, \bibinfo{title}{{Evolution and excitation conditions of outflows in high-mass star-forming regions},} \aap, 557, A94, \dodoi{10.1051/0004-6361/201321589}

\bibitem[{P. {Sanhueza} {et~al.}(2013){Sanhueza}, {Jackson}, {Foster}, {Jimenez-Serra}, {Dirienzo}, \& {Pillai}}]{Sanhueza2013ApJ...773..123S}
{Sanhueza}, P., {Jackson}, J.~M., {Foster}, J.~B., {et~al.} 2013, \bibinfo{title}{{Distinct Chemical Regions in the ``Prestellar'' Infrared Dark Cloud G028.23-00.19},} \apj, 773, 123, \dodoi{10.1088/0004-637X/773/2/123}

\bibitem[{P. {Sanhueza} {et~al.}(2019){Sanhueza}, {Contreras}, {Wu}, {Jackson}, {Guzm{\'a}n}, {Zhang}, {Li}, {Lu}, {Silva}, {Izumi}, {Liu}, {Miura}, {Tatematsu}, {Sakai}, {Beuther}, {Garay}, {Ohashi}, {Saito}, {Nakamura}, {Saigo}, {Veena}, {Nguyen-Luong}, \& {Tafoya}}]{Sanhueza2019ApJ...886..102S}
{Sanhueza}, P., {Contreras}, Y., {Wu}, B., {et~al.} 2019, \bibinfo{title}{{The ALMA Survey of 70 {\ensuremath{\mu}}m Dark High-mass Clumps in Early Stages (ASHES). I. Pilot Survey: Clump Fragmentation},} \apj, 886, 102, \dodoi{10.3847/1538-4357/ab45e9}

\bibitem[{P. {Schilke} {et~al.}(1997){Schilke}, {Walmsley}, {Pineau des Forets}, \& {Flower}}]{Schilke1997A&A...321..293S}
{Schilke}, P., {Walmsley}, C.~M., {Pineau des Forets}, G., \& {Flower}, D.~R. 1997, \bibinfo{title}{{SiO production in interstellar shocks.},} \aap, 321, 293

\bibitem[{F. {Schuller} {et~al.}(2009){Schuller}, {Menten}, {Contreras}, {Wyrowski}, {Schilke}, {Bronfman}, {Henning}, {Walmsley}, {Beuther}, {Bontemps}, {Cesaroni}, {Deharveng}, {Garay}, {Herpin}, {Lefloch}, {Linz}, {Mardones}, {Minier}, {Molinari}, {Motte}, {Nyman}, {Reveret}, {Risacher}, {Russeil}, {Schneider}, {Testi}, {Troost}, {Vasyunina}, {Wienen}, {Zavagno}, {Kovacs}, {Kreysa}, {Siringo}, \& {Wei{\ss}}}]{Schuller2009A&A...504..415S}
{Schuller}, F., {Menten}, K.~M., {Contreras}, Y., {et~al.} 2009, \bibinfo{title}{{ATLASGAL - The APEX telescope large area survey of the galaxy at 870 {\ensuremath{\mu}}m},} \aap, 504, 415, \dodoi{10.1051/0004-6361/200811568}

\bibitem[{Y.~L. {Shirley} {et~al.}(2013){Shirley}, {Ellsworth-Bowers}, {Svoboda}, {Schlingman}, {Ginsburg}, {Rosolowsky}, {Gerner}, {Mairs}, {Battersby}, {Stringfellow}, {Dunham}, {Glenn}, \& {Bally}}]{Shirley2013ApJS..209....2S}
{Shirley}, Y.~L., {Ellsworth-Bowers}, T.~P., {Svoboda}, B., {et~al.} 2013, \bibinfo{title}{{The Bolocam Galactic Plane Survey. X. A Complete Spectroscopic Catalog of Dense Molecular Gas Observed toward 1.1 mm Dust Continuum Sources with 7.{\textdegree}5 <= l <= 194{\textdegree}},} \apjs, 209, 2, \dodoi{10.1088/0067-0049/209/1/2}

\bibitem[{F. {Shu} {et~al.}(1994){Shu}, {Najita}, {Ostriker}, {Wilkin}, {Ruden}, \& {Lizano}}]{Shu1994ApJ...429..781S}
{Shu}, F., {Najita}, J., {Ostriker}, E., {et~al.} 1994, \bibinfo{title}{{Magnetocentrifugally Driven Flows from Young Stars and Disks. I. A Generalized Model},} \apj, 429, 781, \dodoi{10.1086/174363}

\bibitem[{F.~H. {Shu} {et~al.}(1987){Shu}, {Adams}, \& {Lizano}}]{Shu1987ARA&A..25...23S}
{Shu}, F.~H., {Adams}, F.~C., \& {Lizano}, S. 1987, \bibinfo{title}{{Star formation in molecular clouds: observation and theory.},} \araa, 25, 23, \dodoi{10.1146/annurev.aa.25.090187.000323}

\bibitem[{R.~L. {Snell} {et~al.}(1980){Snell}, {Loren}, \& {Plambeck}}]{Snell1980ApJ...239L..17S}
{Snell}, R.~L., {Loren}, R.~B., \& {Plambeck}, R.~L. 1980, \bibinfo{title}{{Observations of CO in L 1551 : evidence for stellar wind driven shocks.},} \apjl, 239, L17, \dodoi{10.1086/183283}

\bibitem[{J.~E. {Staff} {et~al.}(2019){Staff}, {Tanaka}, \& {Tan}}]{Staff2019ApJ...882..123S}
{Staff}, J.~E., {Tanaka}, K. E.~I., \& {Tan}, J.~C. 2019, \bibinfo{title}{{Disk Wind Feedback from High-mass Protostars},} \apj, 882, 123, \dodoi{10.3847/1538-4357/ab36b3}

\bibitem[{Y. {Su} {et~al.}(2019){Su}, {Yang}, {Zhang}, {Gong}, {Wang}, {Zhou}, {Wang}, {Chen}, {Sun}, {Chen}, {Xu}, \& {Jiang}}]{Su2019ApJS..240....9S}
{Su}, Y., {Yang}, J., {Zhang}, S., {et~al.} 2019, \bibinfo{title}{{The Milky Way Imaging Scroll Painting (MWISP): Project Details and Initial Results from the Galactic Longitudes of 25.{\textdegree}8-49.{\textdegree}7},} \apjs, 240, 9, \dodoi{10.3847/1538-4365/aaf1c8}

\bibitem[{D. {Tafoya} {et~al.}(2021){Tafoya}, {Sanhueza}, {Zhang}, {Li}, {Guzm{\'a}n}, {Silva}, {de la Fuente}, {Lu}, {Morii}, {Tatematsu}, {Contreras}, {Izumi}, {Jackson}, {Nakamura}, \& {Sakai}}]{Tafoya2021ApJ...913..131T}
{Tafoya}, D., {Sanhueza}, P., {Zhang}, Q., {et~al.} 2021, \bibinfo{title}{{The ALMA Survey of 70 {\ensuremath{\mu}}m Dark High-mass Clumps in Early Stages (ASHES). III. A Young Molecular Outflow Driven by a Decelerating Jet},} \apj, 913, 131, \dodoi{10.3847/1538-4357/abf5da}

\bibitem[{K. {Takahira} {et~al.}(2014){Takahira}, {Tasker}, \& {Habe}}]{Takahira2014ApJ...792...63T}
{Takahira}, K., {Tasker}, E.~J., \& {Habe}, A. 2014, \bibinfo{title}{{Do Cloud-Cloud Collisions Trigger High-mass Star Formation? I. Small Cloud Collisions},} \apj, 792, 63, \dodoi{10.1088/0004-637X/792/1/63}

\bibitem[{A.~P.~M. {Towner} {et~al.}(2024){Towner}, {Ginsburg}, {Dell'Ova}, {Gusdorf}, {Bontemps}, {Csengeri}, {Galv{\'a}n-Madrid}, {Louvet}, {Motte}, {Sanhueza}, {Stutz}, {Bally}, {Baug}, {Chen}, {Cunningham}, {Fern{\'a}ndez-L{\'o}pez}, {Liu}, {Lu}, {Nony}, {Valeille-Manet}, {Wu}, {{\'A}lvarez-Guti{\'e}rrez}, {Bonfand}, {Di Francesco}, {Nguyen-Luong}, {Olguin}, \& {Whitworth}}]{Towner2024ApJ...960...48T}
{Towner}, A.~P.~M., {Ginsburg}, A., {Dell'Ova}, P., {et~al.} 2024, \bibinfo{title}{{ALMA-IMF. IX. Catalog and Physical Properties of 315 SiO Outflow Candidates in 15 Massive Protoclusters},} \apj, 960, 48, \dodoi{10.3847/1538-4357/ad0786}

\bibitem[{F.~F.~S. {van der Tak} {et~al.}(2007){van der Tak}, {Black}, {Sch{\"o}ier}, {Jansen}, \& {van Dishoeck}}]{Van2007A&A...468..627V}
{van der Tak}, F.~F.~S., {Black}, J.~H., {Sch{\"o}ier}, F.~L., {Jansen}, D.~J., \& {van Dishoeck}, E.~F. 2007, \bibinfo{title}{{A computer program for fast non-LTE analysis of interstellar line spectra. With diagnostic plots to interpret observed line intensity ratios},} \aap, 468, 627, \dodoi{10.1051/0004-6361:20066820}

\bibitem[{K. {Wang} {et~al.}(2011){Wang}, {Zhang}, {Wu}, \& {Zhang}}]{Wang2011ApJ...735...64W}
{Wang}, K., {Zhang}, Q., {Wu}, Y., \& {Zhang}, H. 2011, \bibinfo{title}{{Hierarchical Fragmentation and Jet-like Outflows in IRDC G28.34+0.06: A Growing Massive Protostar Cluster},} \apj, 735, 64, \dodoi{10.1088/0004-637X/735/1/64}

\bibitem[{M. {Wienen} {et~al.}(2012){Wienen}, {Wyrowski}, {Schuller}, {Menten}, {Walmsley}, {Bronfman}, \& {Motte}}]{Wienen2012A&A...544A.146W}
{Wienen}, M., {Wyrowski}, F., {Schuller}, F., {et~al.} 2012, \bibinfo{title}{{Ammonia from cold high-mass clumps discovered in the inner Galactic disk by the ATLASGAL survey},} \aap, 544, A146, \dodoi{10.1051/0004-6361/201118107}

\bibitem[{M. {Wright} {et~al.}(2024){Wright}, {McGuire}, {Ginsburg}, {Hirota}, {Bally}, {Hwangbo}, {Bhadra}, {John}, \& {Dave}}]{Wright2024ApJ...974..150W}
{Wright}, M., {McGuire}, B.~A., {Ginsburg}, A., {et~al.} 2024, \bibinfo{title}{{Accretion and Outflow in Orion-KL Source I},} \apj, 974, 150, \dodoi{10.3847/1538-4357/ad7026}

\bibitem[{B. {Wu} {et~al.}(2017){Wu}, {Tan}, {Nakamura}, {Van Loo}, {Christie}, \& {Collins}}]{Wu2017ApJ...835..137W}
{Wu}, B., {Tan}, J.~C., {Nakamura}, F., {et~al.} 2017, \bibinfo{title}{{GMC Collisions as Triggers of Star Formation. II. 3D Turbulent, Magnetized Simulations},} \apj, 835, 137, \dodoi{10.3847/1538-4357/835/2/137}

\bibitem[{Y. {Wu} {et~al.}(2014){Wu}, {Liu}, \& {Qin}}]{Wu2014ApJ...791..123W}
{Wu}, Y., {Liu}, T., \& {Qin}, S.-L. 2014, \bibinfo{title}{{A Study of Dynamical Processes in the Orion KL Region Using ALMA{\textemdash}Probing Molecular Outflow and Inflow},} \apj, 791, 123, \dodoi{10.1088/0004-637X/791/2/123}

\bibitem[{Y. {Wu} {et~al.}(2004){Wu}, {Wei}, {Zhao}, {Shi}, {Yu}, {Qin}, \& {Huang}}]{Wu2004A&A...426..503W}
{Wu}, Y., {Wei}, Y., {Zhao}, M., {et~al.} 2004, \bibinfo{title}{{A study of high velocity molecular outflows with an up-to-date sample},} \aap, 426, 503, \dodoi{10.1051/0004-6361:20035767}

\bibitem[{F. {Xu} {et~al.}(2024){Xu}, {Wang}, {Liu}, {Tang}, {Evans}, {Palau}, {Morii}, {He}, {Sanhueza}, {Liu}, {Stutz}, {Zhang}, {Chen}, {Li}, {G{\'o}mez}, {V{\'a}zquez-Semadeni}, {Li}, {Mai}, {Lu}, {Liu}, {Chen}, {Li}, {Shi}, {Ren}, {Li}, {Garay}, {Bronfman}, {Dewangan}, {Juvela}, {Lee}, {Zhang}, {Yue}, {Wang}, {Ge}, {Jiao}, {Luo}, {Zhou}, {Tatematsu}, {Chibueze}, {Su}, {Sun}, {Ristorcelli}, \& {Toth}}]{Xu2024ApJS..270....9X}
{Xu}, F., {Wang}, K., {Liu}, T., {et~al.} 2024, \bibinfo{title}{{The ALMA Survey of Star Formation and Evolution in Massive Protoclusters with Blue Profiles (ASSEMBLE): Core Growth, Cluster Contraction, and Primordial Mass Segregation},} \apjs, 270, 9, \dodoi{10.3847/1538-4365/acfee5}

\bibitem[{Y. {Xu} {et~al.}(2024){Xu}, {Wang}, {Liu}, {Li}, {Li}, {Luo}, {Ou}, {Zheng}, \& {Liu}}]{Xu2024AJ....167..285X}
{Xu}, Y., {Wang}, J., {Liu}, S., {et~al.} 2024, \bibinfo{title}{{Dense Outflowing Molecular Gas in Massive Star-forming Regions},} \aj, 167, 285, \dodoi{10.3847/1538-3881/ad47c4}

\bibitem[{A.~Y. {Yang} {et~al.}(2018){Yang}, {Thompson}, {Urquhart}, \& {Tian}}]{Yang2018ApJS..235....3Y}
{Yang}, A.~Y., {Thompson}, M.~A., {Urquhart}, J.~S., \& {Tian}, W.~W. 2018, \bibinfo{title}{{Massive Outflows Associated with ATLASGAL Clumps},} \apjs, 235, 3, \dodoi{10.3847/1538-4365/aaa297}

\bibitem[{K. {Yang} {et~al.}(2025){Yang}, {Lu}, {Zhang}, {Liu}, {Ginsburg}, {Liu}, {Cheng}, {Feng}, {Liu}, {Zhang}, {Mills}, {Walker}, {Inutsuka}, {Battersby}, {Longmore}, {Tang}, {Kauffmann}, {Gu}, {Li}, {Luo}, {Kruijssen}, {Pillai}, {Qiao}, {Qiu}, \& {Shen}}]{Yang2025A&A...694A..86Y}
{Yang}, K., {Lu}, X., {Zhang}, Y., {et~al.} 2025, \bibinfo{title}{{ALMA observations of massive clouds in the central molecular zone: slim filaments tracing parsec-scale shocks},} \aap, 694, A86, \dodoi{10.1051/0004-6361/202453191}

\bibitem[{J. {Yuan} {et~al.}(2017){Yuan}, {Wu}, {Ellingsen}, {Evans}, {Henkel}, {Wang}, {Liu}, {Liu}, {Li}, \& {Zavagno}}]{Yuan2017ApJS..231...11Y}
{Yuan}, J., {Wu}, Y., {Ellingsen}, S.~P., {et~al.} 2017, \bibinfo{title}{{High-mass Starless Clumps in the Inner Galactic Plane: The Sample and Dust Properties},} \apjs, 231, 11, \dodoi{10.3847/1538-4365/aa7204}

\bibitem[{L.~A. {Zapata} {et~al.}(2011){Zapata}, {Loinard}, {Schmid-Burgk}, {Rodr{\'\i}guez}, {Ho}, \& {Patel}}]{Zapata2011ApJ...726L..12Z}
{Zapata}, L.~A., {Loinard}, L., {Schmid-Burgk}, J., {et~al.} 2011, \bibinfo{title}{{Discovery of an Expanding Molecular Bubble in Orion BN/KL},} \apjl, 726, L12, \dodoi{10.1088/2041-8205/726/1/L12}

\bibitem[{L.~A. {Zapata} {et~al.}(2012){Zapata}, {Rodr{\'\i}guez}, {Schmid-Burgk}, {Loinard}, {Menten}, \& {Curiel}}]{Zapata2012ApJ...754L..17Z}
{Zapata}, L.~A., {Rodr{\'\i}guez}, L.~F., {Schmid-Burgk}, J., {et~al.} 2012, \bibinfo{title}{{ALMA Observations of the Outflow from Source I in the Orion-KL Region},} \apjl, 754, L17, \dodoi{10.1088/2041-8205/754/1/L17}

\bibitem[{L.~A. {Zapata} {et~al.}(2019){Zapata}, {Ho}, {Guzm{\'a}n Ccolque}, {Fern{\'a}ndez-Lop{\'e}z}, {Rodr{\'\i}guez}, {Bally}, {Sanhueza}, {Palau}, \& {Saito}}]{Zapata2019MNRAS.486L..15Z}
{Zapata}, L.~A., {Ho}, P. T.~P., {Guzm{\'a}n Ccolque}, E., {et~al.} 2019, \bibinfo{title}{{G5.89: an explosive outflow powered by a proto-stellar merger?},} \mnras, 486, L15, \dodoi{10.1093/mnrasl/slz051}

\bibitem[{L.~A. {Zapata} {et~al.}(2020){Zapata}, {Ho}, {Fern{\'a}ndez-L{\'o}pez}, {Ccolque}, {Rodr{\'\i}guez}, {Reyes-Vald{\'e}s}, {Bally}, {Palau}, {Saito}, {Sanhueza}, {Rivera-Ortiz}, \& {Rodr{\'\i}guez-Gonz{\'a}lez}}]{Zapata2020ApJ...902L..47Z}
{Zapata}, L.~A., {Ho}, P. T.~P., {Fern{\'a}ndez-L{\'o}pez}, M., {et~al.} 2020, \bibinfo{title}{{Confirming the Explosive Outflow in G5.89 with ALMA},} \apjl, 902, L47, \dodoi{10.3847/2041-8213/abbd3f}

\bibitem[{Q. {Zhang} {et~al.}(2005){Zhang}, {Hunter}, {Brand}, {Sridharan}, {Cesaroni}, {Molinari}, {Wang}, \& {Kramer}}]{Zhang2005ApJ...625..864Z}
{Zhang}, Q., {Hunter}, T.~R., {Brand}, J., {et~al.} 2005, \bibinfo{title}{{Search for CO Outflows toward a Sample of 69 High-Mass Protostellar Candidates. II. Outflow Properties},} \apj, 625, 864, \dodoi{10.1086/429660}

\bibitem[{Q. {Zhang} {et~al.}(2015){Zhang}, {Wang}, {Lu}, \& {Jim{\'e}nez-Serra}}]{Zhang2015ApJ...804..141Z}
{Zhang}, Q., {Wang}, K., {Lu}, X., \& {Jim{\'e}nez-Serra}, I. 2015, \bibinfo{title}{{Fragmentation of Molecular Clumps and Formation of a Protocluster},} \apj, 804, 141, \dodoi{10.1088/0004-637X/804/2/141}

\bibitem[{H. {Zinnecker} \& H.~W. {Yorke}(2007){Zinnecker} \& {Yorke}}]{Zinnecker2007ARA&A..45..481Z}
{Zinnecker}, H., \& {Yorke}, H.~W. 2007, \bibinfo{title}{{Toward Understanding Massive Star Formation},} \araa, 45, 481, \dodoi{10.1146/annurev.astro.44.051905.092549}

\end{thebibliography}
\bibliographystyle{aasjournalv7}



\end{document}